\documentclass[3p,times]{elsarticle}

\usepackage{amssymb}
\usepackage{graphics}
\usepackage{eurosym}
\usepackage{amsfonts}
\usepackage{amssymb}
\usepackage{amsmath}
\usepackage{booktabs}
\usepackage{graphicx}
\usepackage{graphicx,epsfig}

\begin{document}
	
	\begin{frontmatter}
		
		
		
		\title{Anisotropic relativistic fluid spheres with a linear equation of state}
		
		
		\author{A. K. Prasad}
		\ead{amitkarun5@gmail.com}
		\author{J. Kumar}
		\ead{jitendark@gmail.com}
		\address{Department of Mathematics, Central University of Jharkhand, Ranchi-835205, India}

\begin{abstract}
 In this work, we present a class of relativistic and well-behaved solution to Einstein's field equations for anisotropic matter distribution. We perform our analysis by using the Buchdahl ansatz for the metric function $ g_{rr} $.   Three different classes of new exact solution are found for anisotropy factor $ \Delta $. We have analyzed our model with various physical aspects such as pressure (radial as well as transverse), energy density, anisotropy
 factor, mass, compactness parameter, adiabatic index $ \gamma $ and  surface redshift. A graphical analysis of energy conditions, TOV equation and causality condition indicates the model are well behaved. The physical  acceptability of the model has verified by considering compact objects with similar mass and radii, such as 4U 1820-30,Vela X-1, PSR J1614-2230, LMC X-4, SMC X-1, 4U 1538-52, Her X-1, Cgy X-2, PSR B1913+16 and PSR J1903+327.
\end{abstract}

\begin{keyword}
	Anisotropic Fluids\sep Compact stars\sep General Relativity
\end{keyword} 

\end{frontmatter}

\section{Introduction}
Due to the highly nonlinear behavior of Einstein's field equations, one of the most important problems in General Relativity is to obtain exact solution of Einstein equations. Exact solution play important role in the development of many areas of gravitational significance such as the study of effects of gravitation in solar system, black hole,star,gravitational collapse,cosmology and dynamic of early universe. Schwarzschild \cite{schwarzschild}, Tolman \cite{35} and Oppenheimer and Volkoff \cite{36} have deal with self-gravitating isotropic fluid spheres, set up  two  approaches that can be followed in order to solve the field equations. The first one consists in making a suitable assumption of the metric functions or energy density. This leads to the finding unknown variables, which are the isotropic pressure and the other metric potential. However, in the framework of this method not always is possible to obtain an exact solution (sometimes one obtains nonphysical pressure-density configurations). The second approach deal with an equation of state which is integrated (iteratively from the center of the compact object) until the pressure vanished indicating that the object surface has been reached. As before, this scheme also presents some drawbacks since it does not always lead to a closed form of the solutions. Many projects are suggested by Ivanov\cite{Ivanov} for constructing charged fluid spheres. The perfect fluid solution was found Gupta and Kumar\cite{2005a}.\par 
On the other hand, a stellar configuration sometimes have not followed  the isotropic condition at all (equal radial $ p_r $ and tangential $ p_t $ pressure). In fact, the theoretical studies by Canuto \cite{Canuto,Canuto1,Canuto2}, Canuto et. al \cite{Canuto3,Canuto4,Canuto5} and Ruderman \cite{Ruderman}, observed that when the matter density is higher than the nuclear density, it may be anisotropic in nature and must be treated relativistically. So, rest the isotropic condition and allow the presence of anisotropy (it leads to unequal radial and tangential pressure $ p_r \neq p_t $ ) within the stellar configuration represent a more realistic situation in the astrophysical sense. Furthermore, the pioneering contribution to local anisotropic properties by Bowers and Liang's  \cite{Bowers} for static spherically symmetric and relativistic configurations, gave rise an
extensive studies within this framework, specifically studies focused on the dynamical incidence of the anisotropies in the arena of equilibrium and stability on collapsed structures \cite{Heintzmann,Cosenza,Cosenza1}. Moreover, as Mak and Harko have contended \cite{Mak}, anisotropy can merge in various settings such as: the presence of a strong center or by the existence of type 3A superfluid \cite{Kippen}, pion condensation \cite{Sawyer} or various types of regime transitions\cite{Sokolov}.\par
In different circumstance, the spherical symmetry also allows a more general anisotropic fluid configuration with an equation of state(EoS).
 If one knows the EoS of the material composition of a compact star, then  one can easily integrate the Tolman-Oppenheimer-Volkoff (TOV) equations to extract the geometrical information of the compact star. For example, Ivanov \cite{Ivanov} used a linear EoS for charged static spherically
symmetric perfect fluid solutions. Sharma and Maharaj \cite{sharma} have been extended this condition   for finding an exact solution to the Einstein field equations with an anisotropic matter distribution. Herrera and Barreto \cite{Barreto} had
considered polytropic stars with anisotropic pressure. Many solutions \cite{Chowdhury,Varela,Ivanov1,Nasim,Isayev,Ivanov2}  of Einstein's equations have been found for anisotropic fluid distribution with different EoS, but, in case the EoS of the material composition of a compact star is not yet known, except some phenomenological assumptions, for example, one can introduce a suitable
metric ansatz for one of the metric functions to analyze the physical features of the star. Initially, this type of method was 
proposed by Vaidya-Tikekar \cite{Vaidya}, and Tikekar \cite{Tikekar} introduced an approach of assigning different geometries with physical 3-spaces (see \cite{Komathiraj,Patel,Mukherjee,2005a} and references therein).
 Finch and Skea \cite{Finch} have also considered such type of metric ansatz that satisfying all criteria of physical acceptability
according to Delgaty and Lake \cite{Delgaty}. This type of  the problem in which to finding the equilibrium configuration of a stellar
structure for anisotropic fluid distribution has been found in \cite{Herrera3,Paul,Paul1,Maharaj1} . Moreover, our stellar model incorporates a family of new
solutions for a static spherically symmetric anisotropic fluid structure with the help of Buchdahl metric potential. In this paper, we consider analogue objects with
similar mass and radii, such as PSR J1614-2230 \cite{Demorest} 4U 1538-52 \cite{Rawls1}, LMC X-4 \cite{Kelley}, SMC X-1 \cite{Rawls1}, Her X-1 \cite{Abubekerov},  Vela X-1 \cite{Rawls1}, PSR J1903+327 \cite{Freire}, PSR B1913+ 16 \cite{Weisberg}, 4U 1820-30 \cite{guver}, Cyg X-1 \cite{casares}.   
\par
The remainder of the article is arranged as follows: Section
2 presents the Einstein's field equations for anisotropic
matter distributions and the approach followed in order to
solve them, in section 4 we match the obtained model with
the exterior spacetime given by Schwarzschild metric, in order
to obtain the constant parameters. In section 5 we analyze the causality condition. Section 6 is devoted to the study of the equilibrium
via Tolman-Oppenheimer-Volkoff (TOV) equation. Sections 7 and 9 describe adiabatic index  and conclusion respectively. 

\section{Field Equation}
Let us consider the static spherically symmetric metric in Schwarzschild coordinates \begin{eqnarray}
 ds^{2}=-e^{\lambda(r)}dr^{2} -r^{2}(d\theta^{2}+\sin^{2}\theta d\phi^{2})+e^{\nu(r)}dt^{2}\label{1}
\end{eqnarray} 
where $\lambda(r)$ and $\nu(r)$ are the functions of radial coordinate $r.$ If (\ref {1}) describes anisotropic matter distribution  then the space-time(\ref {1}) has to satisfy satisfy the energy-momentum tensor  \begin{eqnarray}  T^{i}_{j} = (\rho+p_t)u^{i}u_{j}-p_t\delta^{i}_{j}+(p_r-p_t)\chi^{i}\chi_{j}\label{2}
\end{eqnarray}
with $ u^{i}u_{j}= - \chi^{i}\chi_{j} = 1 $, where $\nu^{i},\chi^{i},\rho,p,p_t$ are the fluid four velocity,unit space-like vector, energy density, radial and transverse  pressures respectively. Thus, the Einstein field equation for line element(\ref{1}) with respect to energy-momentum tensor(\ref{2}) reduce to (Landau and Lifshitz\cite{Landau}) \begin{eqnarray}
\kappa p_r &=&\dfrac{\nu'}{r}e^{-\lambda}-\dfrac{(1-e^{-\lambda})}{r^{2}}\label{3} \\
\kappa p_t&=&\bigg(\dfrac{\nu''}{2}-\dfrac{\lambda'\nu'}{4}+\dfrac{\nu^{'2}}{4}+\dfrac{\nu'-\lambda'}{2r}\bigg)e^{-\lambda}\label{4}\\
\kappa\rho&=&\dfrac{\lambda'}{r}e^{-\lambda}+\dfrac{(1-e^{-\lambda})}{r^{2}}\label{5}
\end{eqnarray}
where $ \kappa=8\pi $ and the prime denotes the derivative with respect to $r$. Here we will work on $ c=G=1 $, where $ c $ is the speed of light and $ G $ is the gravitational constant.Consequently, $ \Delta = p_t-p_r $ is denoted as the
anisotropy factor according to Herrera and Leon \cite{Leon},and it measures the pressure anisotropy of the fluid. It is to be noted that at the origin of the stellar configuration
$ \Delta=0 $, i.e., $ p_t=p_r $ is a particular case of an isotropic pressure. Using eqs.(\ref{3}) and (\ref{4}), one can obtain the simple form of anisotropic factor($ \Delta=0 $) , which yields
\begin{equation}
\Delta =\kappa \, (p_{t} -\, p_{r} ) = e^{-\lambda } \left[\frac{\nu''}{2} -\frac{\lambda' \nu'}{4} +\frac{{\nu'}^{2} }{4} -\frac{\nu'+\lambda '}{2r} -\frac{1}{r^{2} } \right]\,  +\frac{1}{r^{2} }. \label{6}
\end{equation}
The anisotropic pressure is repulsive if $ p_t > p_r $ and attractive if $ p_t < p_r $ of the stellar model.In the system of equs.(\ref{3})-(\ref{5}), we have three equations with five unknowns.Thus, the system of equations is undetermined. Hence,to solve these equation we have to reduce the
number of unknown functions.
\section{Exact Solution of the Model for Anisotropic Stars} 
 In this section we solve the system of equs.(\ref{3})-(\ref{5}). For this we employ a well known metric ansatz\cite{Buchdahl} that encompasses almost all the known solutions to the static Einstein equations with a perfect fluid source which is given by \begin{eqnarray} e^{\lambda}=\frac{K\,(1+Cr^2)}{K+Cr^2}, ~~~~~ \textrm{with} ~~~ K < 0~~~ \textrm{and}~~ C > 0,\label{7} \end{eqnarray} where $ K $ and $ C $ are constant parameters that characterize the geometry of the star. Initially the Buchdahl\cite{Buchdahl} have considered above metric potential to study a relativistic compact star. Note that above metric potential is free from singularity at $ r = 0 $ and the metric coefficient is $ e^{\lambda(0)} = 1.$ Moreover it provides a  monotonic decreasing energy density as follows
 \begin{eqnarray}
   \frac{\kappa\rho}{C} = \dfrac{(K-1)(3+Cr^{2})}{K(1+Cr^{2})^{2}}\label{8}\end{eqnarray} 
   \par
\begin{figure}[]
 \begin{center}
 \includegraphics[width=5cm]{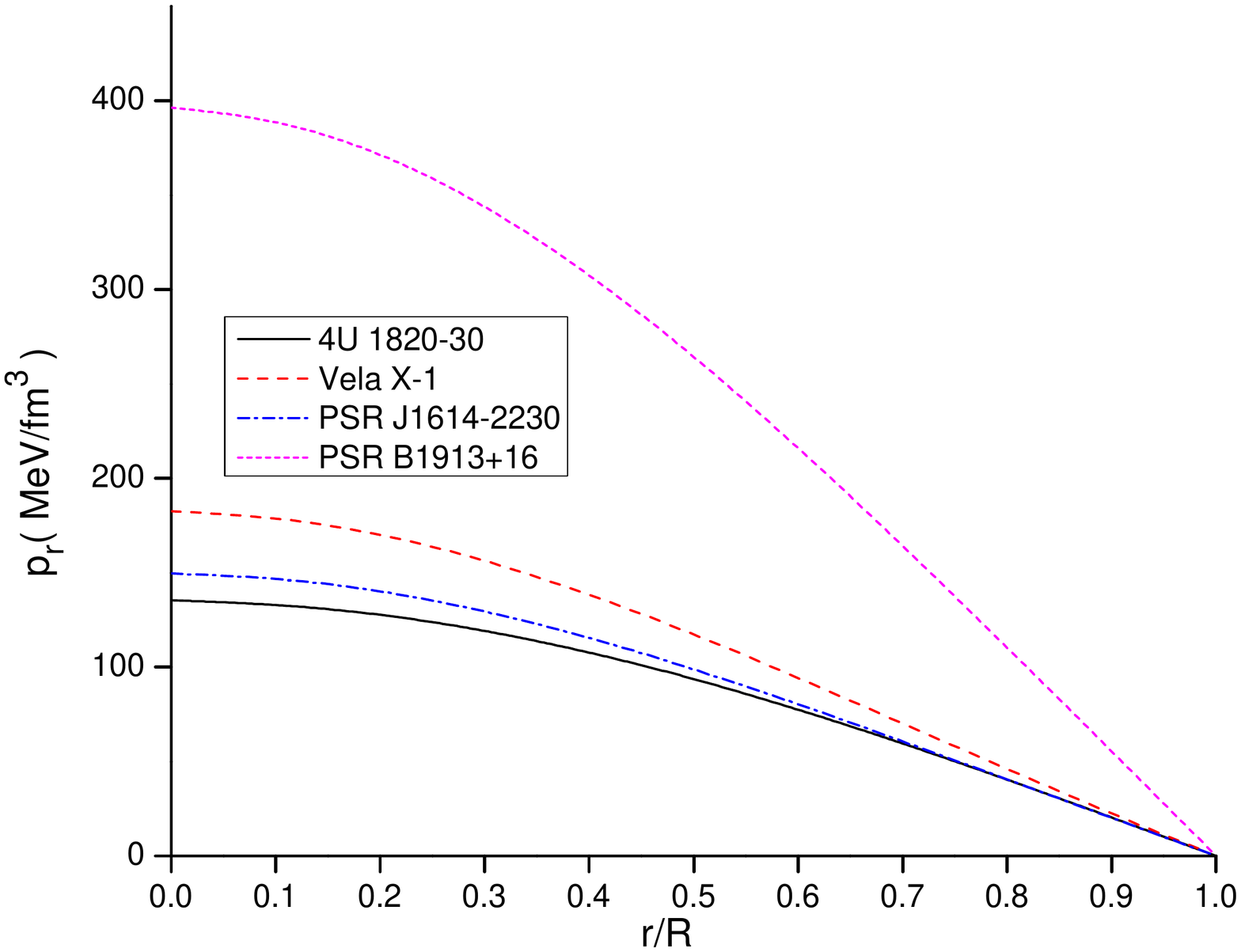}\includegraphics[width=5cm]{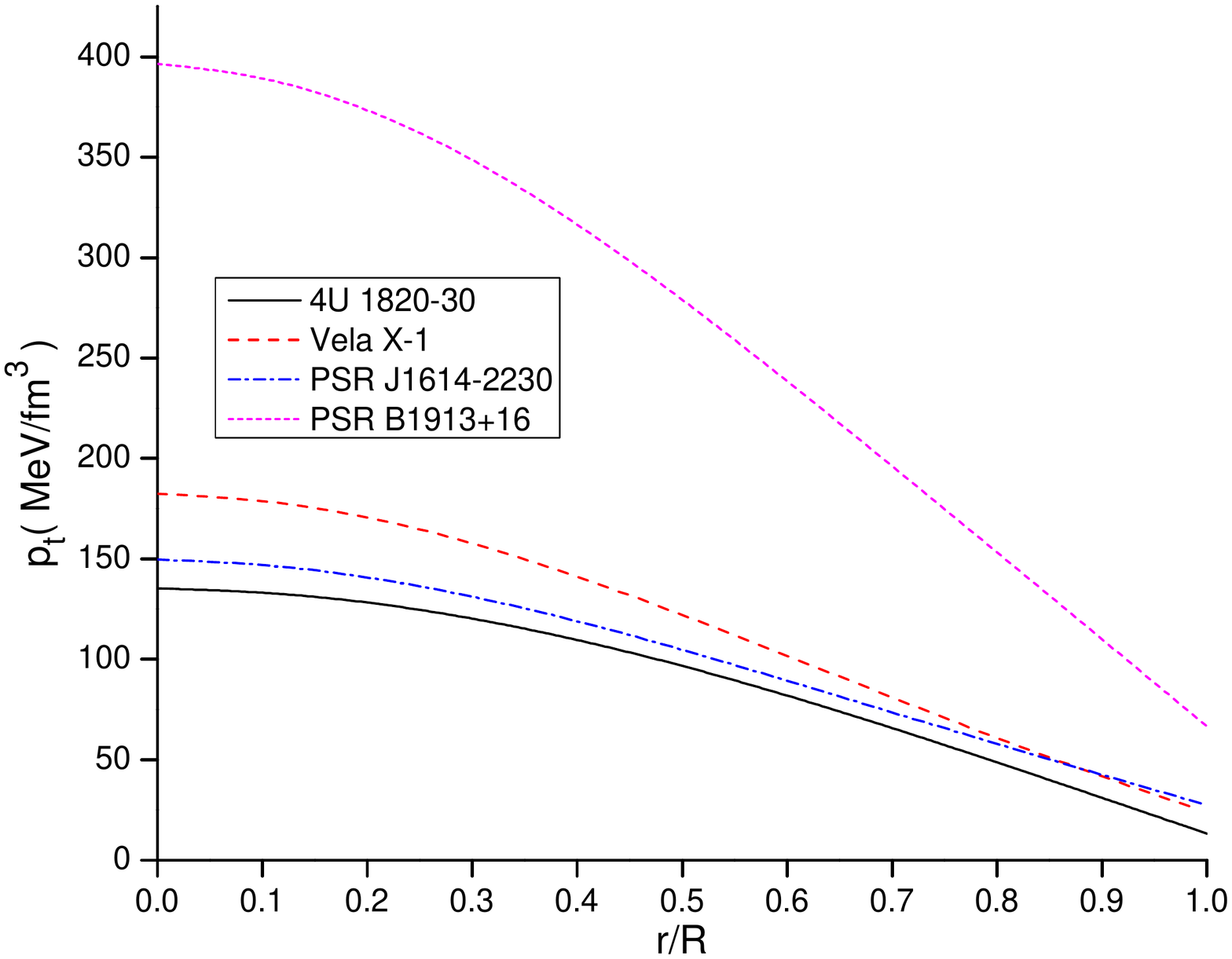}\includegraphics[width=5cm]{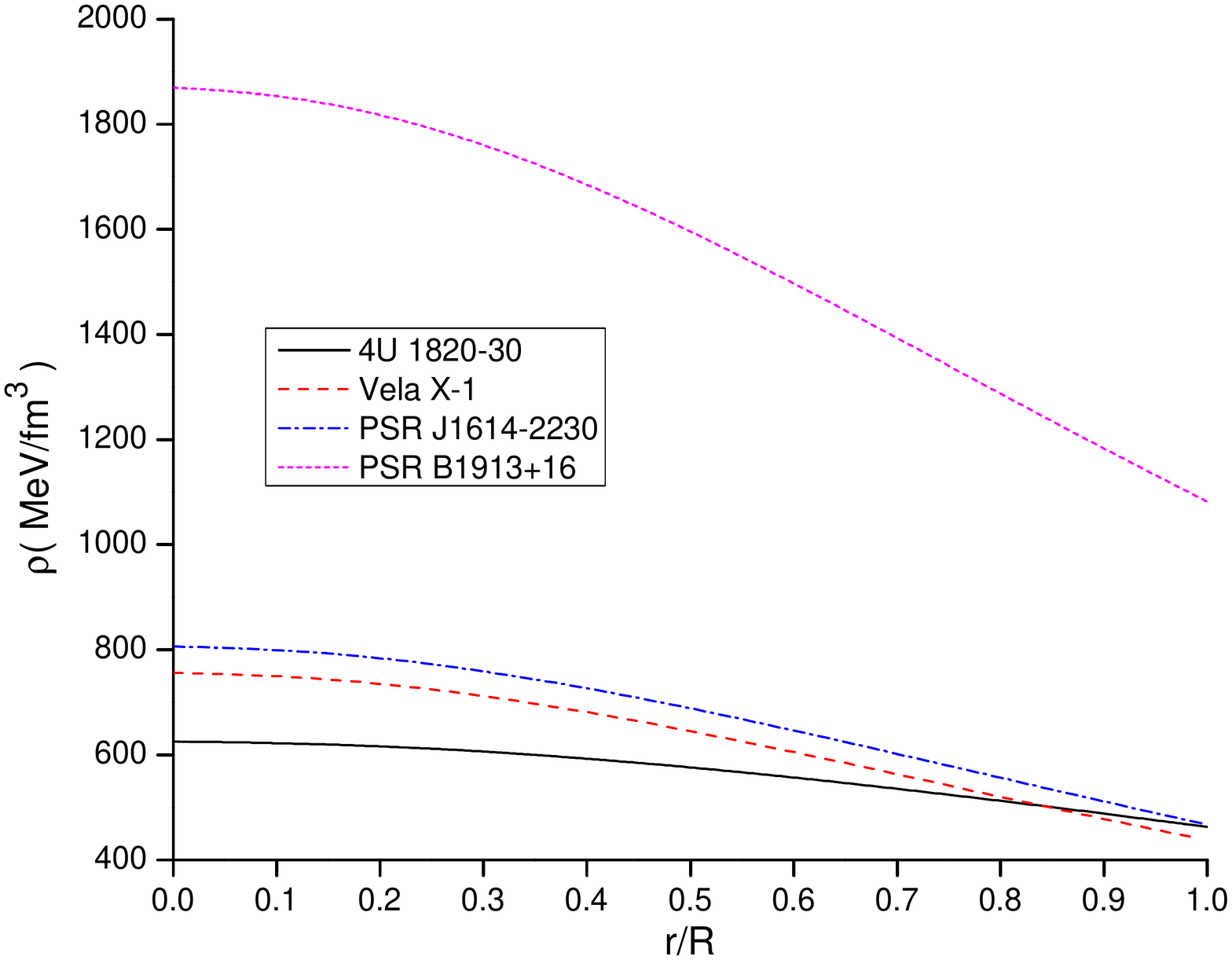}\\\includegraphics[width=5cm]{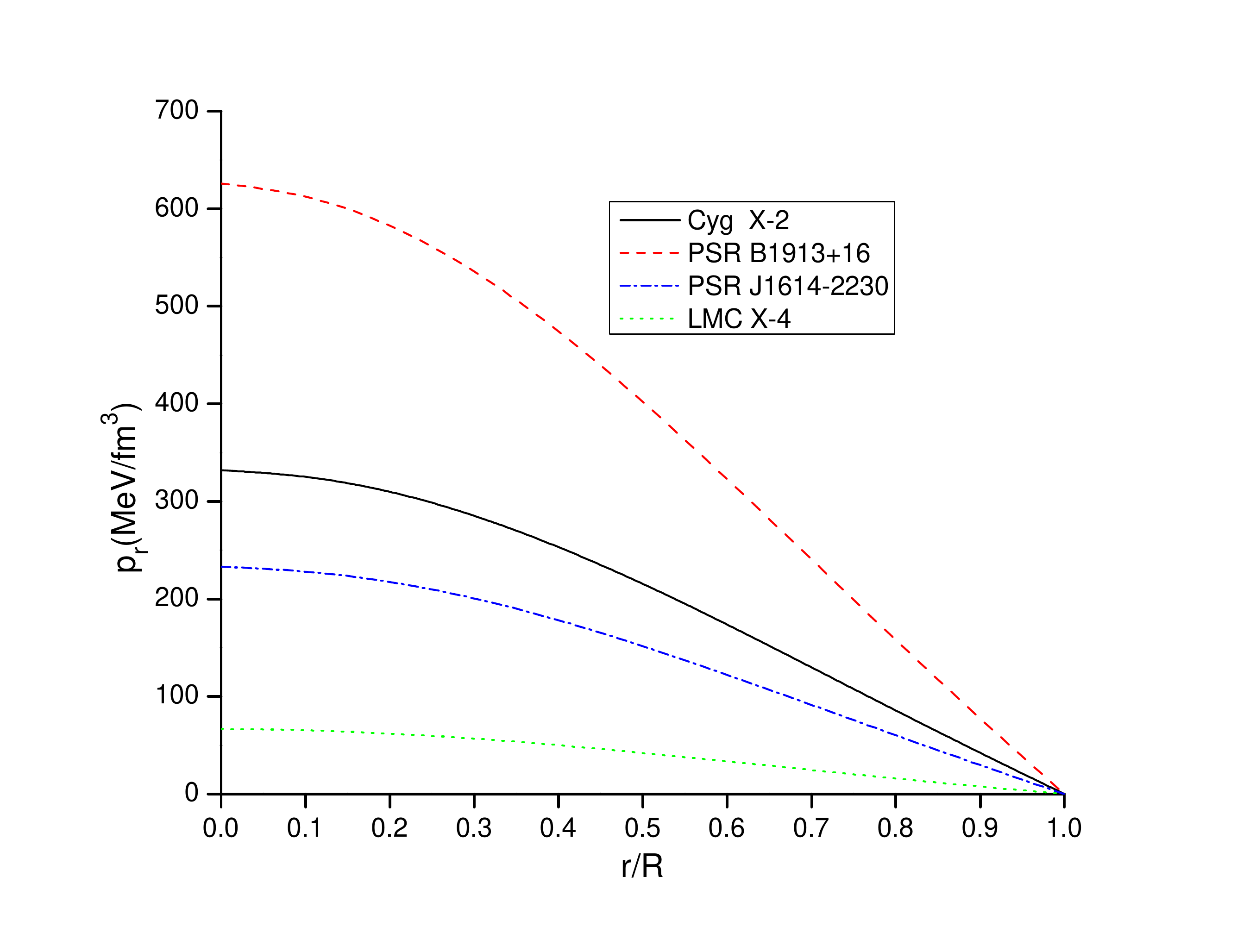}\includegraphics[width=5cm]{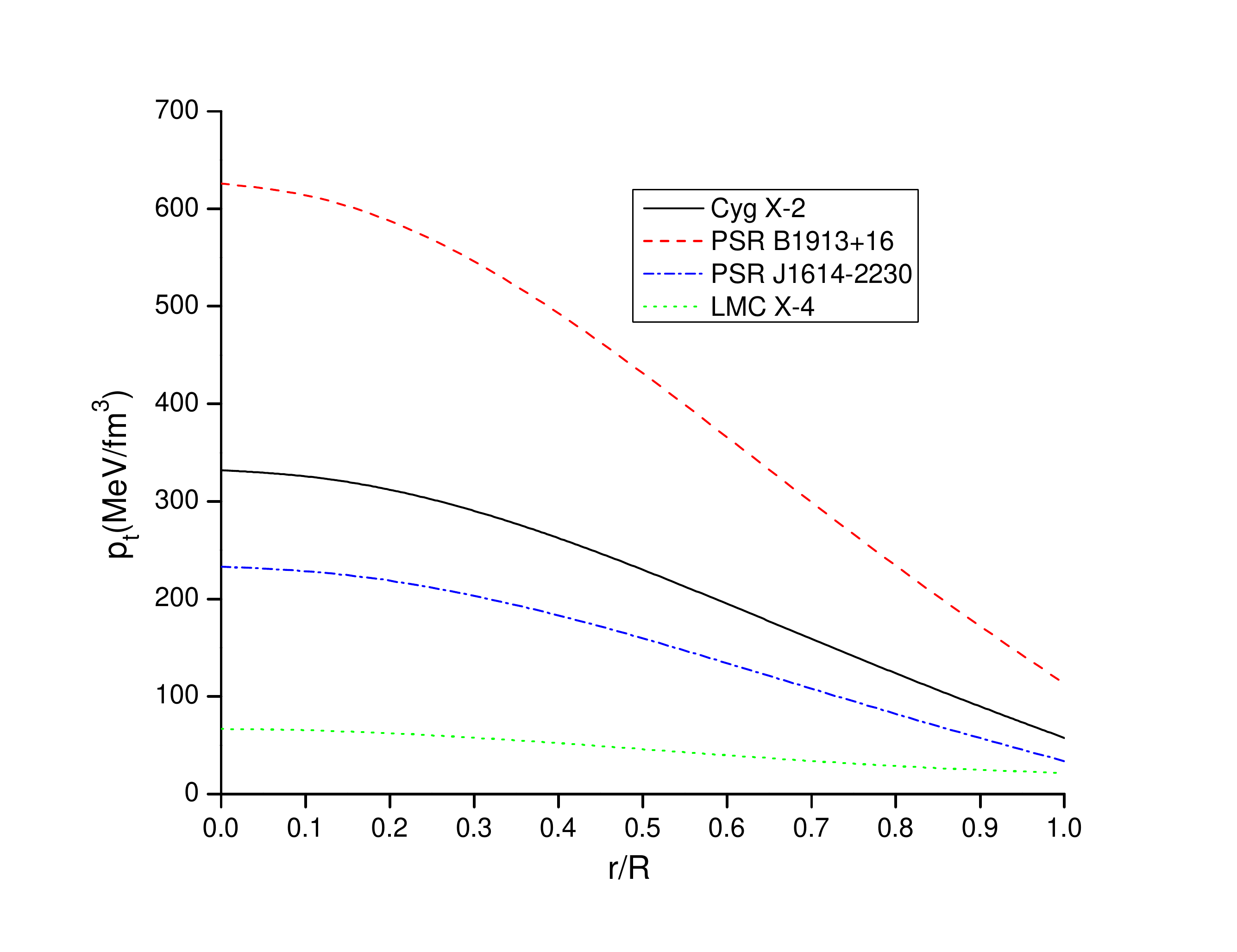}\includegraphics[width=5cm]{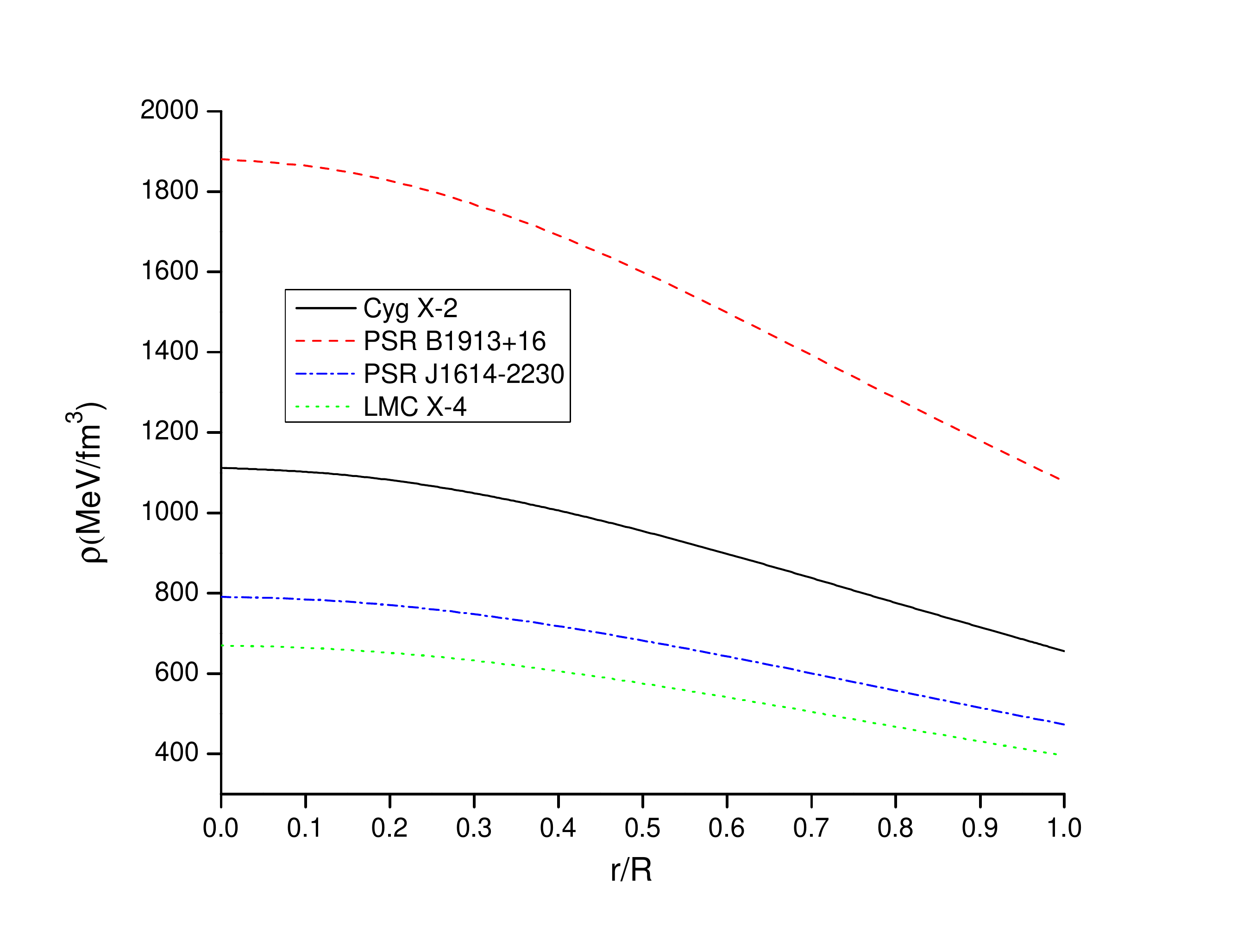}\\\includegraphics[width=5cm]{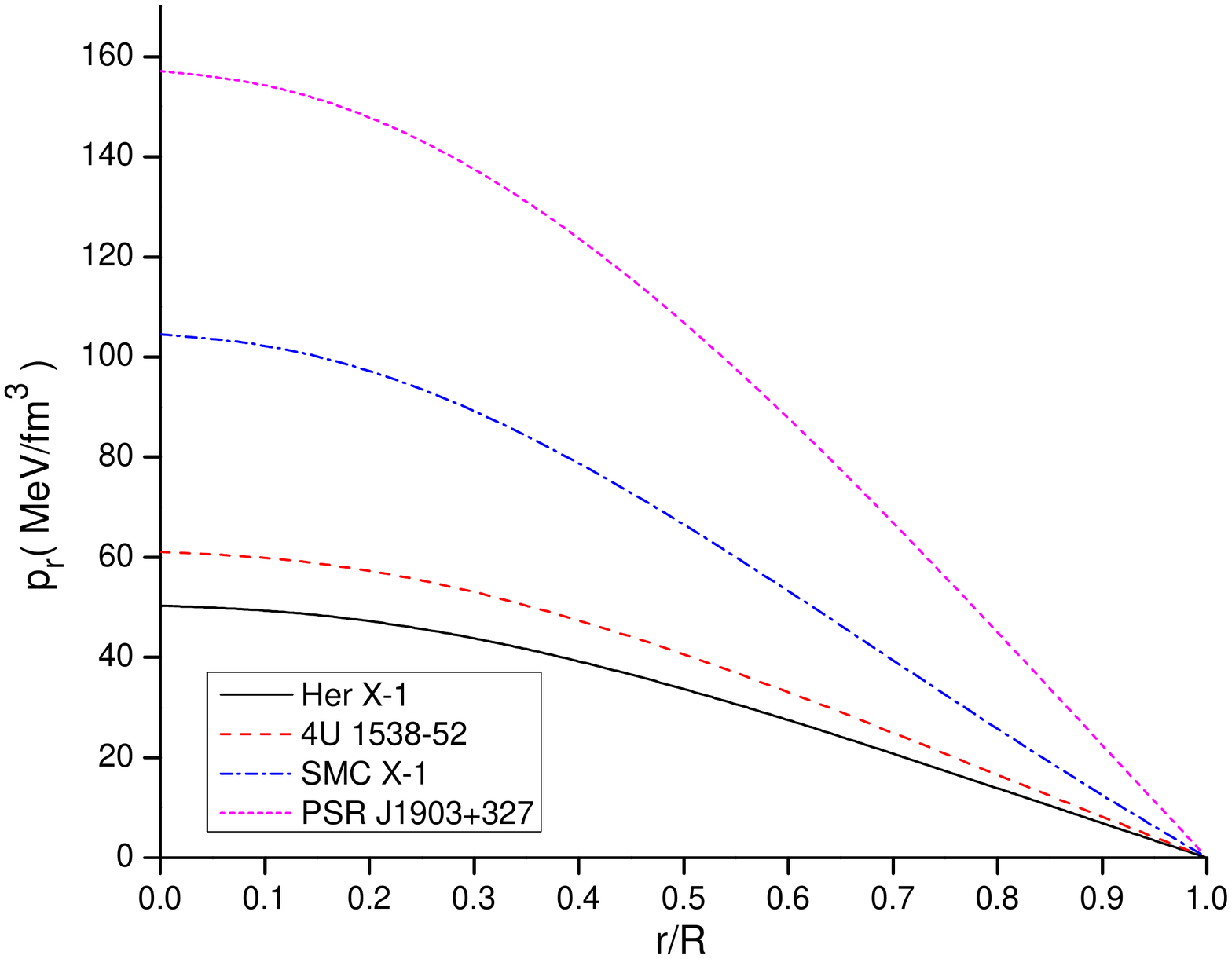}\includegraphics[width=5cm]{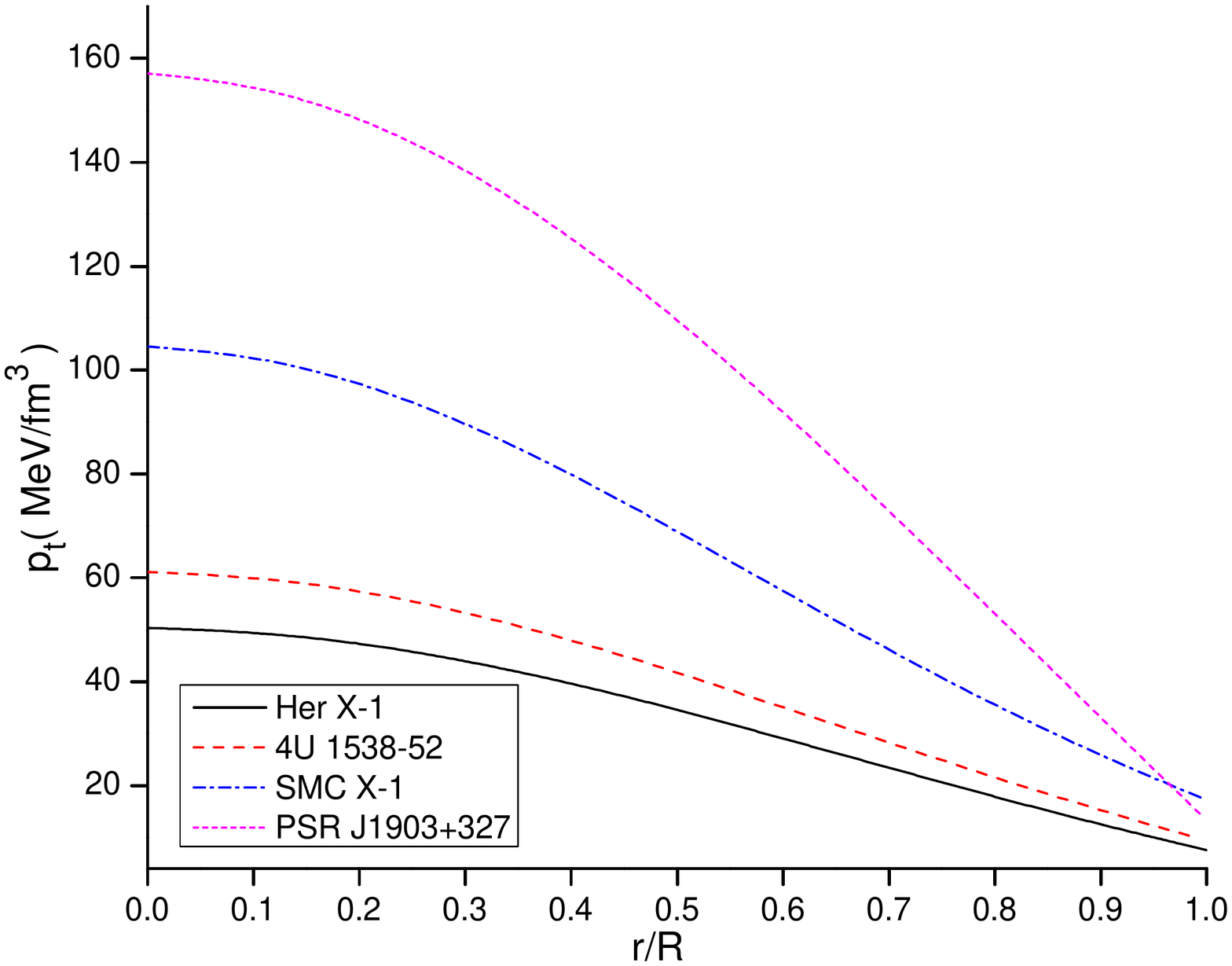}\includegraphics[width=5cm]{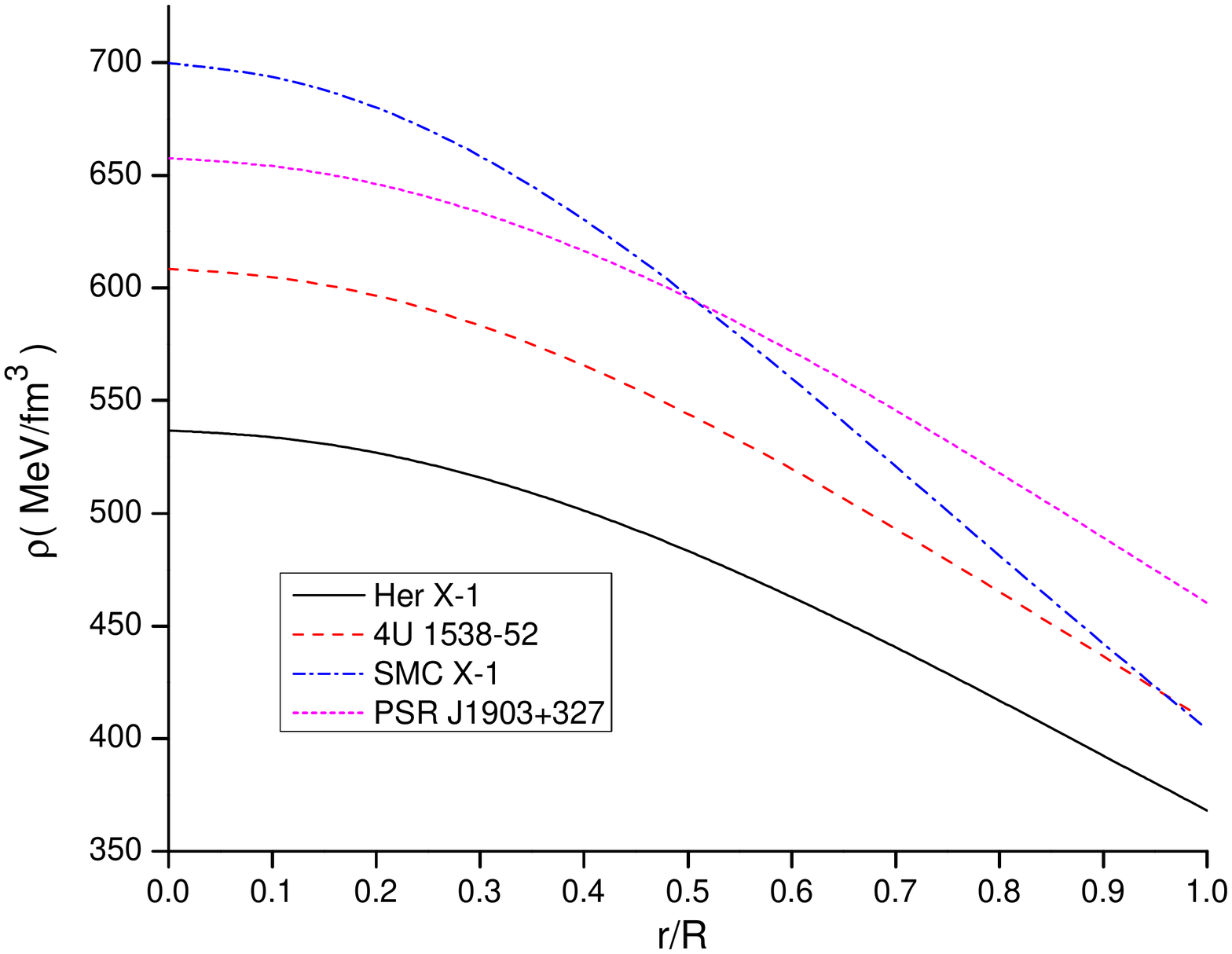}
 \caption{Behaviour of radial pressure, tangential pressure and density vs. fractional radius $ r/R $  for case I(first row), case II(second row) and case III(third row).
 The values of the parameter which we have used for graphical presentation are:(i) $K = -0.4995,~ C = 0.0023km^{-2}$ for 4U 1820-30 (case I) ; (ii) K = -1.097, $C = 0.0044 km^{-2}$  for Vela X-1 (case I); (iii) K = -0.901, $C = 0.0042km^{-2}$ for PSR J1614-2230 (case I); (iv) K = -0.798, $C = 0.0092km^{-2}$ for PSR B1913+16 (case I); (v) K = -.8196, $C = 0.0055km^{-2},\,\,\alpha = -0.64$  for Cgy X-2 (case II); (vi) K =-0.82, $C = 0.0094km^{-2}, \,\,\,\alpha = -0.81 $  for PSR B1913+16 (case II); (vii) K = -0.8202, $C = 0.0039km^{-2}, \,\,\,\alpha = -0.36$  for PAR J1614-2230 (case II); (viii) K = -2.977, $C = 0.0055km^{-2},\,\,\,\alpha = -2.25$ for LMC X-4 1937+21 (case II); (ix) K=-1.9779, $C = 0.0039 km^{-2},\,\,\,\alpha =0.0081 $ for Her X-1 (case III); (x) K=-1.9779, $C = 0.0045 km^{-2}, \,\,\,\alpha =0.0064 $ for 4U1538-52 (case III); (xi) K=-1.9776, $C = 0.0052 km^{-2},\,\,\,\alpha =0.0049 $ for SMC X-1 (case III); (xii)  K=-0.6, $C = 0.0027 km^{-2},\,\,\,\alpha =0.16 $ for PSR J1903+327 (case III).See Table \ref{t1} for more details}\label{f1}
 \end{center}
 \end{figure}

\begin{figure}[]
  \begin{center}
  \includegraphics[width=5cm]{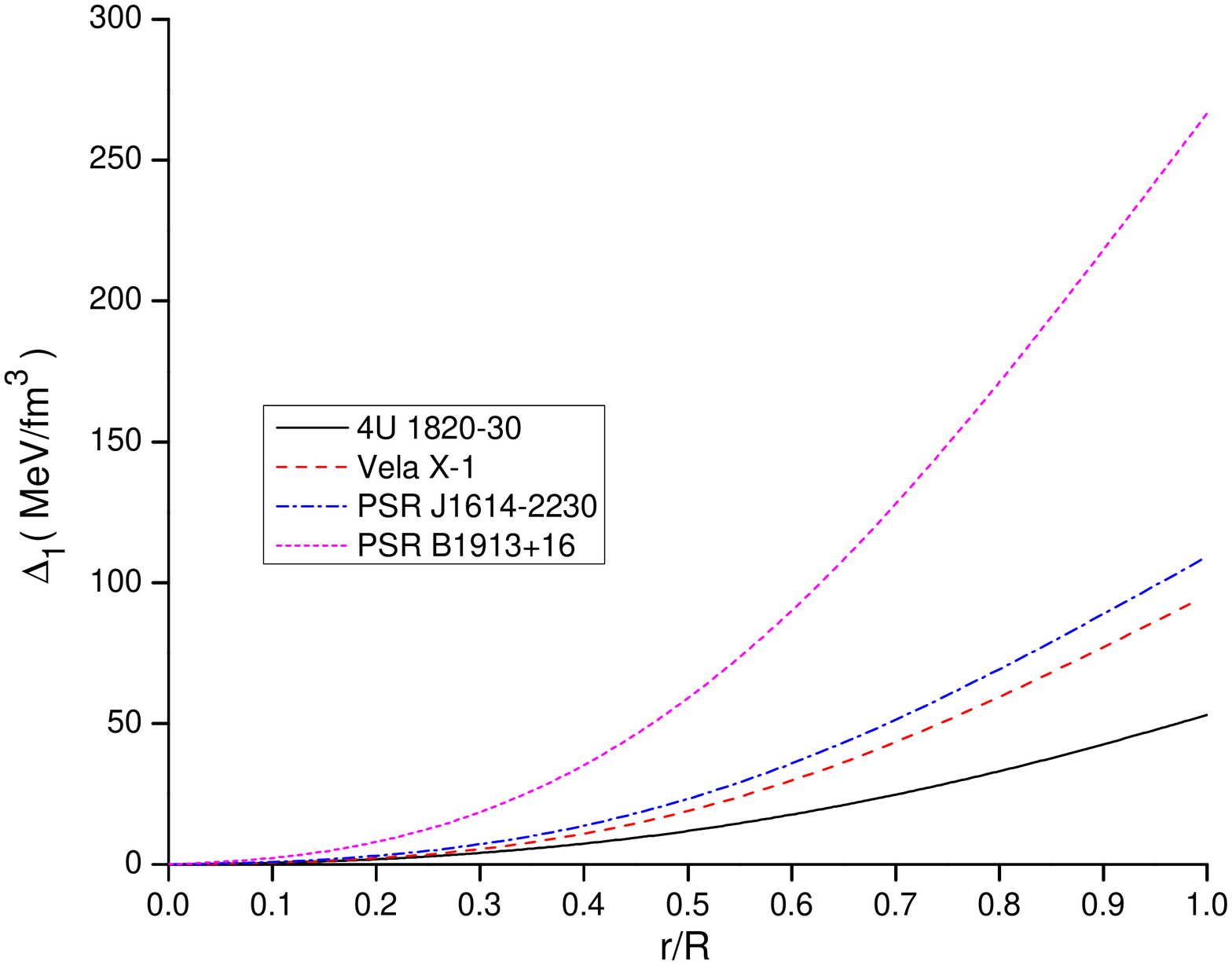}\includegraphics[width=5cm]{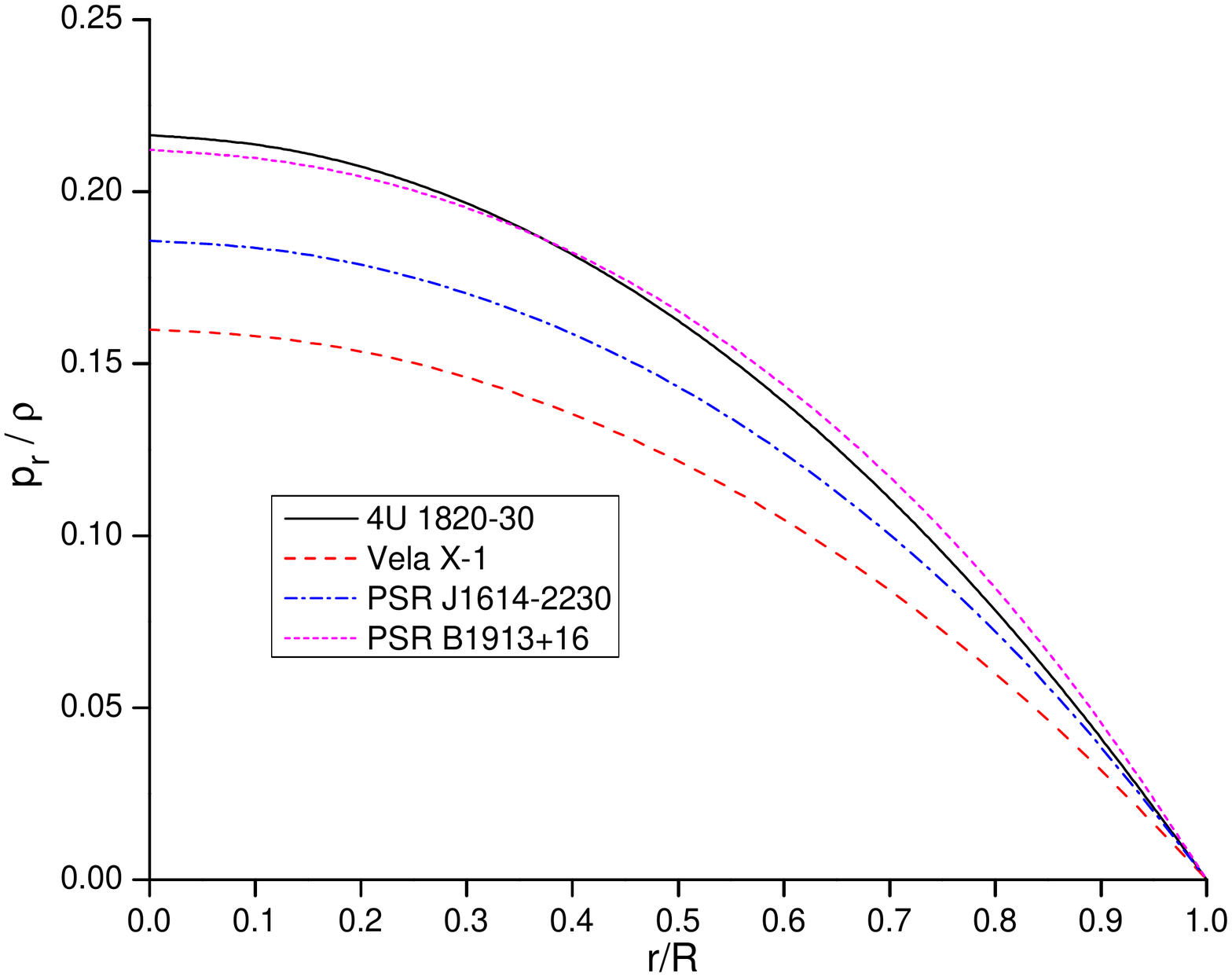}\includegraphics[width=5cm]{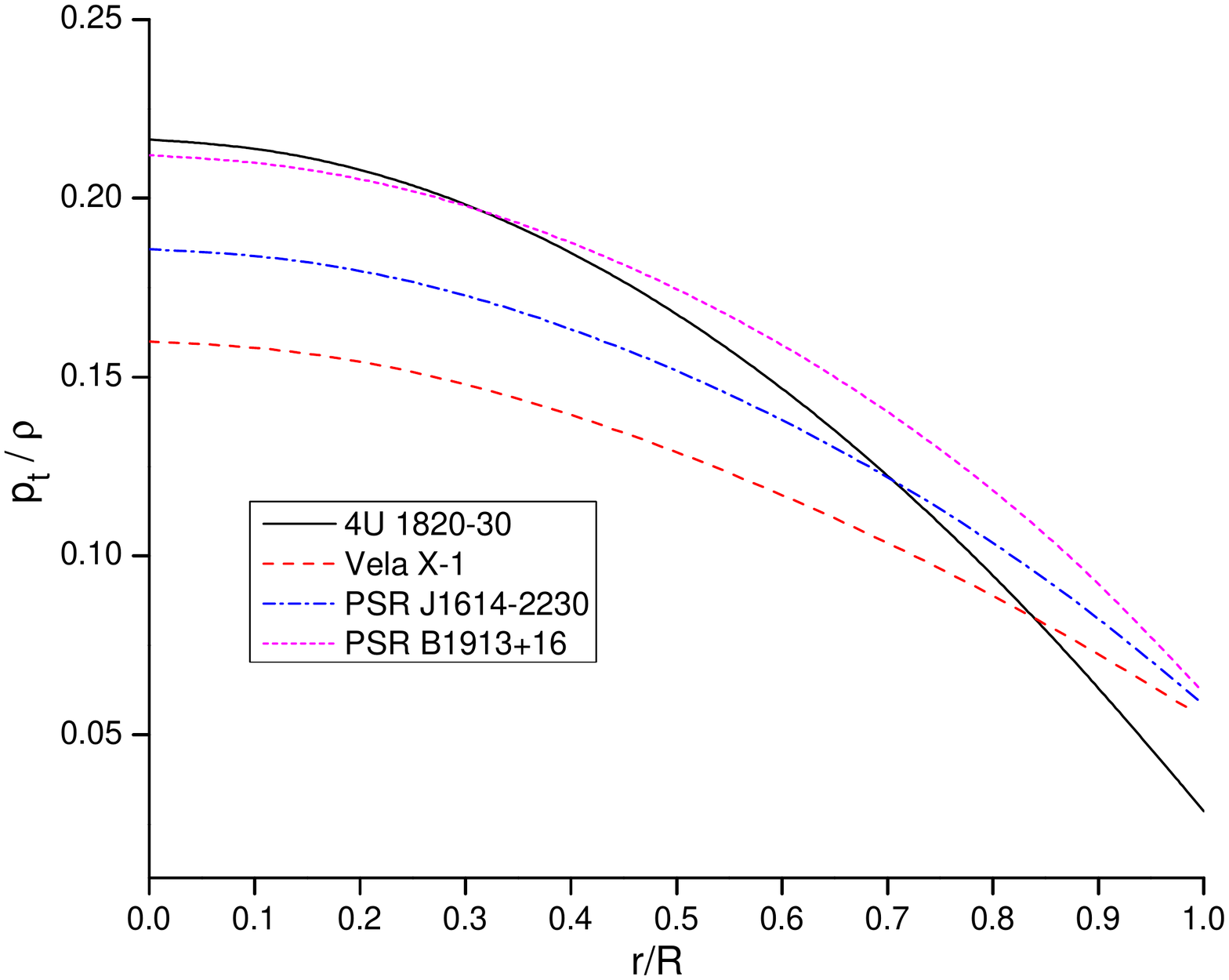}\\\includegraphics[width=5cm]{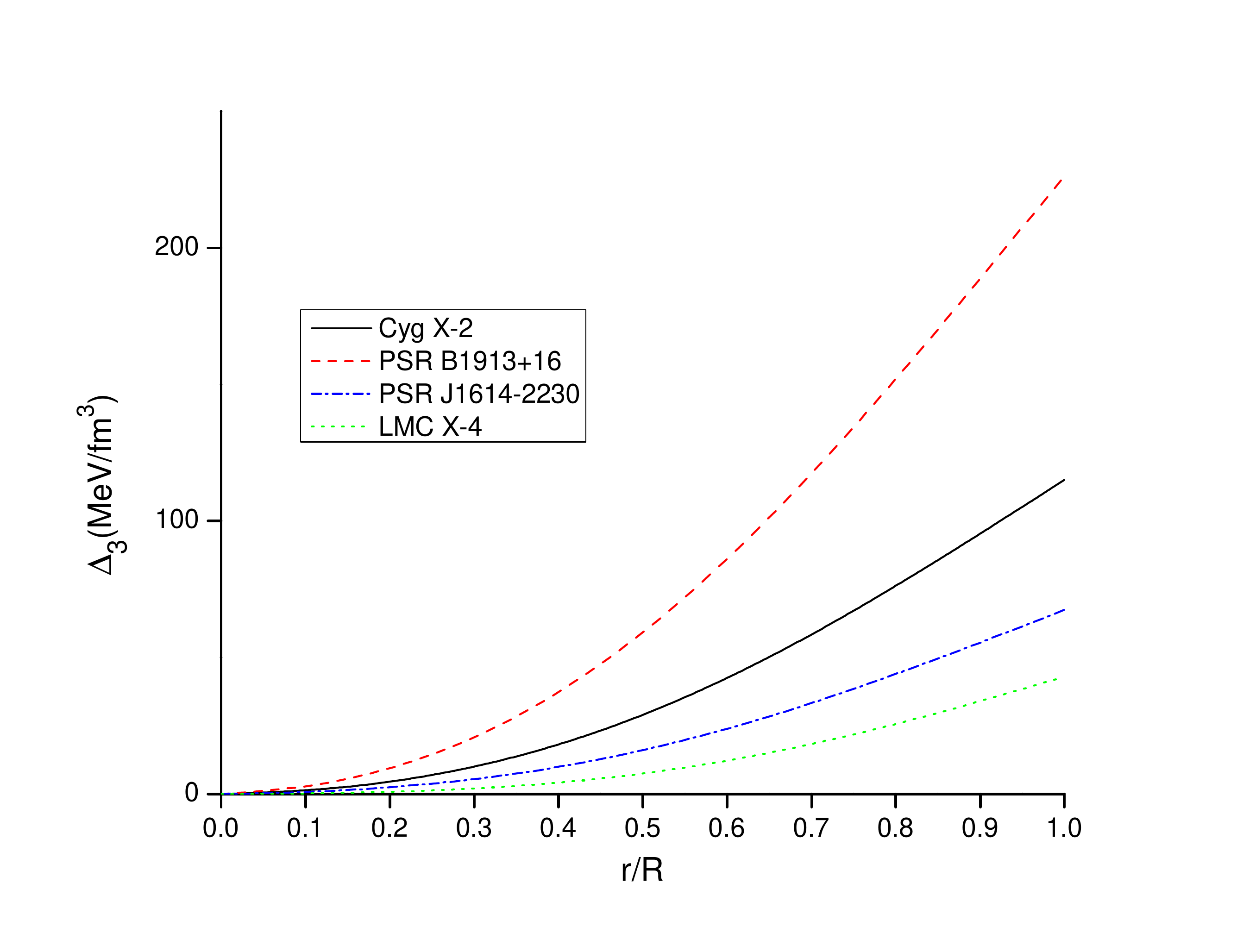}\includegraphics[width=5cm]{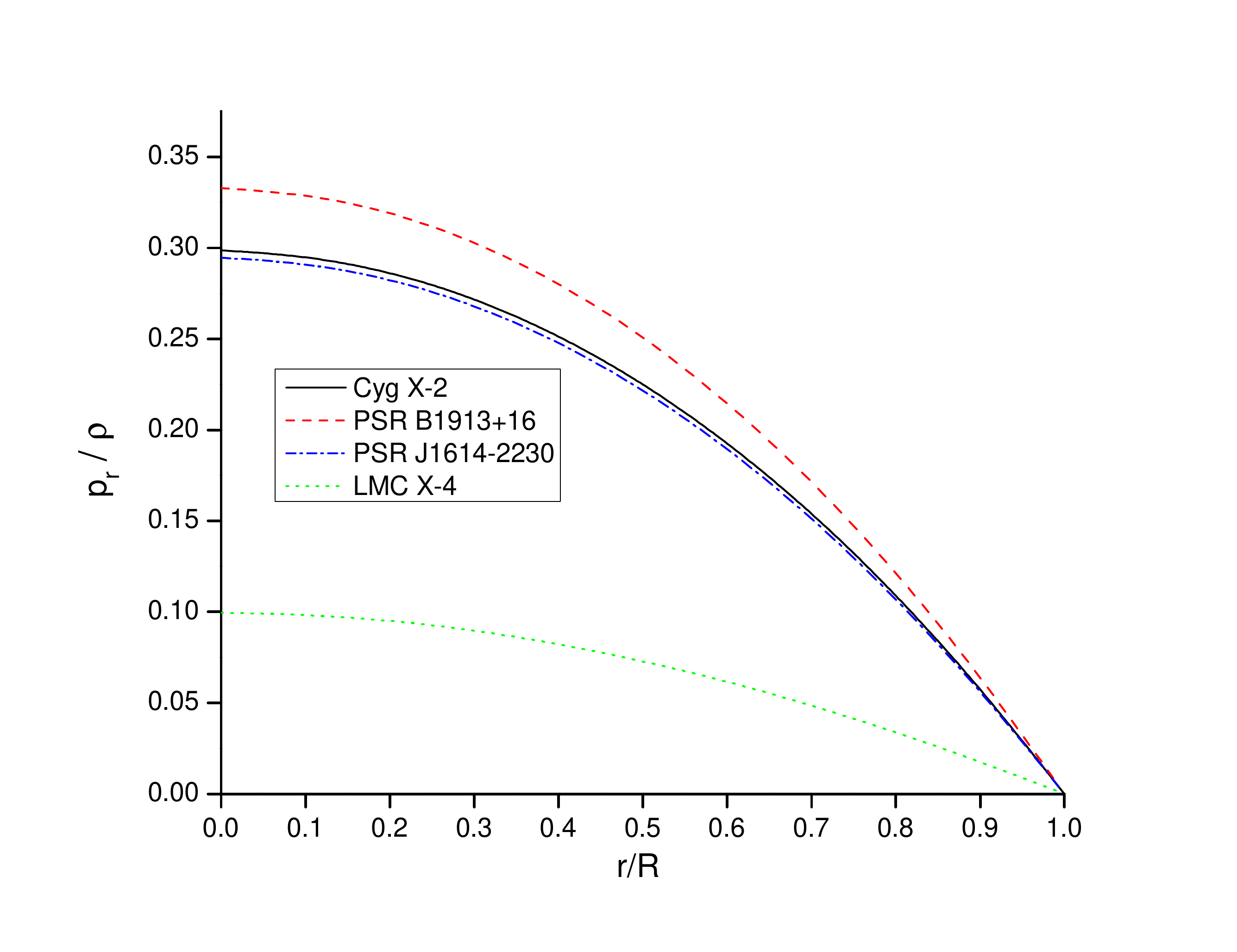}\includegraphics[width=5cm]{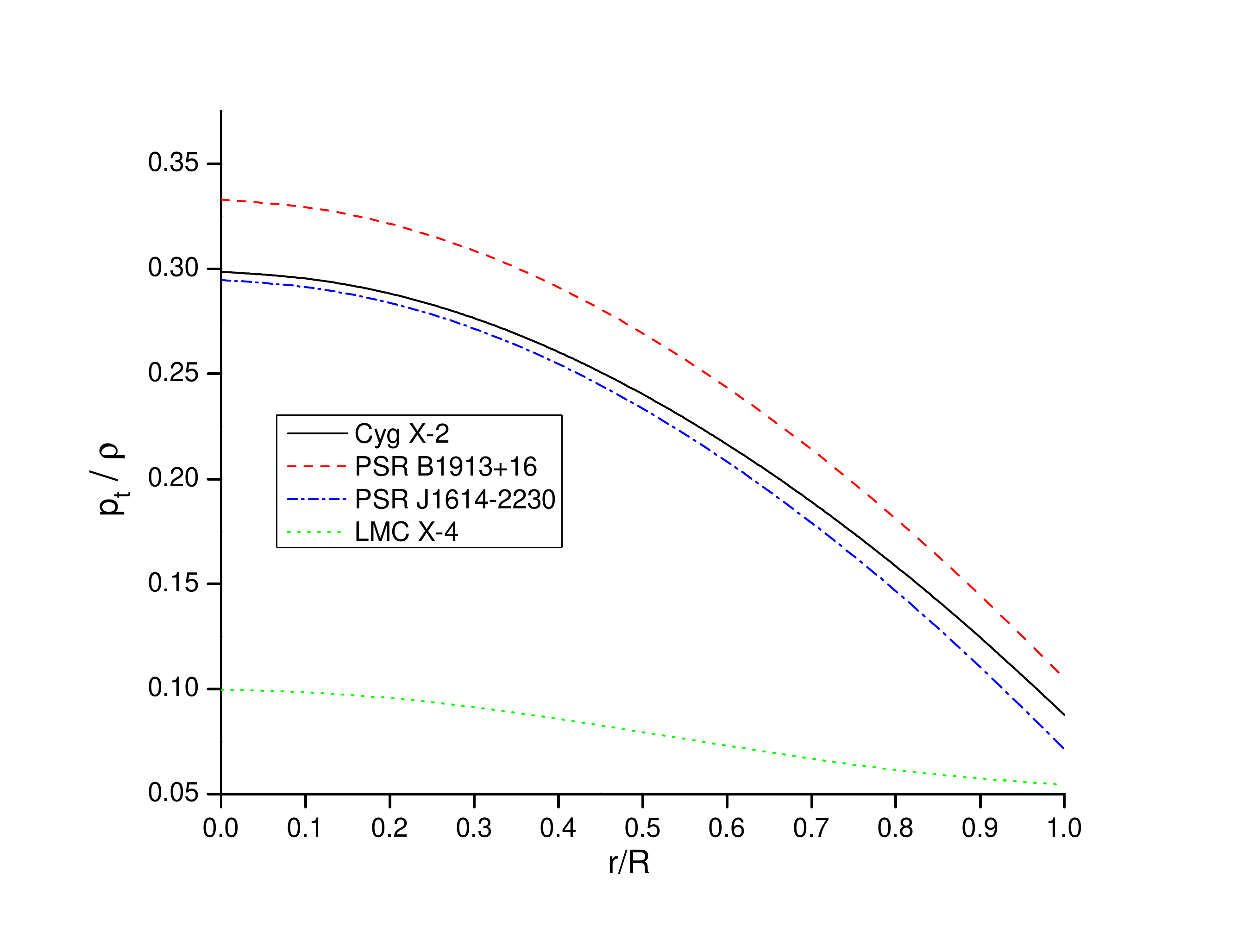}
  \\\includegraphics[width=5cm]{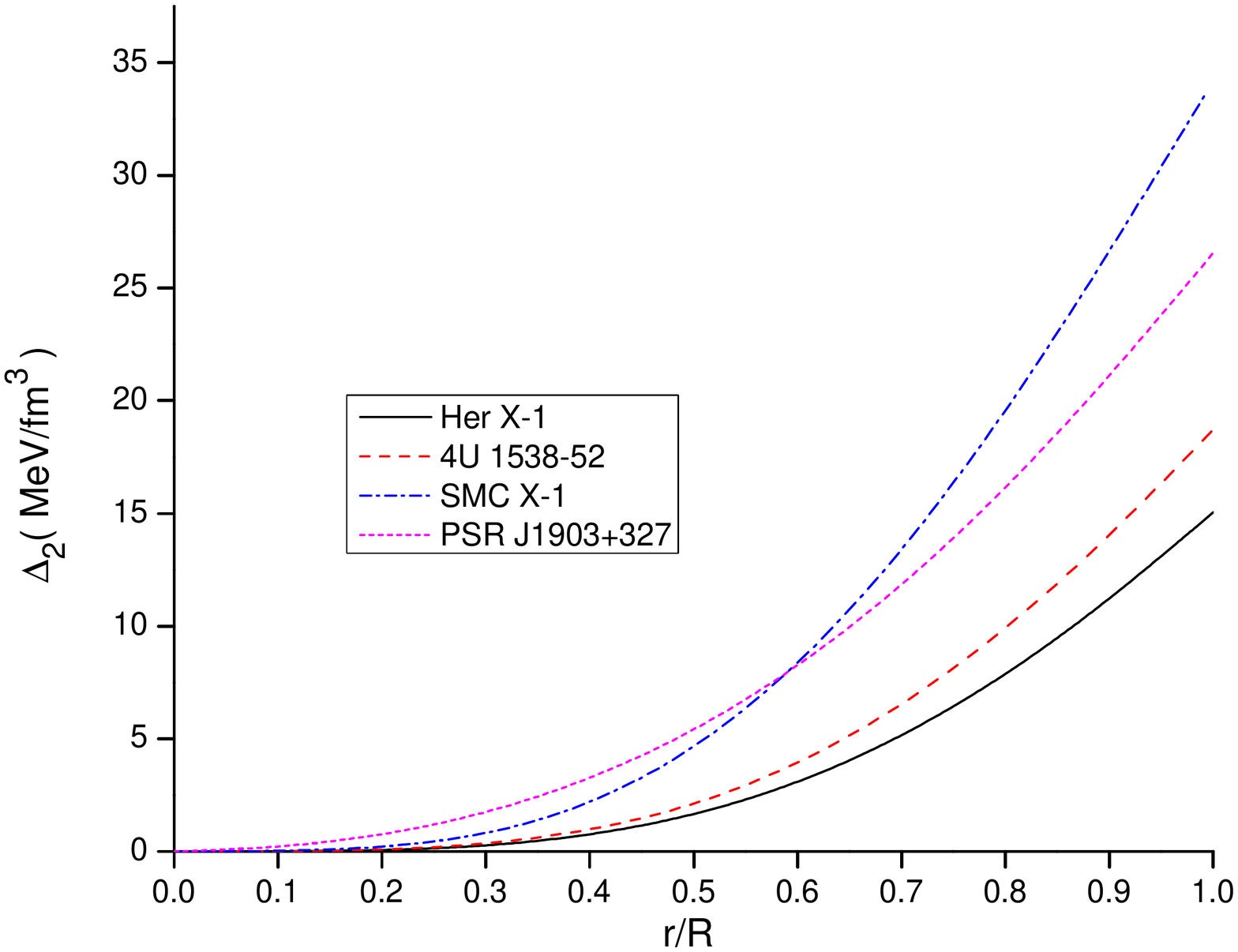}\includegraphics[width=5cm]{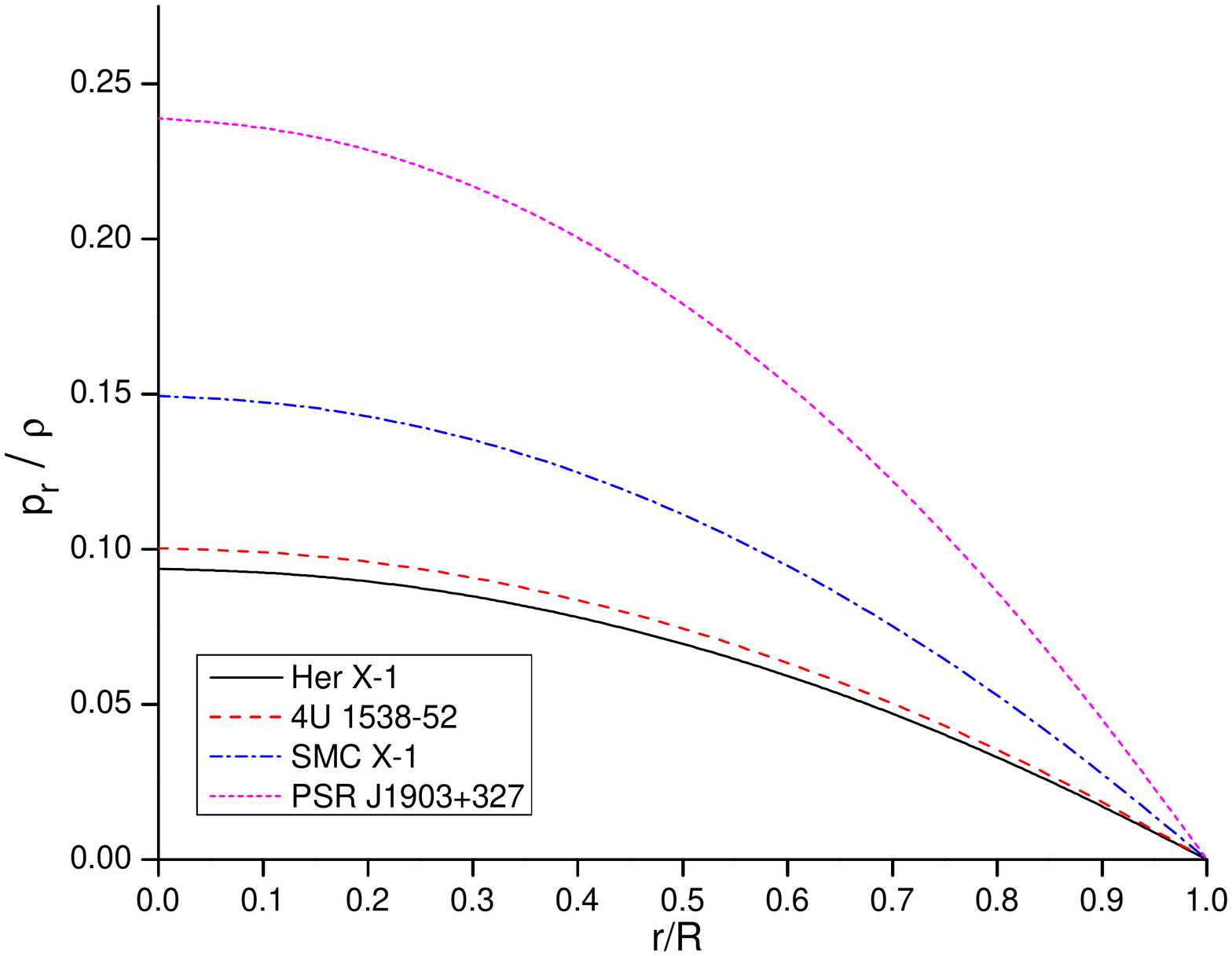}\includegraphics[width=5cm]{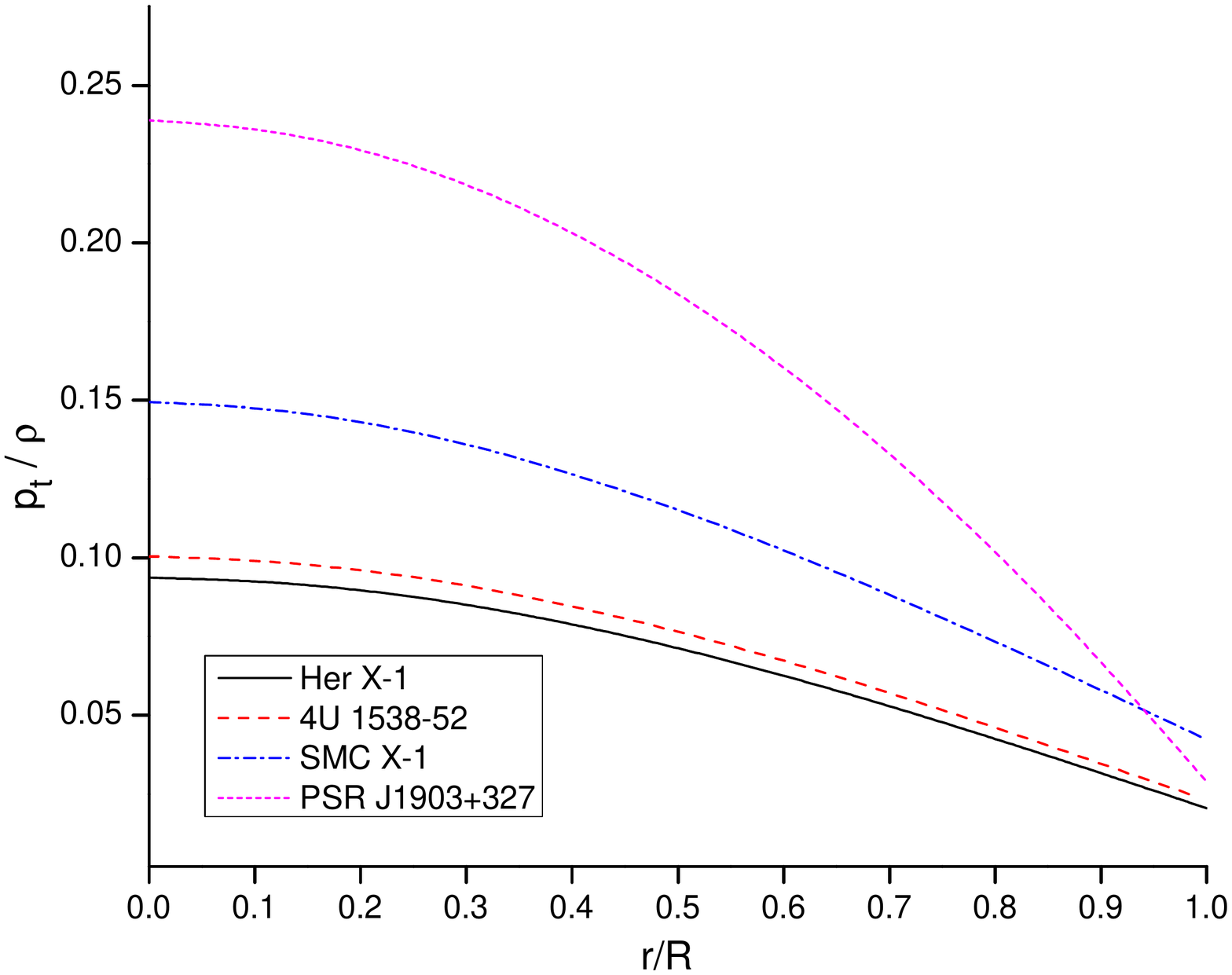}
  \caption{Behaviour of anisotropic factor and pressure-density ratio vs. fractional radius $ r/R $ for case I(first row), case II(second row) and case III(third row).For plotting this figure we have employed data set values of physical parameters and constants which are the same as used in Fig.\ref{f1}.}\label{f2}
  \end{center}
  \end{figure}

 Now we introduce the transformation $ e^{\nu} = Z^2(r) $ and substituting the value of $ e^{\lambda} $ in equ.(\ref{6}), we get
 \begin{equation}
 \frac{d^2Z}{dr^2}-\left[\frac{K+2\,K\,Cr^2+C^2r^4}{r\,(K+Cr^2)\,(1+Cr^2)} \right]\,\frac{dZ}{dr}+\left[\frac{(1-K)\,C^2r^4}{r^2\,(K+Cr^2)\,(1+Cr^2)}-\frac{\Delta~K\,(1+Cr^2)}{(K+Cr^2)}\,\right]\,Z = 0. \label{9}
 \end{equation}
 In order to solve equ.(\ref{9}) for $ Z $, we will use Mak and Harko's approach\cite{Mak}. Therefore we are taking an appropriate transformation $ X=\sqrt{\dfrac{K+Cr^{2}}{K-1}}$ and $ Z=(1-X^2)^{1/4}\,Y $, equ.(\ref{9}) transform to a differential equation of the form
\begin{equation}
\frac{d^2Y}{dX^2}+\frac{1}{(1-X^2)}\Bigg[1-K+\frac{\Delta\,K(K-1)^2\,(X^2-1)^2}{((K-1)\,(X^2-1)-1)^2}-\frac{2+3X^2}{4(1-X^2)}\Bigg]Y =0 \label{10}
\end{equation} 
Here, our aim is to solve the equ.(\ref{10}) by using Bijalwan and gupta's approach\cite{2011 b}.In this framework, we choose the expression for anisotropy 
 \begin{eqnarray}
\Delta=C^2r^2\Big[\frac{(1-K)(4K\,Cr^2 -7\,Cr^2-K-2)+\alpha(1+Cr^2)^2}{4K(1-K)(1+Cr^2)^3}\Big]\label{11}
 \end{eqnarray}
 where $ \alpha $ is an arbitrary constant that can take positive, negative or zero value.In equ.(\ref{11}), $ \Delta=0 $ at the center ,i.e. the radial pressure is equal to the tangential pressure at the center.Now using the value of $ \Delta $ in equ.(\ref{10}),we have 
 \begin{eqnarray}
 \frac{d^2Y}{dX^2} + \alpha Y =0\label{12}
 \end{eqnarray}  
 which is a second order differential equation in Y. Now, we are classifying each solution of equation(\ref{12}), briefly
\begin{subequations}
\begin{align}
\textrm{Case~ I:}~~~  Y&=A_{1}\,X+B_{1},~~~~~~~~~~~~~~~\textrm{if} ~~~\alpha=0\label{12a}\\
\textrm{Case~ II:}~~~  Y&=A_{2}\,e^{\Phi\,X} + B_{2}\,e^{-\Phi\,X},~~~~~~~~~~~~~\textrm{if} ~~~ \alpha=-\Phi^2\label{12b}\\
\textrm{Case~ III:}~~~  Y&=A_{3}\,\cos(\Phi\,X)+B_{3}\,\sin(\Phi\,X),~~~~~~~~~~\textrm{if} ~~~ \alpha=\Phi^2 \label{12c}
 \end{align}
\end{subequations}
where $A_{1}$, $B_{1}$, $A_{2}$, $B_{2}$, $A_{3}$, and $B_{3}$ are arbitrary constant of integration.Now the expression of radial and tangential pressure can be derived as follows in original coordinates,\\
\textbf{Case I:} 
 \begin{eqnarray}
Z(r) &=& A_{1}\Bigg(\frac{1+Cr^2}{1-K}\Bigg)^{1/4}\Bigg(\sqrt{\frac{K+Cr^2}{K-1}}+\frac{B_1}{A_1}\Bigg)\label{14}\\
\frac{\kappa p_r}{C}&=& \frac{K+Cr^2}{K(1+Cr^2)^2}+\frac{2\sqrt{K+Cr^2}}{K(1+Cr^2)(\sqrt{K+Cr^2}+B_1/A_1\sqrt{K-1})}+\frac{1-K}{K(1+Cr^2)}\label{15}\\
\frac{\kappa p_t}{C}&=& \frac{K+Cr^2}{K(1+Cr^2)^2}+\frac{2\sqrt{K+Cr^2}}{K(1+Cr^2)(\sqrt{K+Cr^2}+(B_1/A_1)\sqrt{K-1})}+\frac{1-K}{K(1+Cr^2)}+ \Delta_1\label{16}\\
\textrm{where}~~~~ \Delta_1&=&\frac{Cr^2}{4K(1+Cr^2)^3}\Big[4K\,Cr^2 -7\,Cr^2-K-2\Big]\nonumber    
 \end{eqnarray}
 \textbf{Case II:} 
  \begin{eqnarray}
 Z(r) &=& A_{2}\Bigg(\frac{1+Cr^2}{1-K}\Bigg)^{1/4}\Bigg(e^{\Phi\sqrt{\frac{K+Cr^2}{K-1}}}+(B_2/A_2)\,e^{-\Phi\sqrt{\frac{K+Cr^2}{K-1}}}\Bigg)\label{17}\\
 \frac{\kappa p_r}{C}&=& \frac{K+Cr^2}{K(1+Cr^2)^2}+\frac{2\sqrt{K+Cr^2}}{K\sqrt{K-1}(1+Cr^2)}\left(\frac{e^{\Phi\sqrt{\frac{K+Cr^2}{K-1}}}-(B_2/A_2)\,e^{-\Phi\sqrt{\frac{K+Cr^2}{K-1}}}}{e^{\Phi\sqrt{\frac{K+Cr^2}{K-1}}}+(B_2/A_2)\,e^{-\Phi\sqrt{\frac{K+Cr^2}{K-1}}}}\right)+\frac{1-K}{K(1+Cr^2)}\label{18}\\
 \frac{\kappa p_t}{C}&=& \frac{K+Cr^2}{K(1+Cr^2)^2}+\frac{2\sqrt{K+Cr^2}}{K\sqrt{K-1}(1+Cr^2)}\left(\frac{e^{\Phi\sqrt{\frac{K+Cr^2}{K-1}}}-(B_2/A_2)\,e^{-\Phi\sqrt{\frac{K+Cr^2}{K-1}}}}{e^{\Phi\sqrt{\frac{K+Cr^2}{K-1}}}+(B_2/A_2)\,e^{-\Phi\sqrt{\frac{K+Cr^2}{K-1}}}}\right)+\frac{1-K}{K(1+Cr^2)}+ \Delta_2\nonumber\label{19}\\ \\
 \textrm{where}~~~~ \Delta_2&=&\frac{Cr^2}{K(1+Cr^2)^2}\Bigg[\frac{4K\,Cr^2 -7\,Cr^2-K-2}{4(1+Cr^2)}-\frac{\Phi(1+Cr^2)}{1-K}\Bigg]\nonumber    
  \end{eqnarray}
   \textbf{Case III:} 
    \begin{eqnarray}
   Z(r) &=& A_{3}\Bigg(\frac{1+Cr^2}{1-K}\Bigg)^{1/4}\Bigg[\cos\Big(\Phi\sqrt{\frac{K+Cr^2}{K-1}}\Big)+(B_3/A_3)\,\sin\Big(\Phi\sqrt{\frac{K+Cr^2}{K-1}}\Big)\Bigg]\label{20}\\
   \frac{\kappa p_r}{C}&=& \frac{K+Cr^2}{K(1-K)(1+Cr^2)}\left[\frac{1-K}{1+Cr^2}-2\Phi\sqrt{\frac{K-1}{K+Cr^2}}F(r)\right]+\frac{1-K}{K(1+Cr^2)}\label{21}\\ 
    \frac{\kappa p_r}{C}&=& \frac{K+Cr^2}{K(1-K)(1+Cr^2)}\left[\frac{1-K}{1+Cr^2}-2\Phi\sqrt{\frac{K-1}{K+Cr^2}}F(r)\right]+\frac{1-K}{K(1+Cr^2)}+\Delta_3\label{22}\\ 
    \textrm{where}~~~~ \Delta_3&=&\frac{Cr^2}{K(1+Cr^2)^2}\Bigg[\frac{4K\,Cr^2 -7\,Cr^2-K-2}{4(1+Cr^2)}+\frac{\Phi(1+Cr^2)}{1-K}\Bigg]\nonumber\\F(r)&=&\left(\frac{(B_3/A_3)\,\cos\Big(\Phi\sqrt{\frac{K+Cr^2}{K-1}}\Big)-\sin\Big(\Phi\sqrt{\frac{K+Cr^2}{K-1}}\Big)}{\cos\Big(\Phi\sqrt{\frac{K+Cr^2}{K-1}}\Big)+(B_3/A_3)\,\sin\Big(\Phi\sqrt{\frac{K+Cr^2}{K-1}}\Big)}\right)\nonumber  
    \end{eqnarray}
 In a realistic scenario, the following physical character of above solutions is required:
 \begin{enumerate}[(i)]
 \item the energy density is positive definite and decreasing monotonically within $ 0<r<R, $ i.e., its gradient is negative within $ 0<r<R. $ 
 \item the pressure is positive definite and decreasing monotonically within $ 0<r<R, $ i.e., its gradient is negative within $ 0<r<R. $
 \item the ratio of pressure and energy density is less than unity within the stellar model.
 \item the radial pressure should be vanished at the boundary but the tangential pressure needs not necessarily vanish. However, the radial pressure is equal to the tangential pressure  at the center, i.e, $ \Delta(0)=0 $\cite{Ivanov,Bowers}.
 \end{enumerate}
 To examine the positive density and pressure of realistic star, we plot this in Fig.(\ref{f1}). In Fig.(\ref{f2}), we can show that the anisotropy($\Delta$) vanishes at the center and increasing monotonically within stellar model which implies tangential pressure is always greater than radial pressure. Also ratio of pressure and density, i.e, $ \frac{p_r}{\rho} $ and $ \frac{p_t}{\rho} $ lie between $ 0 $ to $ 1 $, which state that the underlying fluid distribution is real and nonexotic in nature \cite{Rahaman1}.
 \section{Boundary condition} 
 In order to find the arbitrary constant $A_{1}$, $B_{1}$, $A_{2}$, $B_{2}$, $A_{3}$, and $B_{3}$ we match our interior spacetime(\ref{1}) to the exterior Schwarzschild line element at the boundary of the star($ r=R $). The expression of exterior Schwarzschild line element is as follows
  \begin{eqnarray}
  ds^{2}=-\bigg(1-\frac{2M}{r}\bigg)^{-1}dr^{2}-r^{2}(d\theta^{2}+\sin^{2}\theta d\phi^{2})+\bigg( 1-\frac{2M}{r}\bigg)dt^{2} \label{23}
 \end{eqnarray}
  where $ M $ is the gravitational mass of the distribution such that \begin{eqnarray*} M(r)=\dfrac{\kappa}{2}\int_{0}^{r}\rho r^{2}dr,\label{78}\end{eqnarray*}  
  Using the Darmois-Israel boundary condition\cite{Israel}, we match line elements (\ref{1}) and (\ref{23}) at the boundary ($ r=R $) which are tantamount to the following two conditions 
   \begin{eqnarray} e^{-\lambda}=1-\dfrac{2M}{R},~~~\textrm{and}~~~e^{\nu}=y^{2}=1-\dfrac{2M}{R} \label{24}\\ p_{r}(R)=0.\label{25}
 \end{eqnarray}
Now using the conditions (\ref{24}) and (\ref{25}), we can compute the values of arbitrary constants. Thus the expression of arbitrary constants $ A_1 $ to $ A_3 $ and $ B_1 $ to $ B_3 $ as follows:
\begin{eqnarray*}
\frac{B_1}{A_1} &=& -\sqrt{\frac{K+CR^2}{K-1}}\Bigg[1+\frac{2(1+CR^2)}{2\,CR^2+1-K\,CR^2}\Bigg]\\
\frac{B_2}{A_2} &=&\Bigg[ \frac{2\,CR^2+1-K\,CR^2+2\Phi\sqrt{\frac{K+CR^2}{K-1}}}{2\Phi\sqrt{\frac{K+CR^2}{K-1}}-2\,CR^2-1+K\,CR^2}\Bigg]e^{\Big(\Phi\sqrt{\frac{K+Cr^2}{K-1}}\Big)}\\
\frac{B_3}{A_3} &=& \frac{G(R)\cos\Big(\Phi\sqrt{\frac{K+CR^2}{K-1}}\Big)+G_{1}(R)\sin\Big(\Phi\sqrt{\frac{K+CR^2}{K-1}}\Big)}{G_{1}(R)\sin\Big(\Phi\sqrt{\frac{K+CR^2}{K-1}}\Big)-G(R)\cos\Big(\Phi\sqrt{\frac{K+CR^2}{K-1}}\Big)}\\\nonumber
\textrm{where}~~~ G(R)&=&\frac{2\,CR^2+1-K\,CR^2}{K(1+CR^2)^2},~~~~~~G_{1}(R)=\frac{2\Phi(K+CR^2)}{K(K-1)(1+CR^2)}\sqrt{\frac{K-1}{K+CR^2}}
\end{eqnarray*}
One can decide the absolute mass M of the star for a given radius R. It might be referenced here that limits on excellent structures, including the mass-radius ratio as proposed by the Buchdahl-Bondi condition\cite{Buchdahl,28}, exist and are given by $ \frac{2M}{R}\leq \frac{8}{9} $. This fills in as an upper bound on the absolute compactness of the static spherically symmetric isotropic compact stars. However, this bound has been refreshed in the presence of charged gravitational fields\cite{29,30}, for a nonzero cosmological constant\cite{31} and use of the thin-shell formalism\cite{joao}.  The effect of the mass-radius ratio on the equation of state has been considered by Carvalho et al. \cite{32} for white dwarf and neutron stars and for the nuclear center by Lattimer\cite{33}. Here, we have find specific masses and radii(see in Table-\ref{t1}) of the compact star estimated by Gangopadhyay et al.\cite{Gangopadhyay}.\par
 The gravitational redshift from the surface of the star as
measured by a distant observer $ (g_{tt} \rightarrow -1) $ is given by 
\begin{equation*}
z_s = e^{\lambda(R)} -1 = \dfrac{1-\sqrt{1-\frac{2M}{R}}}{\sqrt{1-\frac{2M}{R}}}
\end{equation*}
where $g_{tt}(R) = e^{\nu(R)} = (1-\frac{2M}{R}) $ is the metric function. Buchdahl\cite{Buchdahl} have observed that the gravitational
redshift is $ z_s < 2 $ for spherically
symmetric distribution of a prefect fluid. However, Karmakar and Barraco \cite{Karmakar,Barraco} suggested that the gravitational redshift of anisotropy star models is 3.84. Therefore, the resulting surface redshift (see Fig.\ref{f6}) from the solution also compatible
with this discussion.
 \begin{figure}[]
 \begin{center}
 \includegraphics[width=5cm]{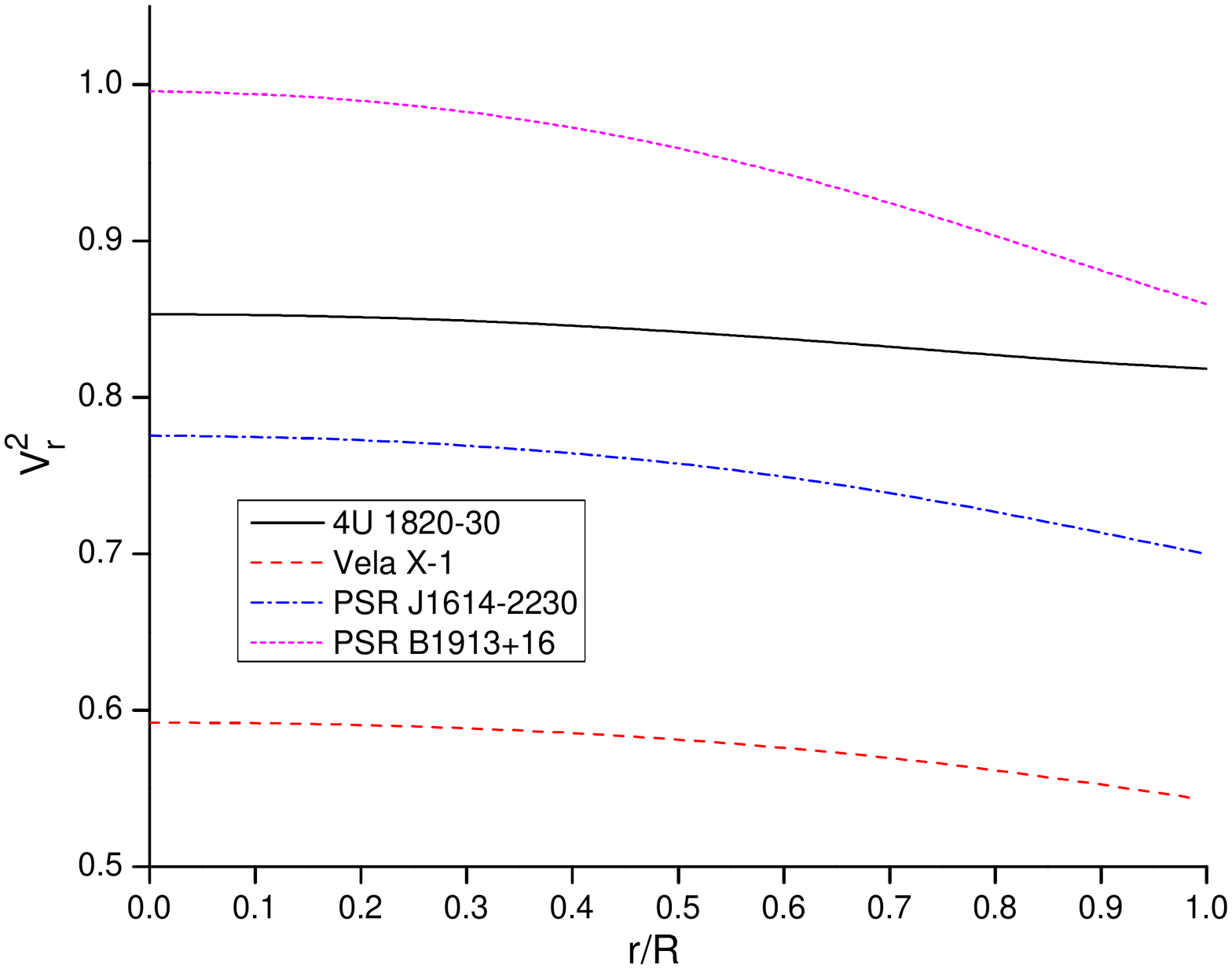}\includegraphics[width=5cm]{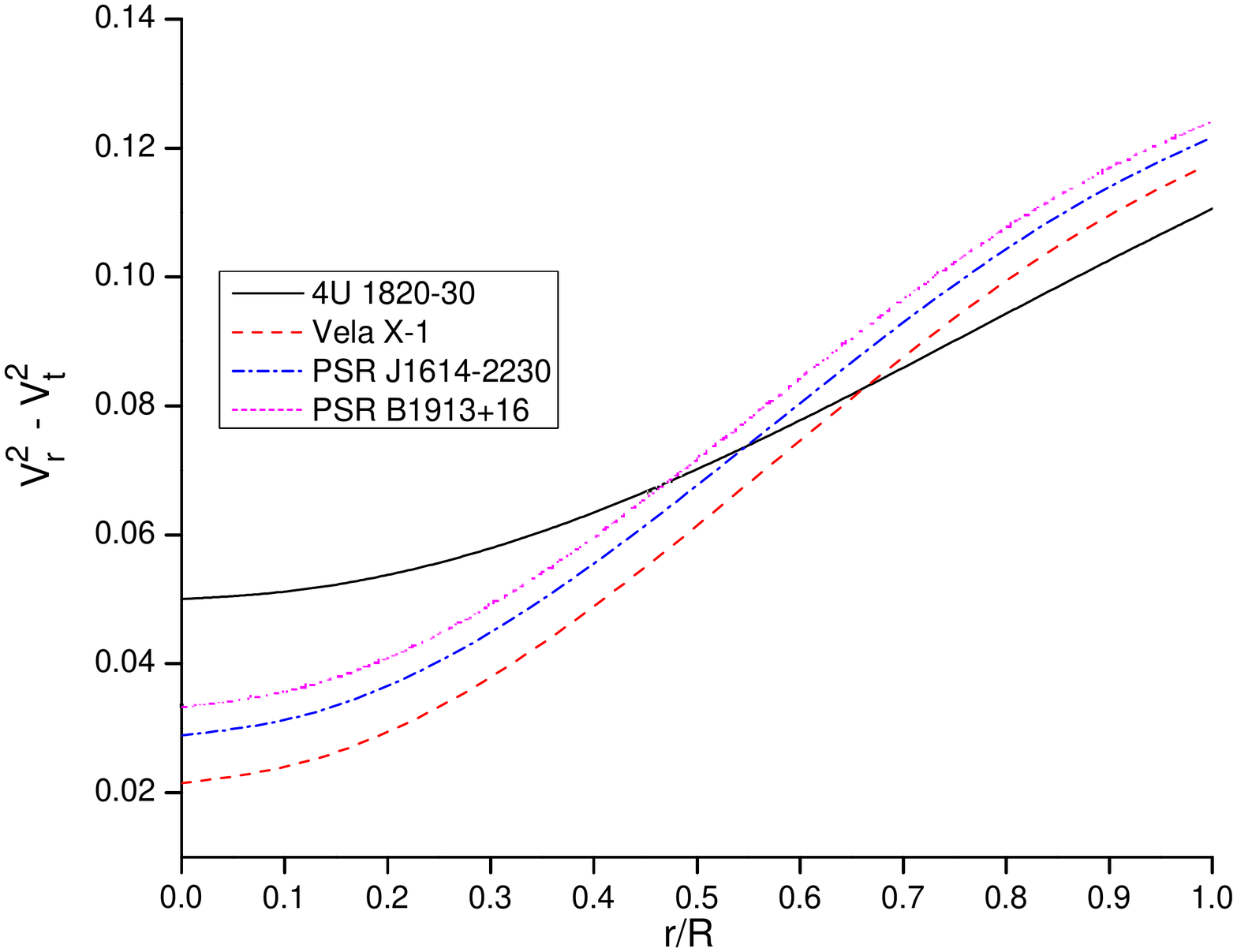}\includegraphics[width=5cm]{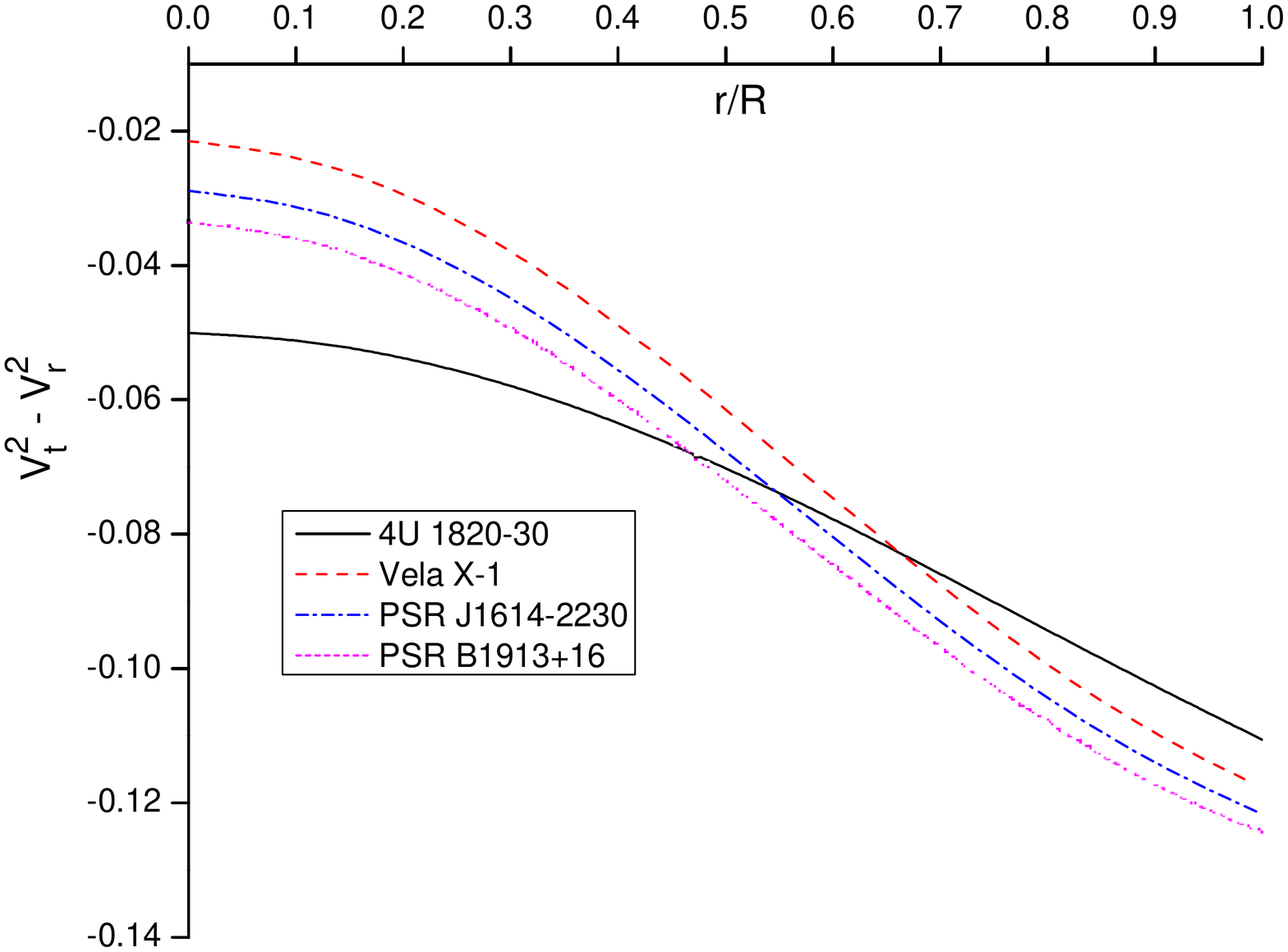}\\\includegraphics[width=5cm]{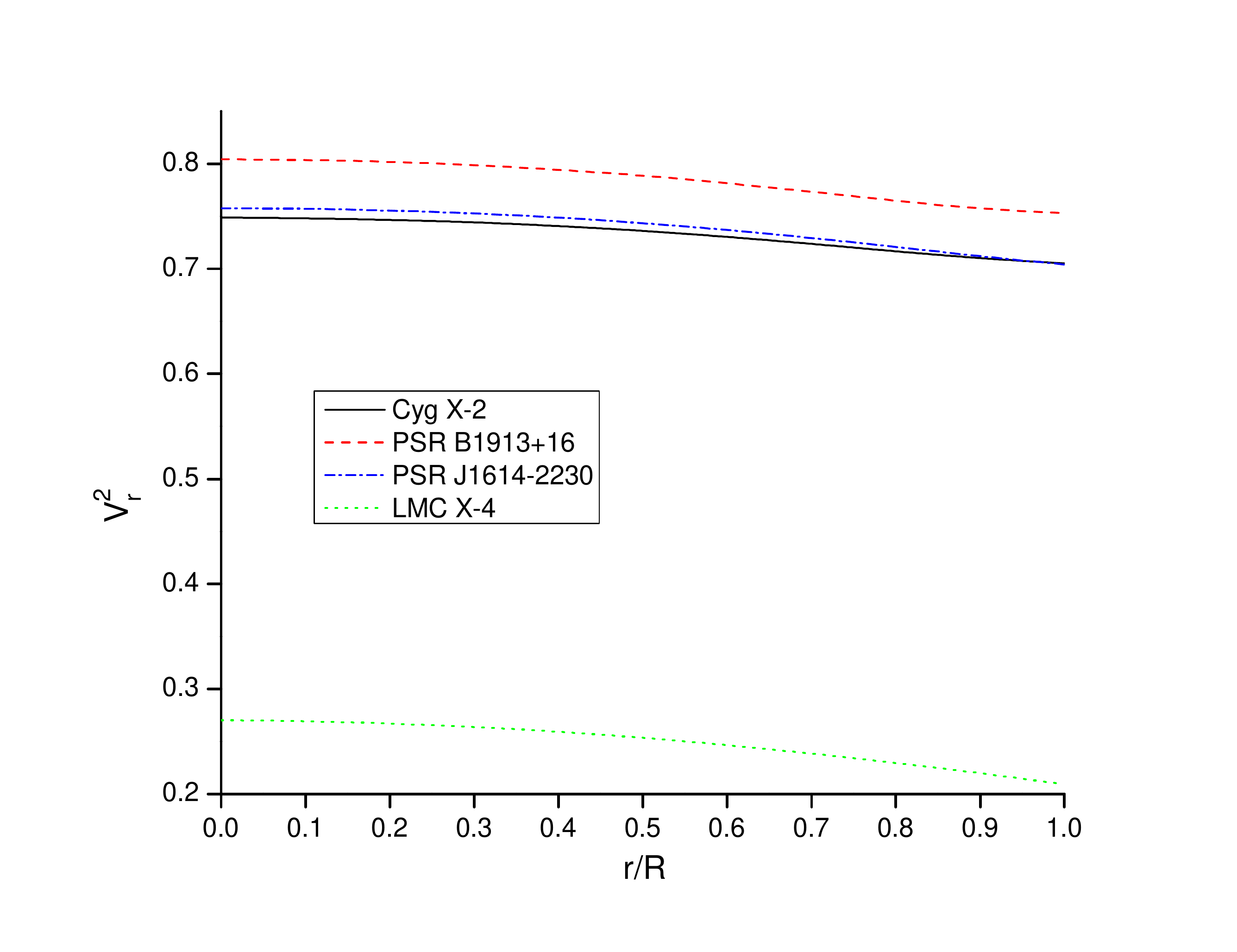}\includegraphics[width=5cm]{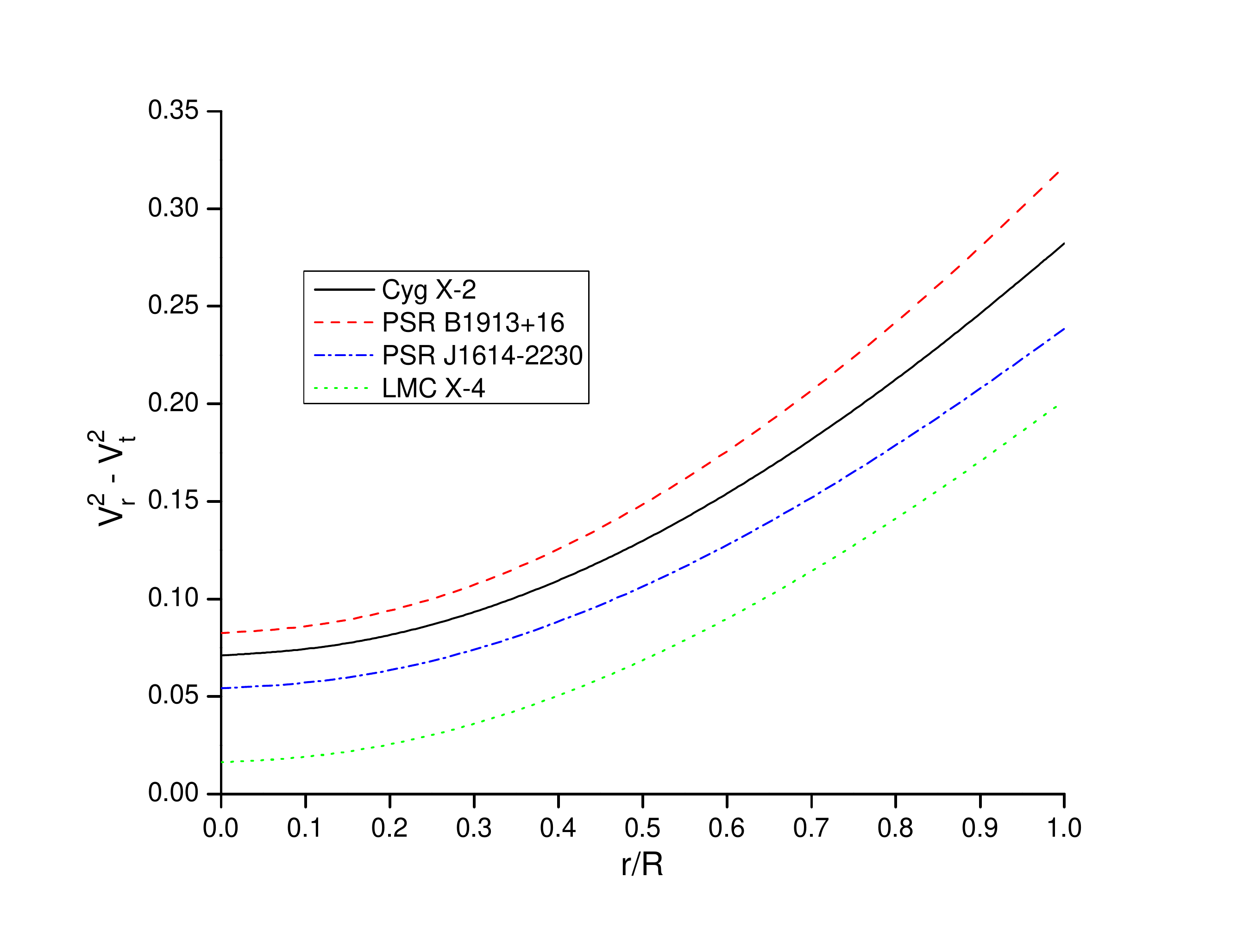}\includegraphics[width=5cm]{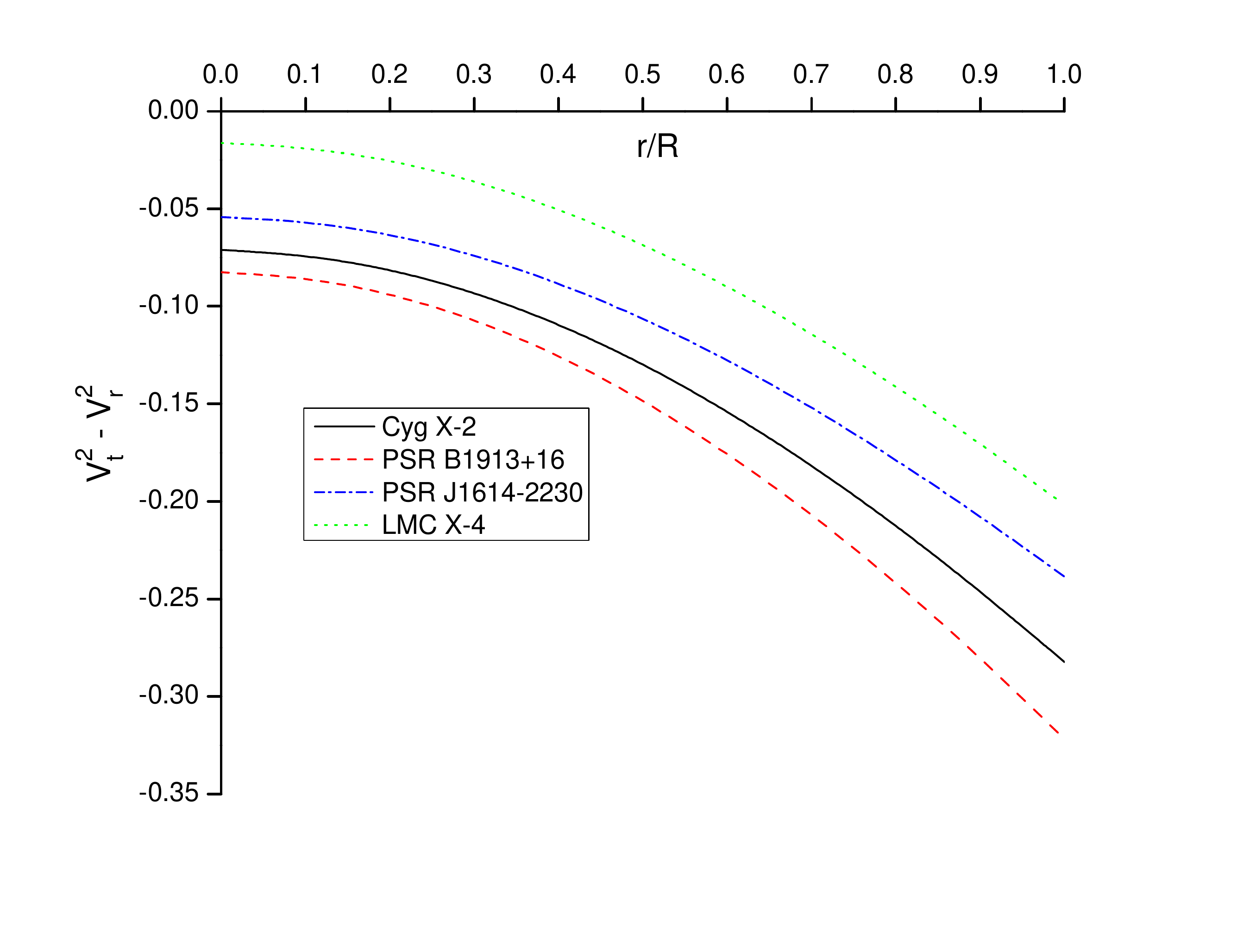}
 \\\includegraphics[width=5cm]{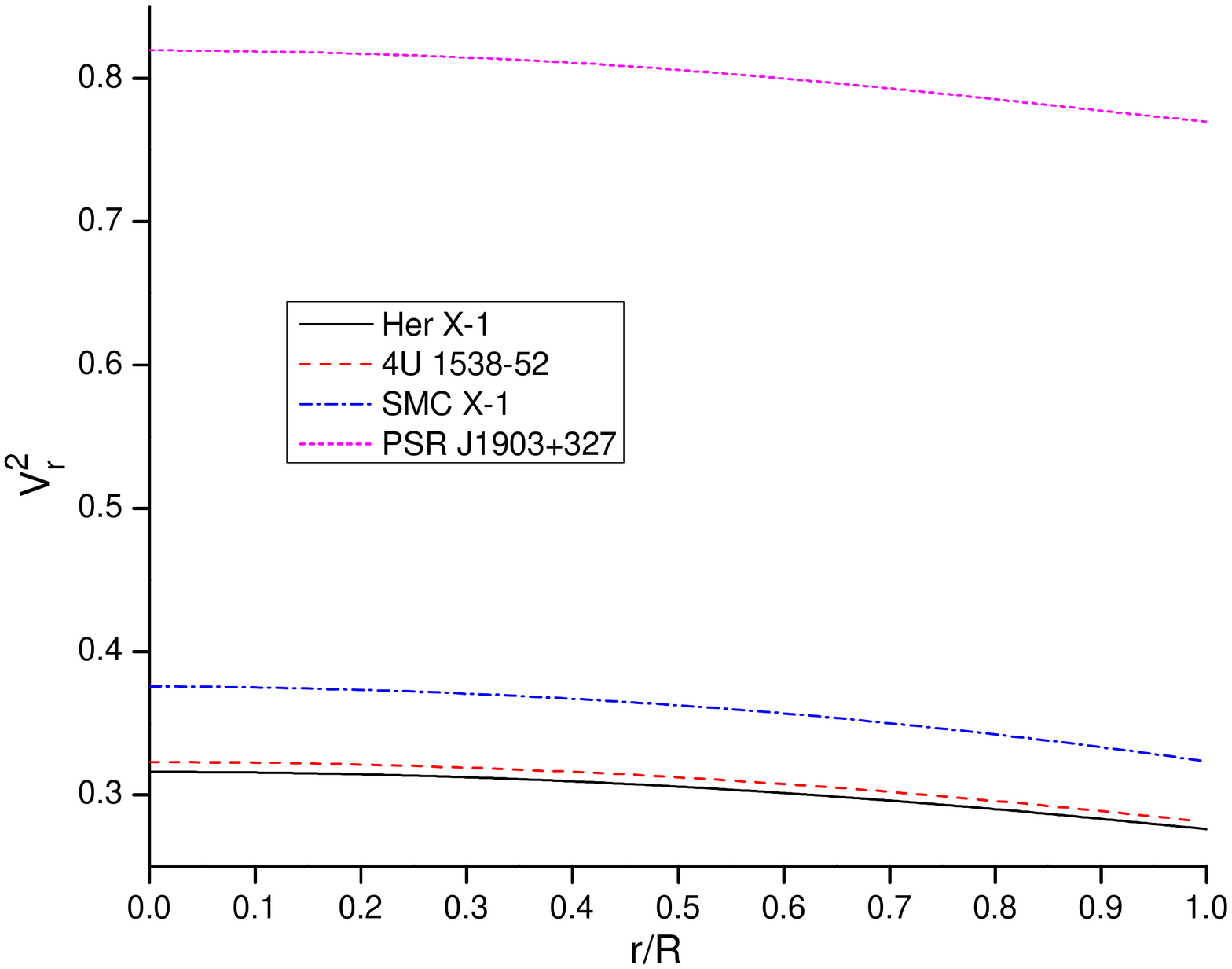}\includegraphics[width=5cm]{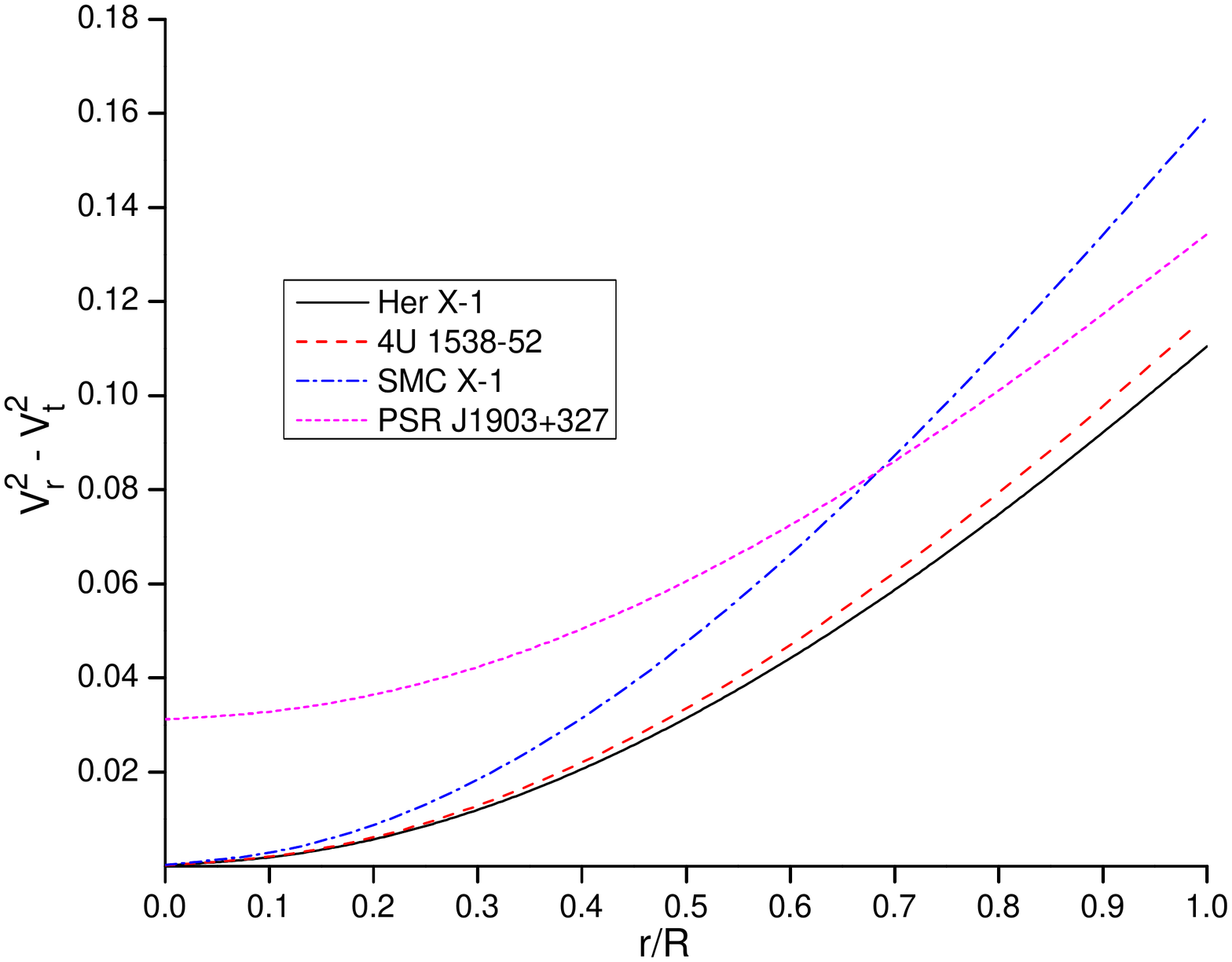}\includegraphics[width=5cm]{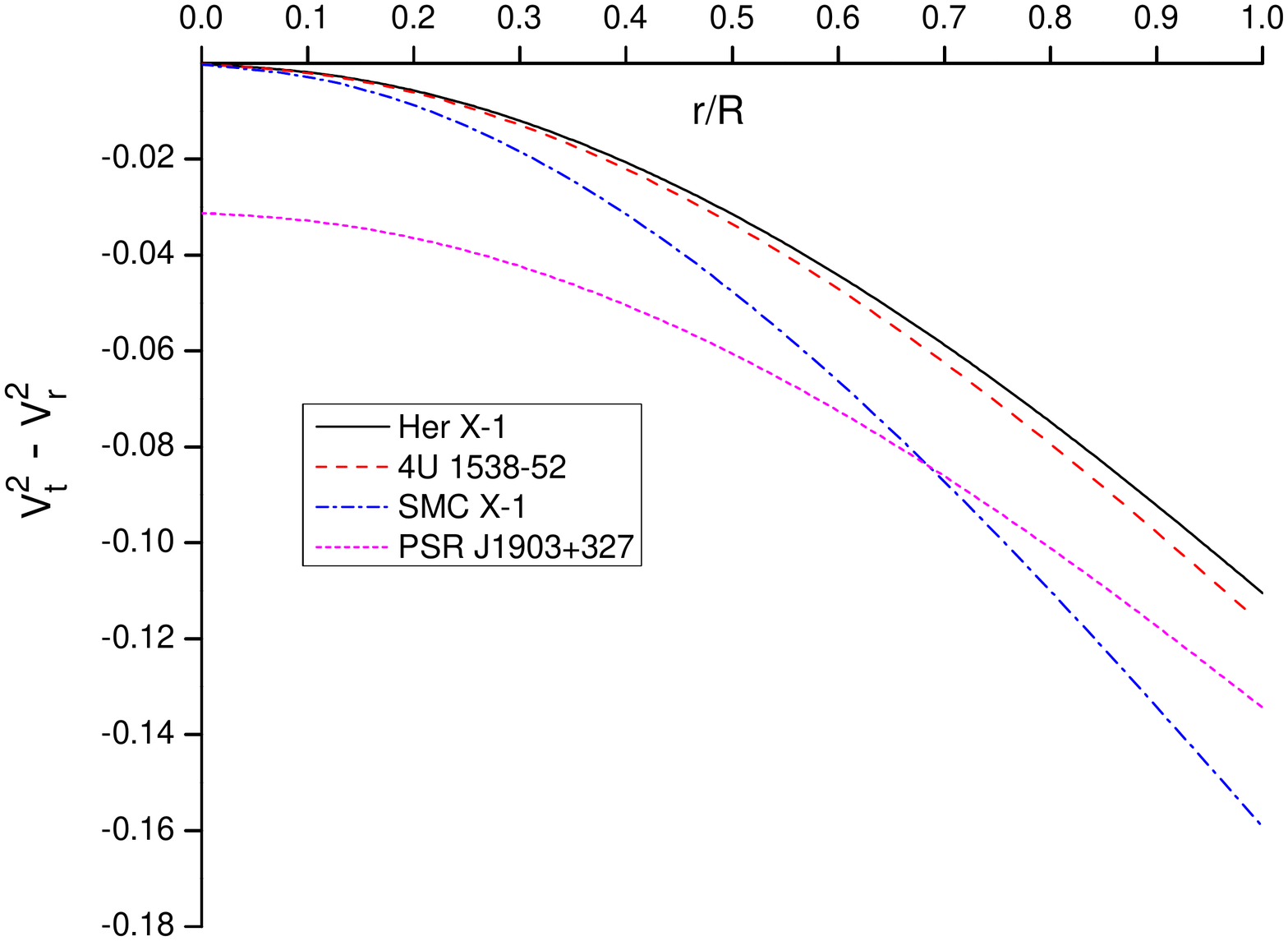}
 \caption{Behaviour of speed of sound vs. fractional radius $ r/R $ for case I(first row), case II(second row) and case III(third row). For plotting this figure we have employed data set values of physical parameters and constants which are the same as used in Fig.\ref{f1}.}\label{f3}
 \end{center}
 \end{figure}
\section{Mass function and Compactness}
In this section, we have discussed the mass function $ M(r) $ and  mass-radius relationship i.e, compactness parameter $ u(r) $.  Buchdahl\cite{Buchdahl} suggested that the mass-radius ratio of a relativistic static spherically symmetric fluid stellar model should be $ \frac{M}{R} \leq \frac{4}{9} $. In this context, Mak and Harko \cite{Mak} have obtained a generalized formula for the mass-radius ratio. In our model, we have obtained the mass function and compactness parameter of relativistic compact stars as follows 
\begin{figure}[]
	\begin{center}
		\includegraphics[width=6cm]{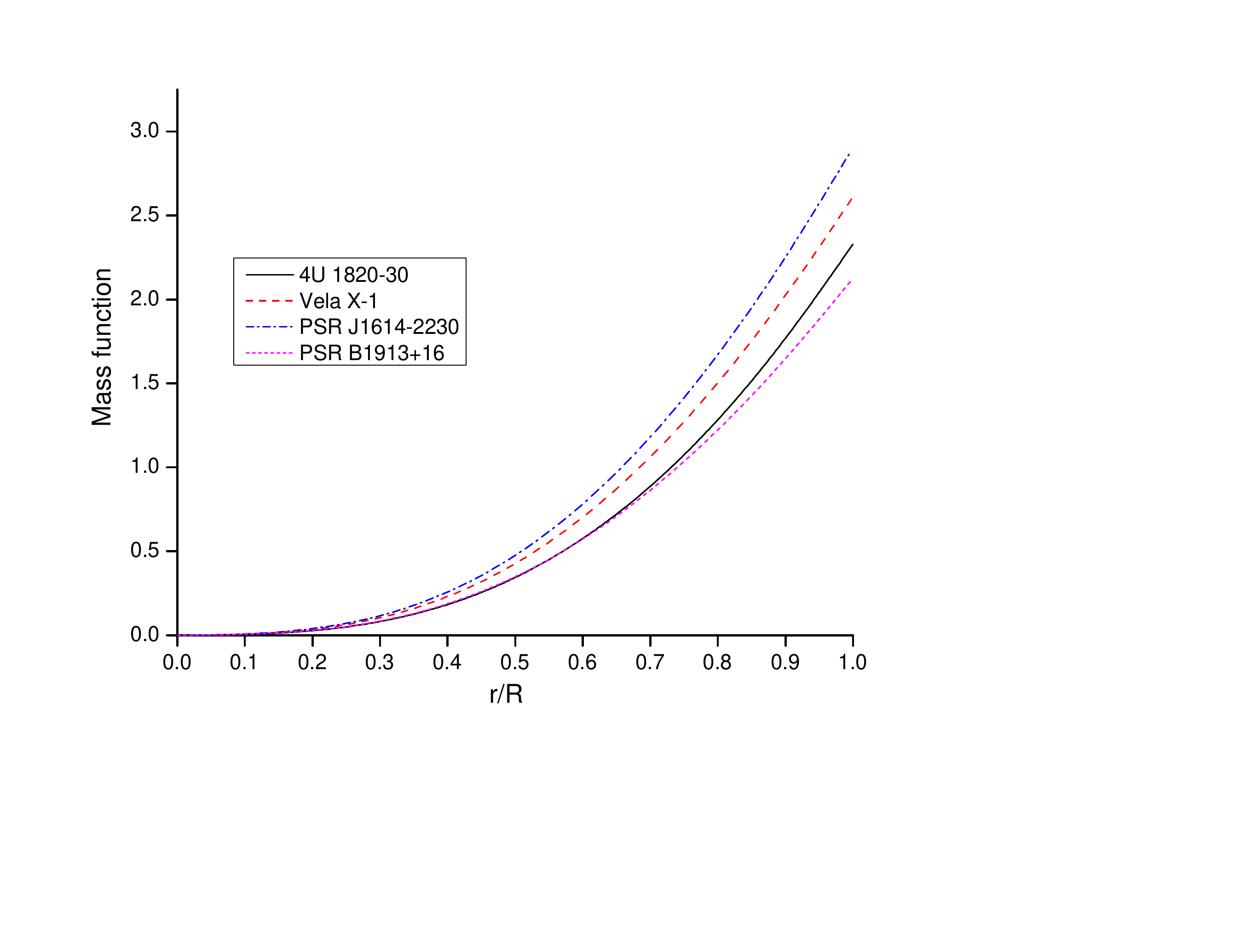}\includegraphics[width=6cm]{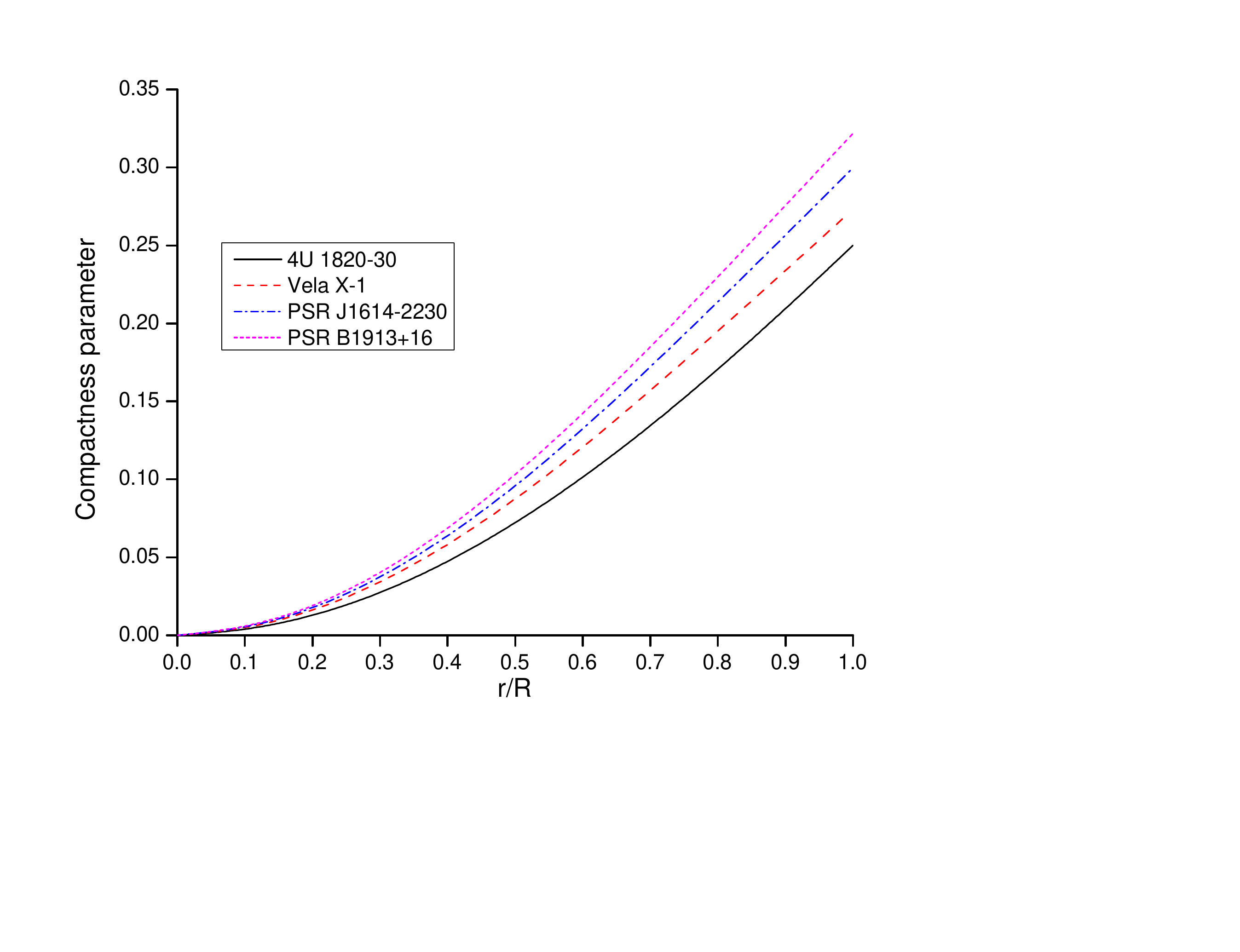}\\\includegraphics[width=6cm]{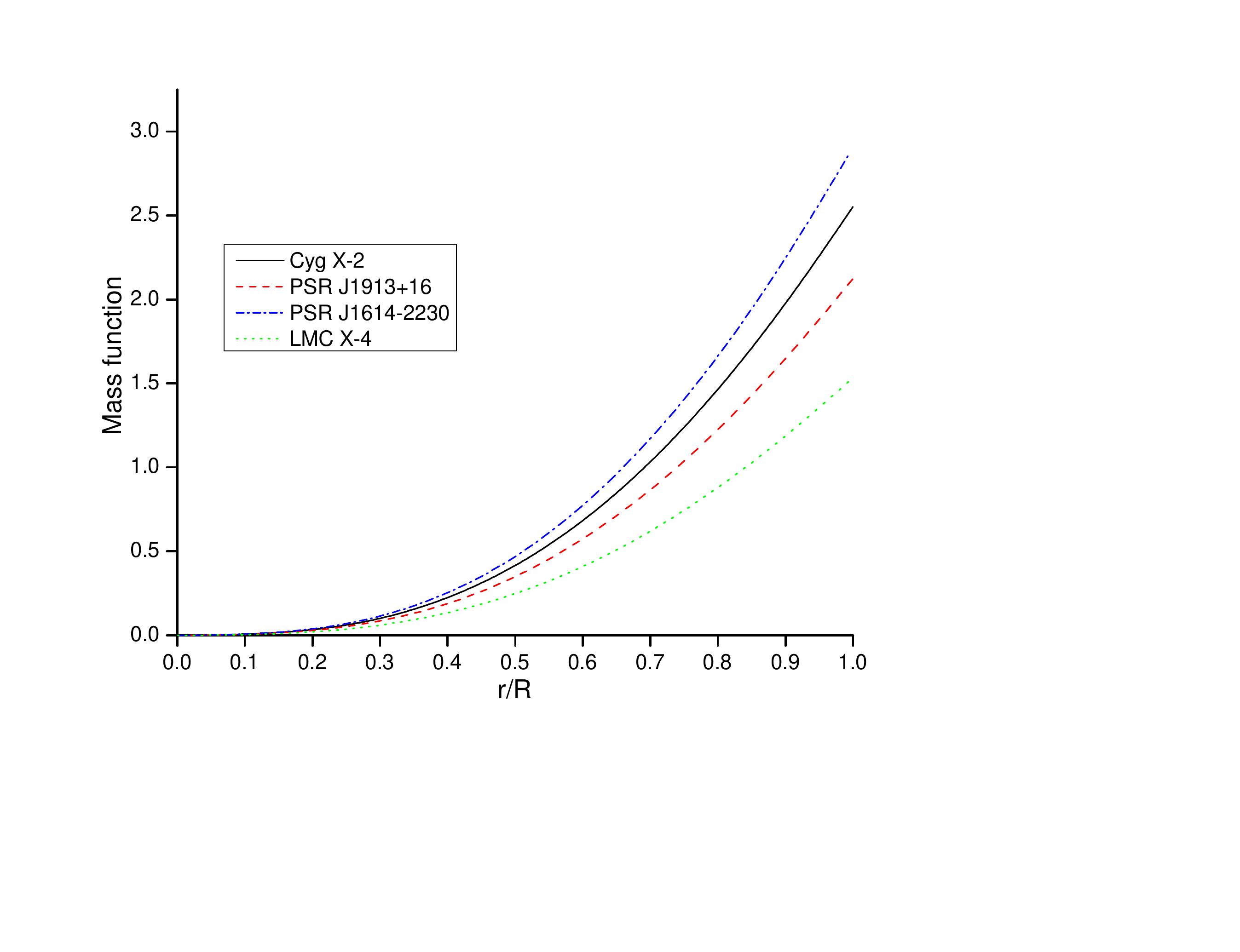}\includegraphics[width=6cm]{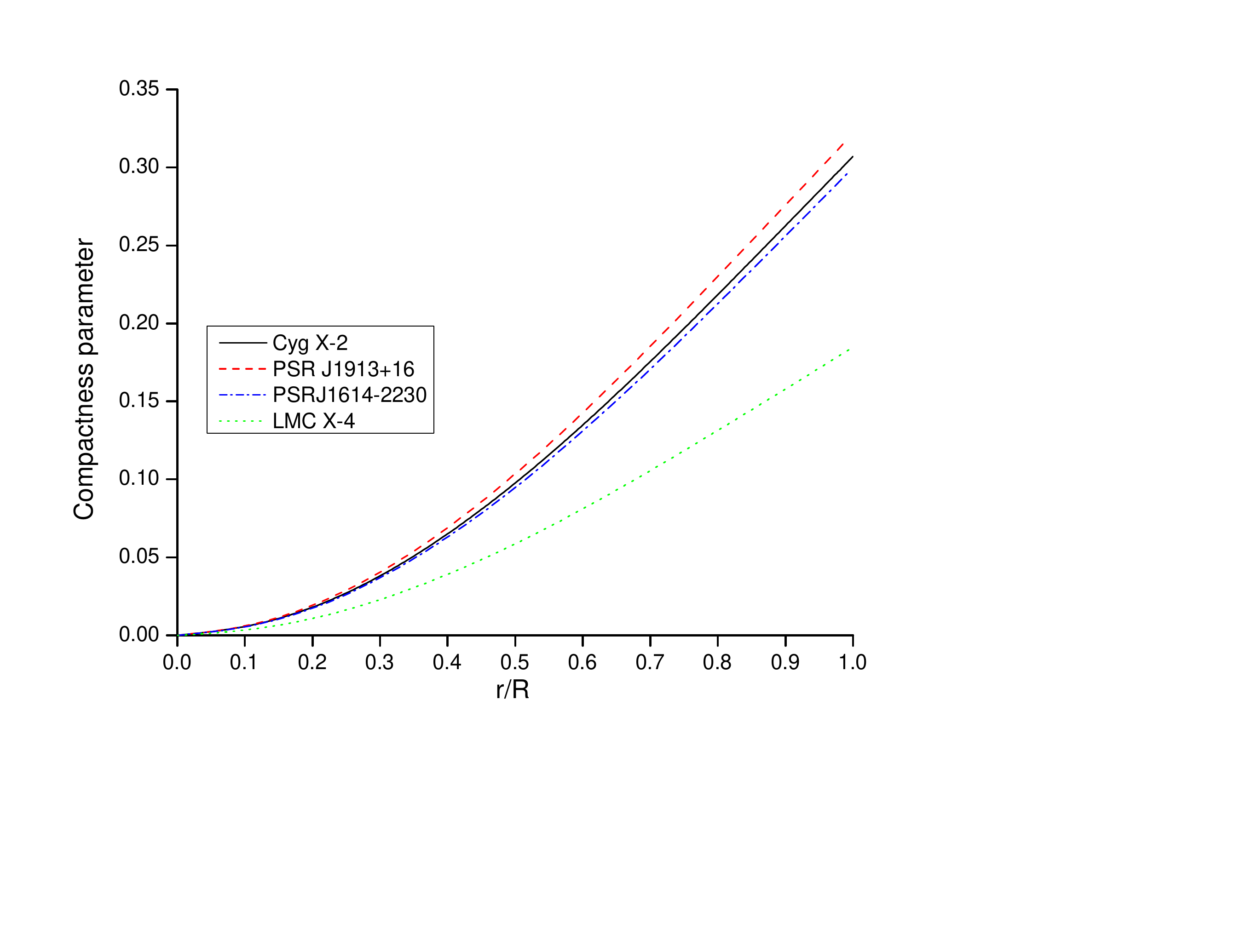}\\\includegraphics[width=6cm]{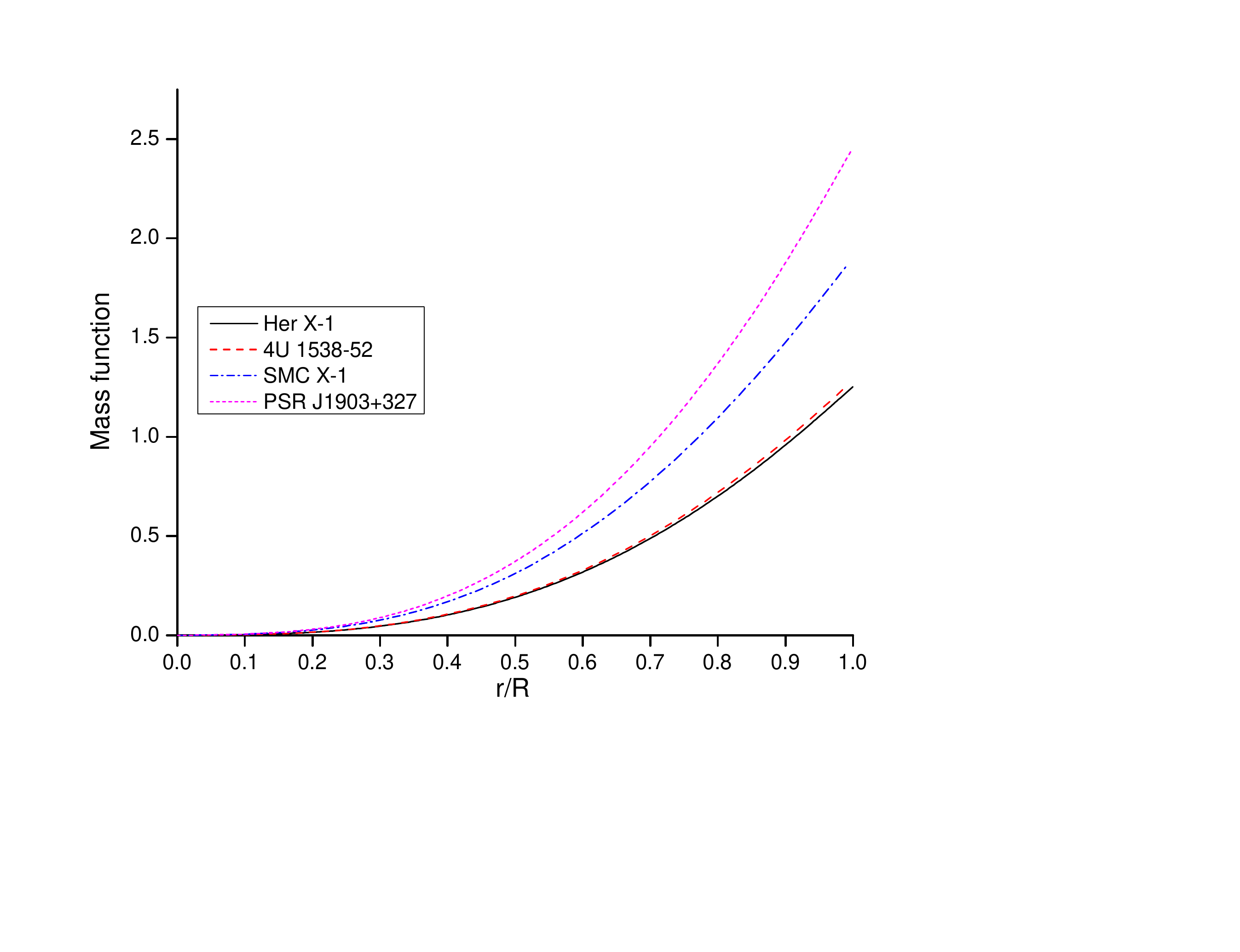}\includegraphics[width=6cm]{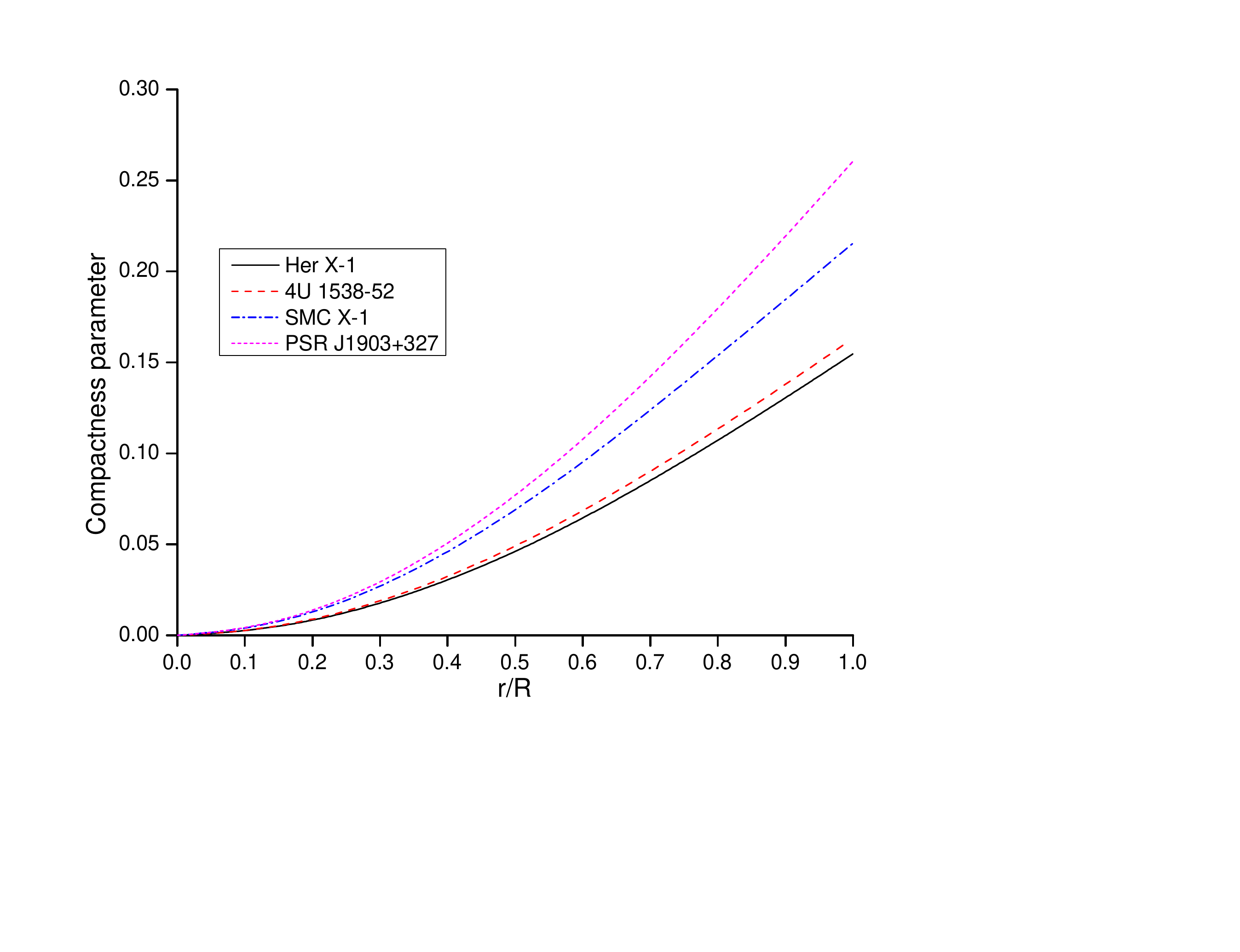}
		\caption{Behaviour of mass function and compactness parameter vs. fractional radius $ r/R $ for case I(first row), case II(second row) and case III(third row). For plotting this figure we have employed data set values of physical parameters and constants which are the same as used in Fig.\ref{f1}.}\label{fe8}
	\end{center}
\end{figure}
\begin{eqnarray} 
	M(r)= \dfrac{Cr^3(K-1)}{2K\,(1+Cr^2)},\label{e35}
\end{eqnarray}
\begin{eqnarray}
u = \dfrac{M}{r} =	\dfrac{Cr^2(K-1)}{2K\,(1+Cr^2)}\label{e36}
\end{eqnarray}
Through compactness parameter, we can analyze many extreme properties to compact objects, such as, the emission of x-rays and gamma-rays, which makes them relevant for the high energy astrophysics.
The profile of $ M(r) $ and  $ u(r) $ has shown in Fig.\ref{fe8}. we can see that the mass function is regular at the center of the stars.  Also it is monotonic increasing function of $ r $ and positive inside the relativistic compact stars and the compactness parameter $ u(r) $  increases with increase $ r/R $. This shows that the compactness of compact stars lies in the expected range of Buchdahl limit\cite{Buchdahl}. We have calculated this bound for  compact stars in our model and have shown
it in Table \ref{t1}.
\section{Causality condition and Herrera’s cracking method}
For any relativistic star, radial and transverse velocity of sound should satisfy $v_{r}^{2},v_{t}^{2}\leq 1$ and this is known as the causality condition. On the other hand, Le Chatelier's principle demands that the speed of sound be positive, i.e.,$v_{r}^{2},v_{t}^{2}\geq 0$. Combining the above two inequalities, one gets $0 < 
v_{r}^{2},v_{t}^{2}\leq1$, where the radial and transverse velocity of sound are obtained as
\begin{eqnarray}
v_{r}^{2} =\frac{dp_r}{d\rho}=\frac{dp_r}{dr}/\frac{d\rho}{dr}\label{26}\\v_{t}^{2} =\frac{dp_t}{d\rho}=\frac{dp_t}{dr}/\frac{d\rho}{dr}\label{27}
\end{eqnarray}
where $ \dfrac{d\rho}{dr}=\dfrac{2Cr\,(1-K)(5+Cr^2)}{K(1+Cr^2)^3} $, the expressions of $ =\frac{dp_r}{dr}~~\textrm{and}~~\frac{d\rho}{dr} $ are as follows\\
\textbf{Case I} 
\begin{eqnarray}
\frac{dp_r}{dr}&=&\frac{2Cr}{K}\left[ \frac{ K(Cr^{ 2 }-1)-2Cr^{2}}{\left( 1 + Cr^{2}\right)^{3}}+\frac{1 + Cr^{ 2 }} {\sqrt{K+Cr^{2}}}-2\sqrt{ K +Cr^{2}}-\frac{\left(1 +Cr^{ 2 } \right)} {(\sqrt{ K + Cr^2 }+(B_1/A_1) \sqrt{K - 1})}\right]\label{28}\\
\frac{dp_t}{dr}&=&\frac{dp_r}{dr}+\frac{Cr}{2K(1+Cr^2)^3}\left[ (4K-7)Cr^2 +\frac{1-2Cr^2(4KCr^2-7Cr^2-K-2)}{1+Cr^2} \right]\label{29}   
\end{eqnarray}
\textbf{Case II}
\begin{eqnarray}
\frac{dp_r}{dr}&=&\frac{2Cr}{K}\Bigg[\frac{ K(Cr^{ 2 }-1)-2Cr^{2}}{\left( 1 + Cr^{2}\right)^{3}}+ \frac{\Phi\left(1-2K-Cr^{2 }\right)}{\sqrt{K-1}\sqrt{K+Cr^{2}}\left(1+Cr^{ 2}\right)^{2}}\left(\frac{e^{\Phi\sqrt{\frac{K+Cr^2}{K-1}}}-(B_2/A_2)\,e^{-\Phi\sqrt{\frac{K+Cr^2}{K-1}}}}{e^{\Phi\sqrt{\frac{K+Cr^2}{K-1}}}+(B_2/A_2)\,e^{-\Phi\sqrt{\frac{K+Cr^2}{K-1}}}}\right)\nonumber\\&& +\frac{4\Phi^2(B_2/A_2)}{(K-1)(1+Cr^2)\Big(e^{\Phi\sqrt{\frac{K+Cr^2}{K-1}}}+(B_2/A_2)\,e^{-\Phi\sqrt{\frac{K+Cr^2}{K-1}}}\Big)}\Bigg]\label{30}\\
\frac{dp_t}{dr}&=&\frac{dp_r}{dr}+\frac{2Cr}{K(1+Cr^2)^2}\left[ \frac{(K-1)5Cr^2}{4(1+Cr^2)^2} +\frac{1-2Cr^2(4KCr^2-7Cr^2-K-2)}{4(1+Cr^2)}-\frac{\Phi^2(1+Cr^2)}{1-K} \right] \label{31} 
\end{eqnarray}
\textbf{Case III}
\begin{eqnarray}
\frac{dp_r}{dr}&=&\frac{2Cr}{K}\Bigg[\frac{ K(Cr^{ 2 }-1)-2Cr^{2}}{\left( 1 + Cr^{2}\right)^{3}}+ \frac{\Phi\left(1-2K-Cr^{2 }\right)}{\sqrt{K-1}\sqrt{K+Cr^{2}}\left(1+Cr^{ 2}\right)^{2}}F(r)\nonumber\\&& +\frac{-\Phi^2(1+(B_3/A_3)^2)}{(K-1)(1+Cr^2)\Big(\cos\Big(\Phi\sqrt{\frac{K+Cr^2}{K-1}}\Big)+(B_3/A_3)\,\sin\Big(\Phi\sqrt{\frac{K+Cr^2}{K-1}}\Big)\Big)}\Bigg]\label{32} \\
\frac{dp_t}{dr}&=&\frac{dp_r}{dr}+\frac{2Cr}{K(1+Cr^2)^2}\left[ \frac{(K-1)5Cr^2}{4(1+Cr^2)^2} +\frac{1-2Cr^2(4KCr^2-7Cr^2-K-2)}{4(1+Cr^2)}+\frac{\Phi^2(1+Cr^2)}{1-K} \right]\label{33}   
\end{eqnarray}
 With the help of graphical representation we have shown that our model satisfies the causality condition (please see Fig.\ref{3}(left)). Now we are interested in checking the
stability of the anisotropic stars under the radial perturbations using the concept of Herrera\cite{Herrera}. By using the principle
of Herrera \cite{Herrera}, Abreu et al.\cite{Abreu} introduced the concept of "cracking", which states that the region of an anisotropic
fluid sphere, where $ -1\leq v_{t}^{2}-v_{r}^{2}\leq 0 $, is potentially stable but the region where $ 0< v_{t}^{2}-v_{r}^{2}\leq 1 $ is potentially unstable. From Fig.\ref{3}(right) it is clear that our model is potentially stable. Moreover $ 0<v_{r}^{2}<1 $ and $ 0<v_{r}^{2}<1 $  therefore,
according to Andreasson\cite{30}, $|v_{r}^{2}-v_{t}^{2}| < 1$ which is also shown in Fig.\ref{3}(left).
\section{Tolman-Oppenheimer-Volkoff (TOV) equations}
\begin{figure}[]
 \begin{center}
 \includegraphics[width=5cm]{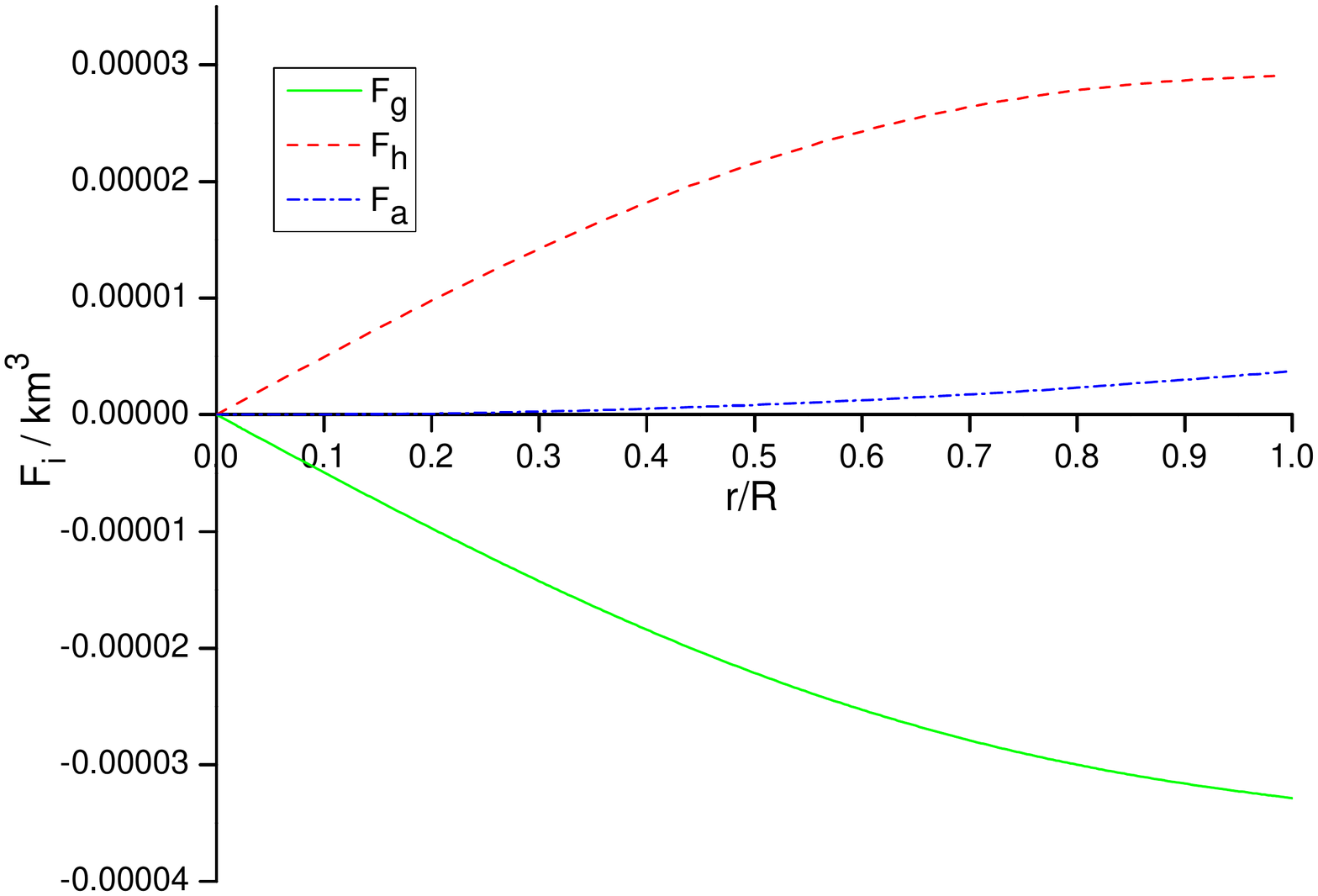}\includegraphics[width=5cm]{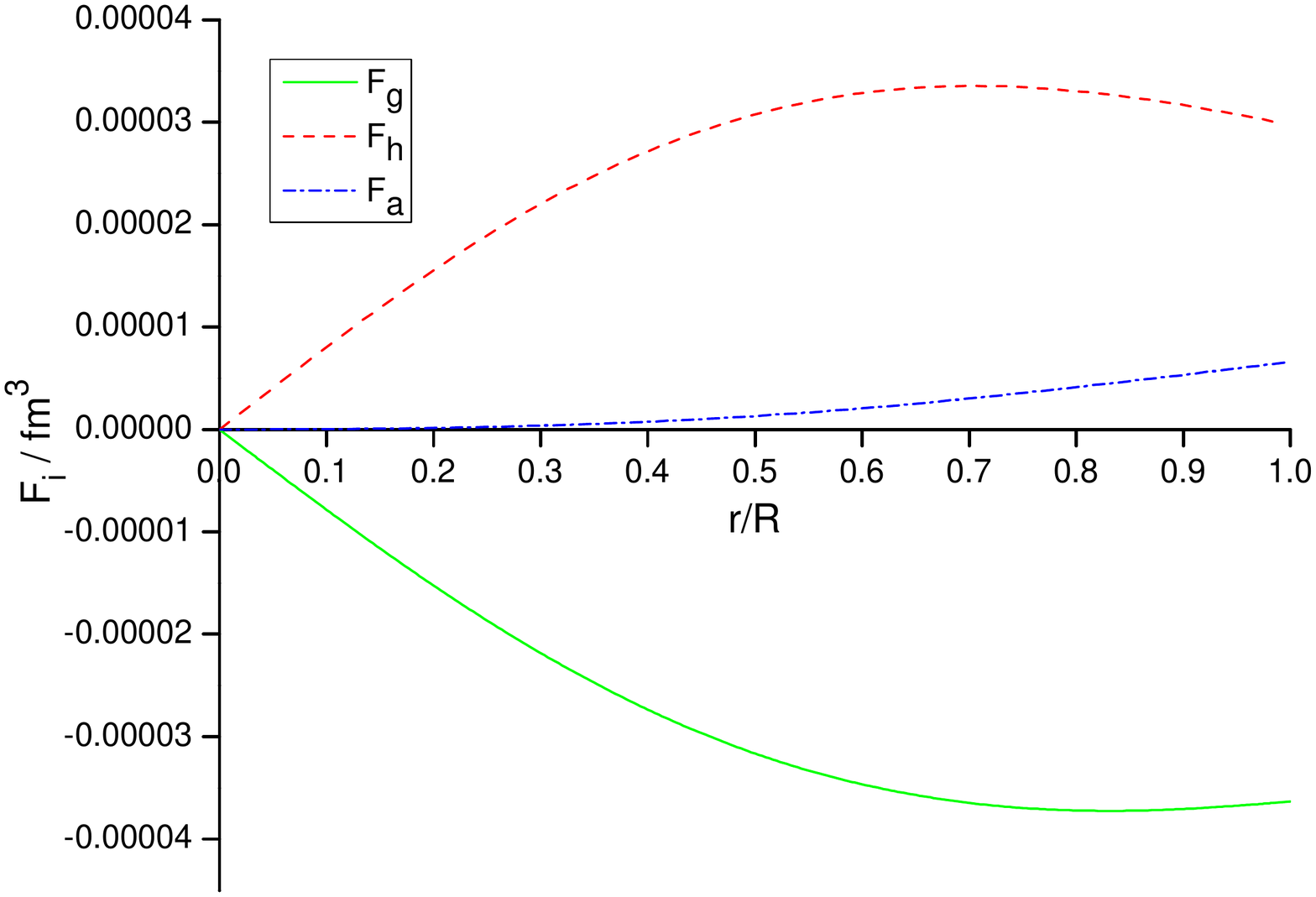}\includegraphics[width=5cm]{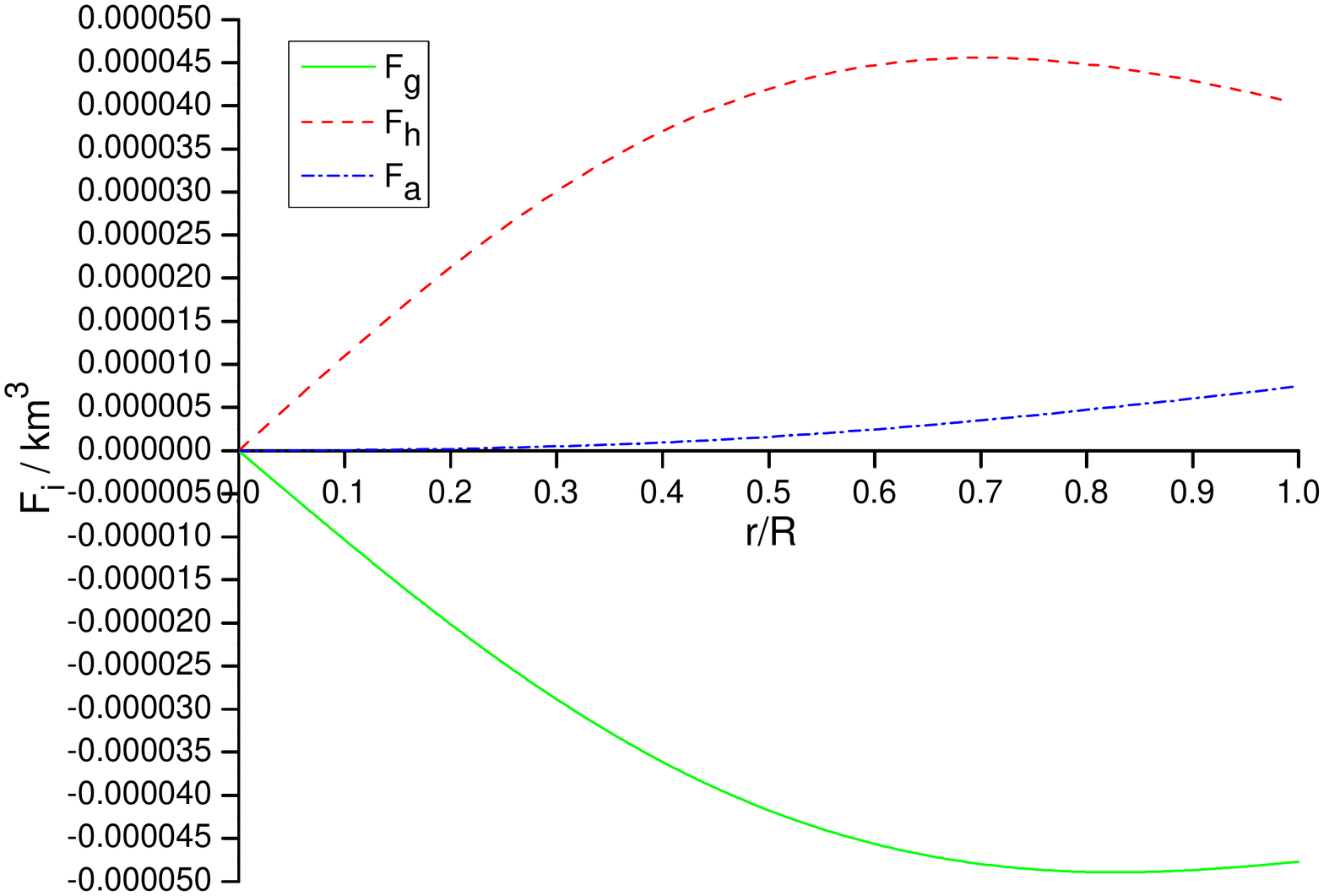}\\\includegraphics[width=5cm]{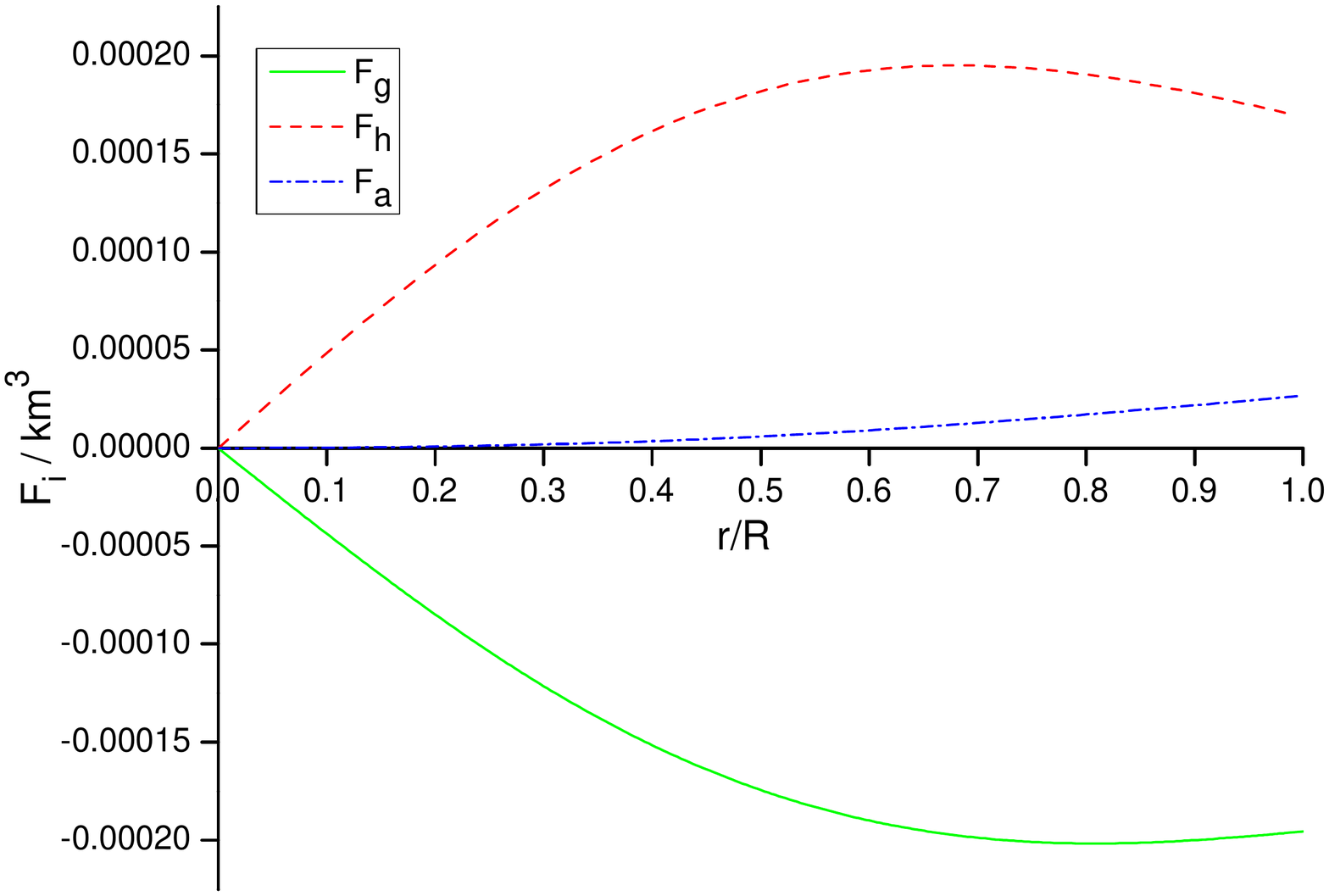}\includegraphics[width=5cm]{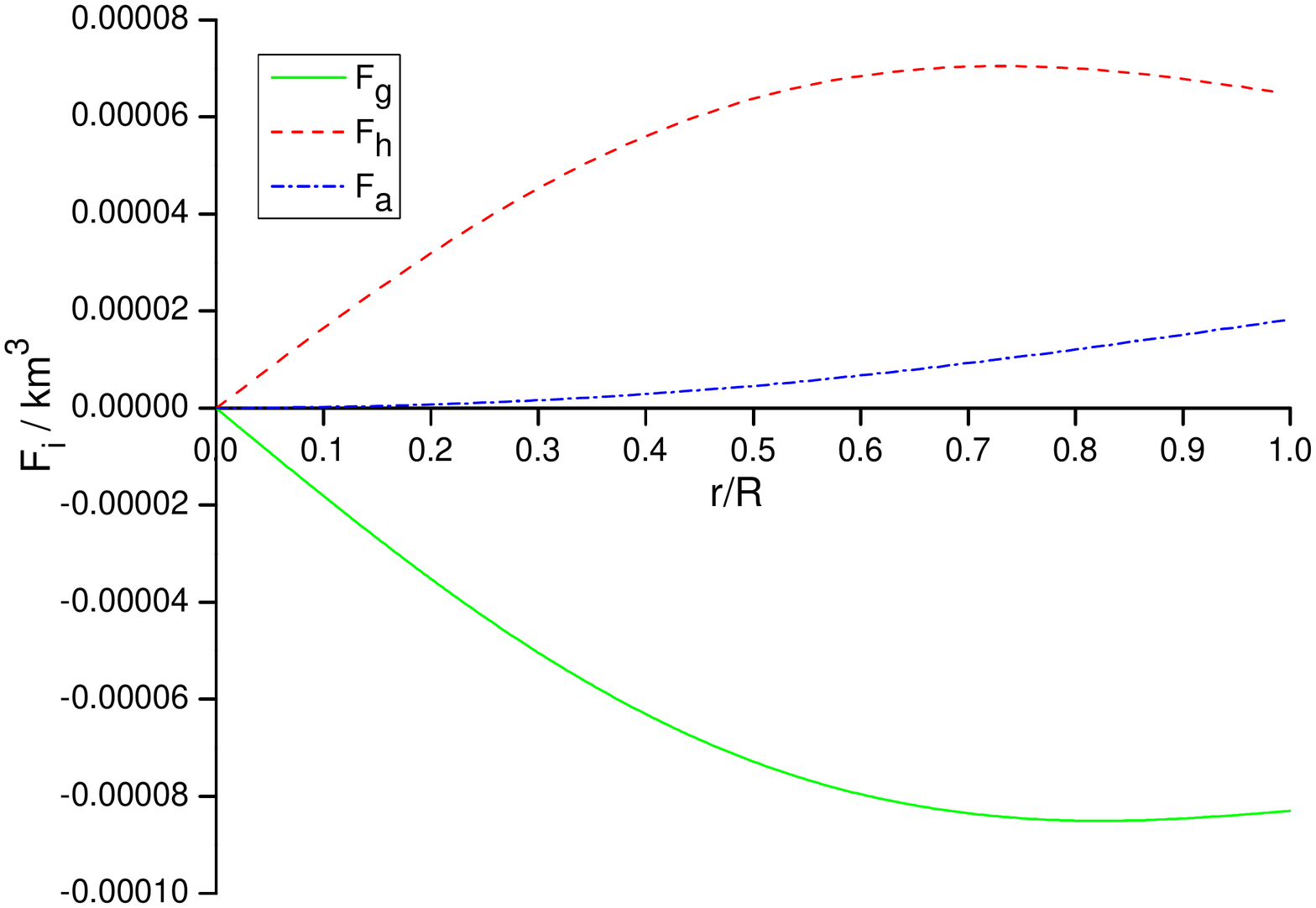}\includegraphics[width=5cm]{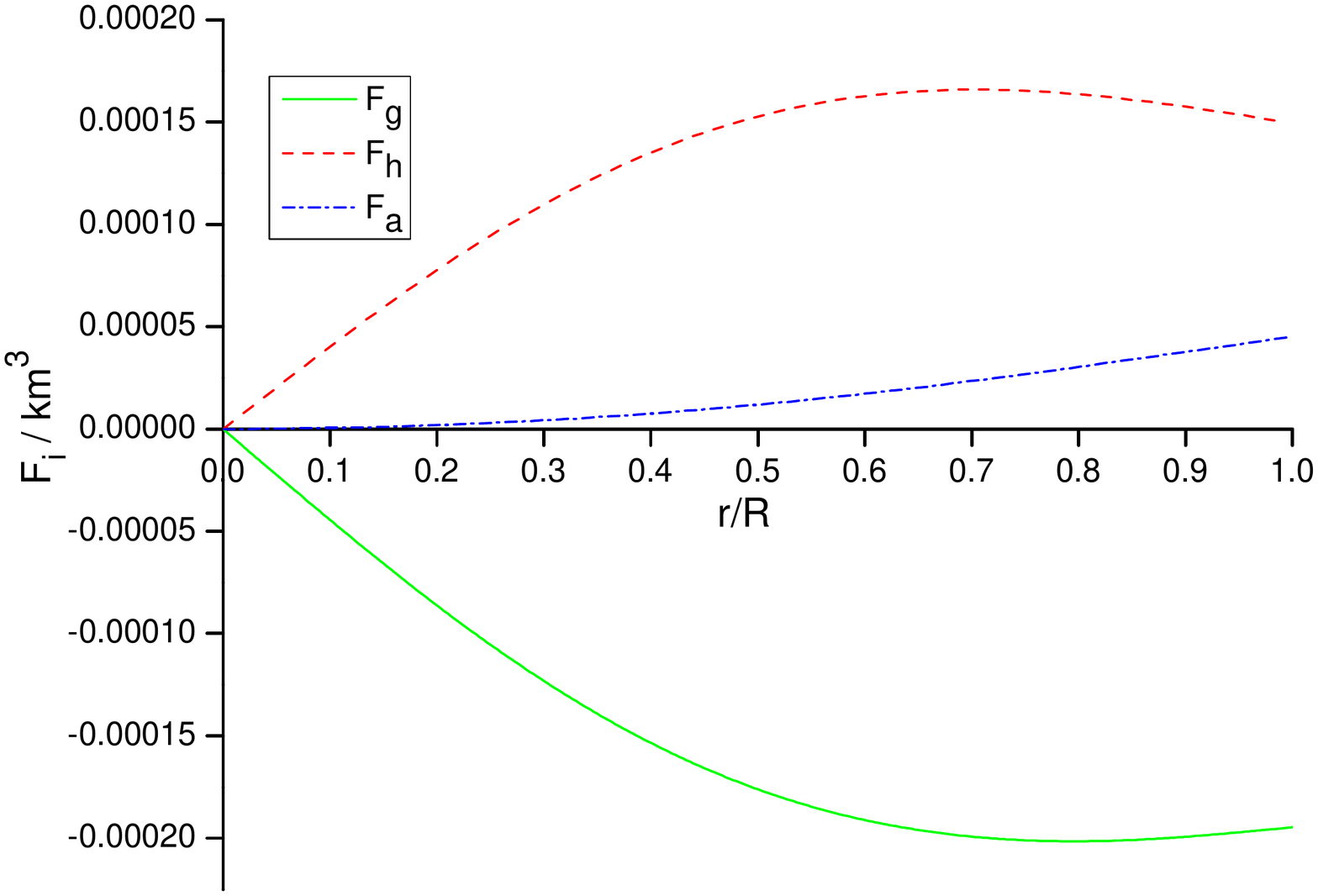}\\\includegraphics[width=5cm]{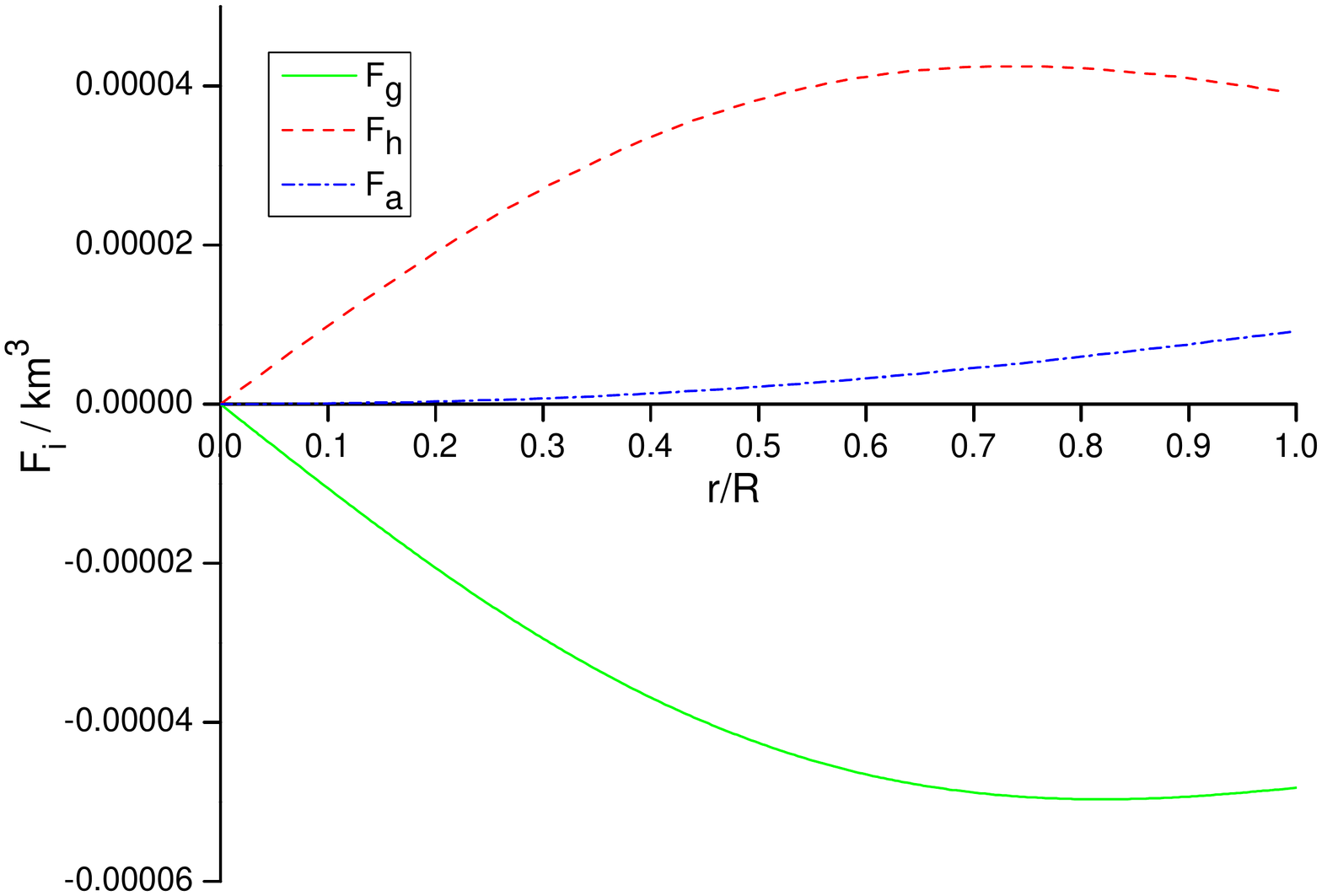}\includegraphics[width=5cm]{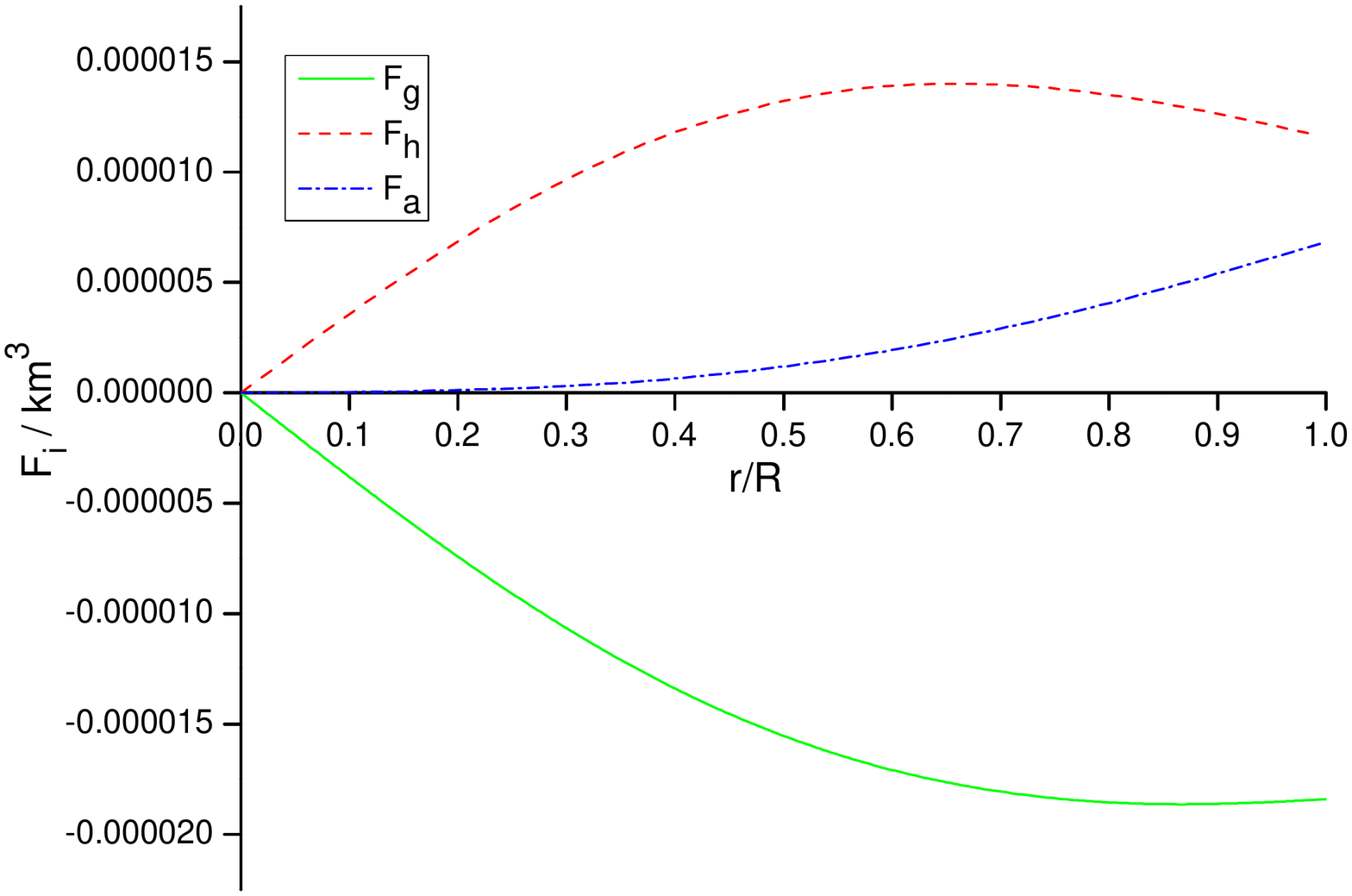}
 \includegraphics[width=5cm]{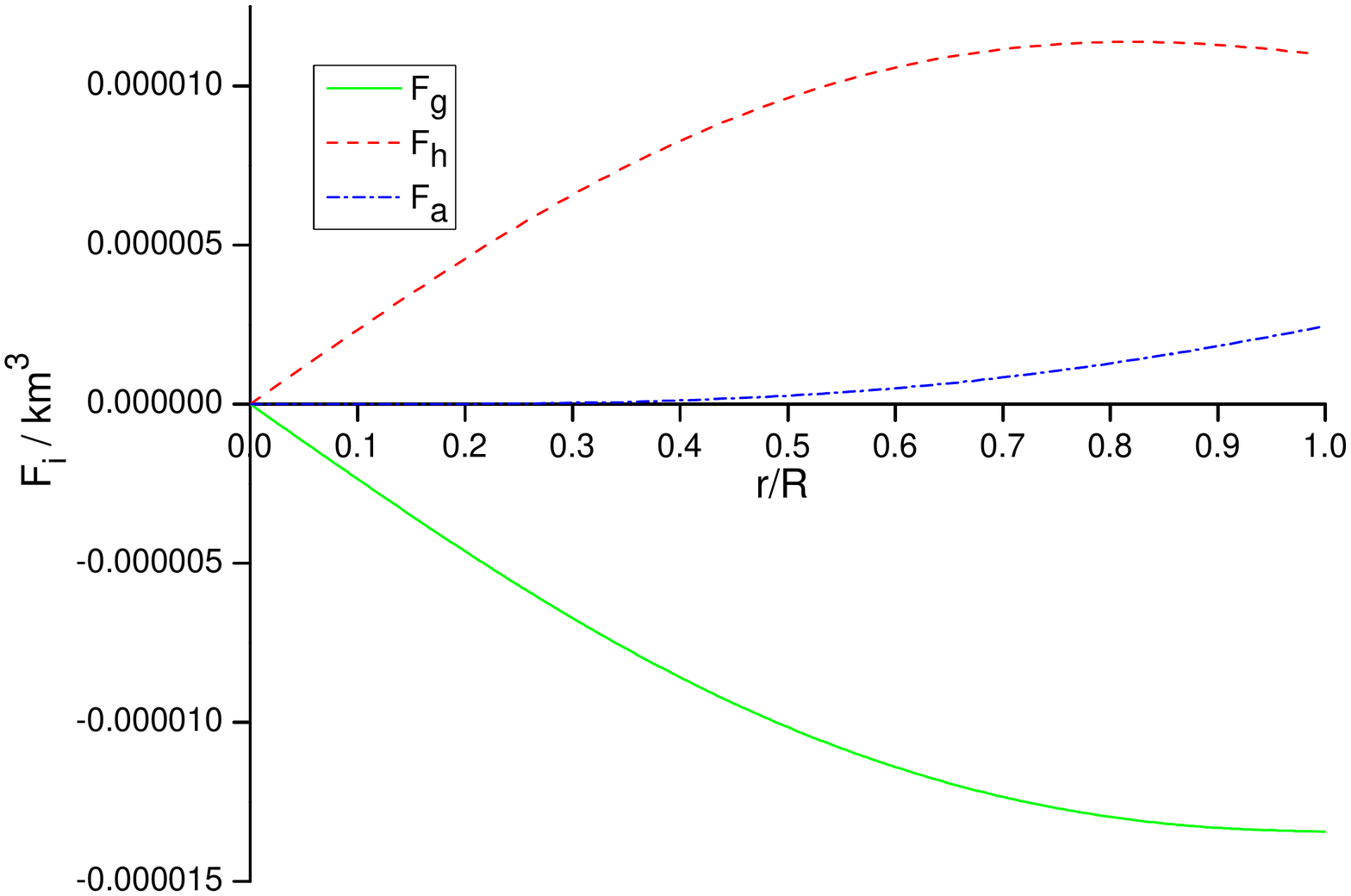}\\\includegraphics[width=5cm]{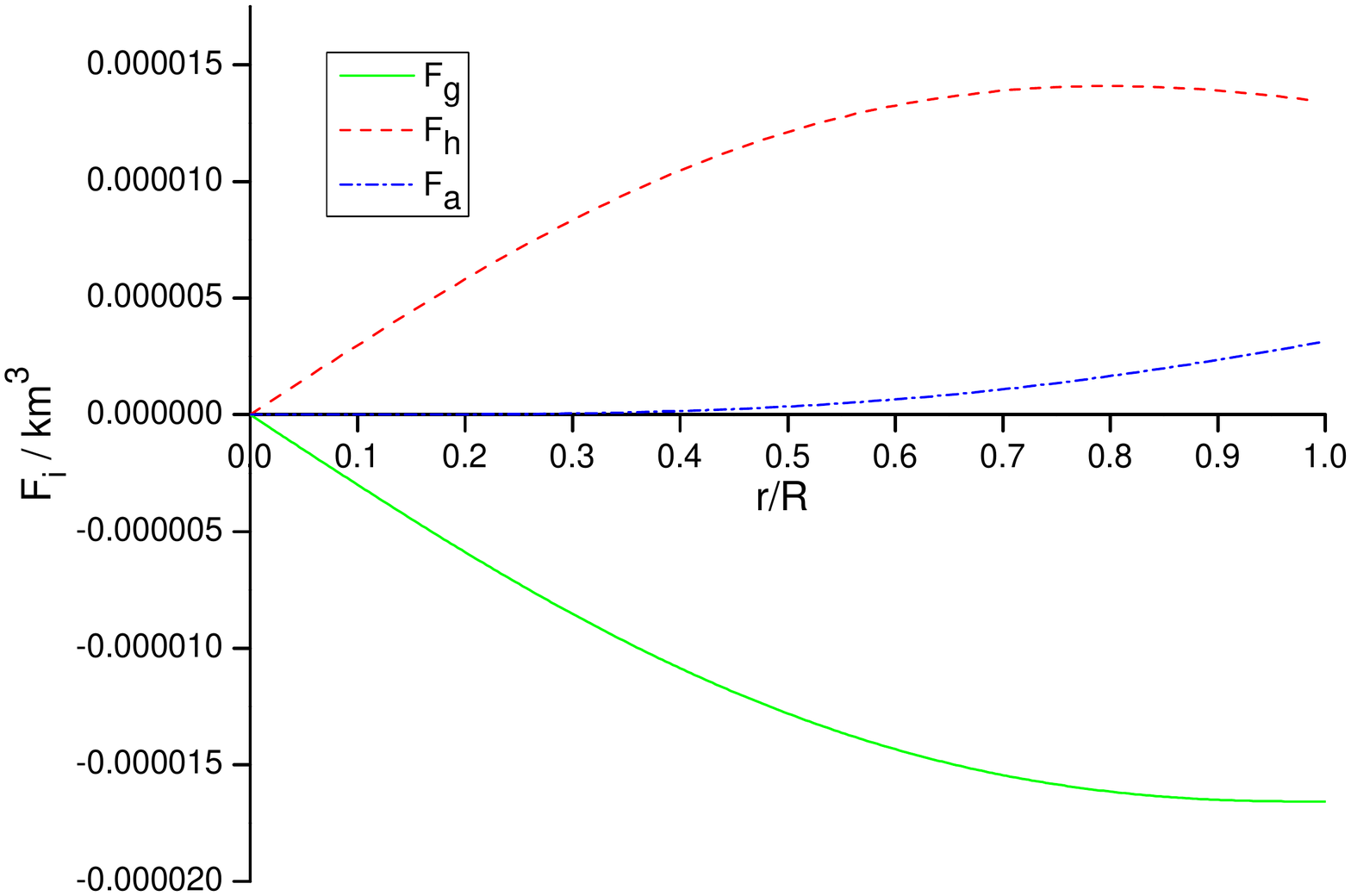}\includegraphics[width=5cm]{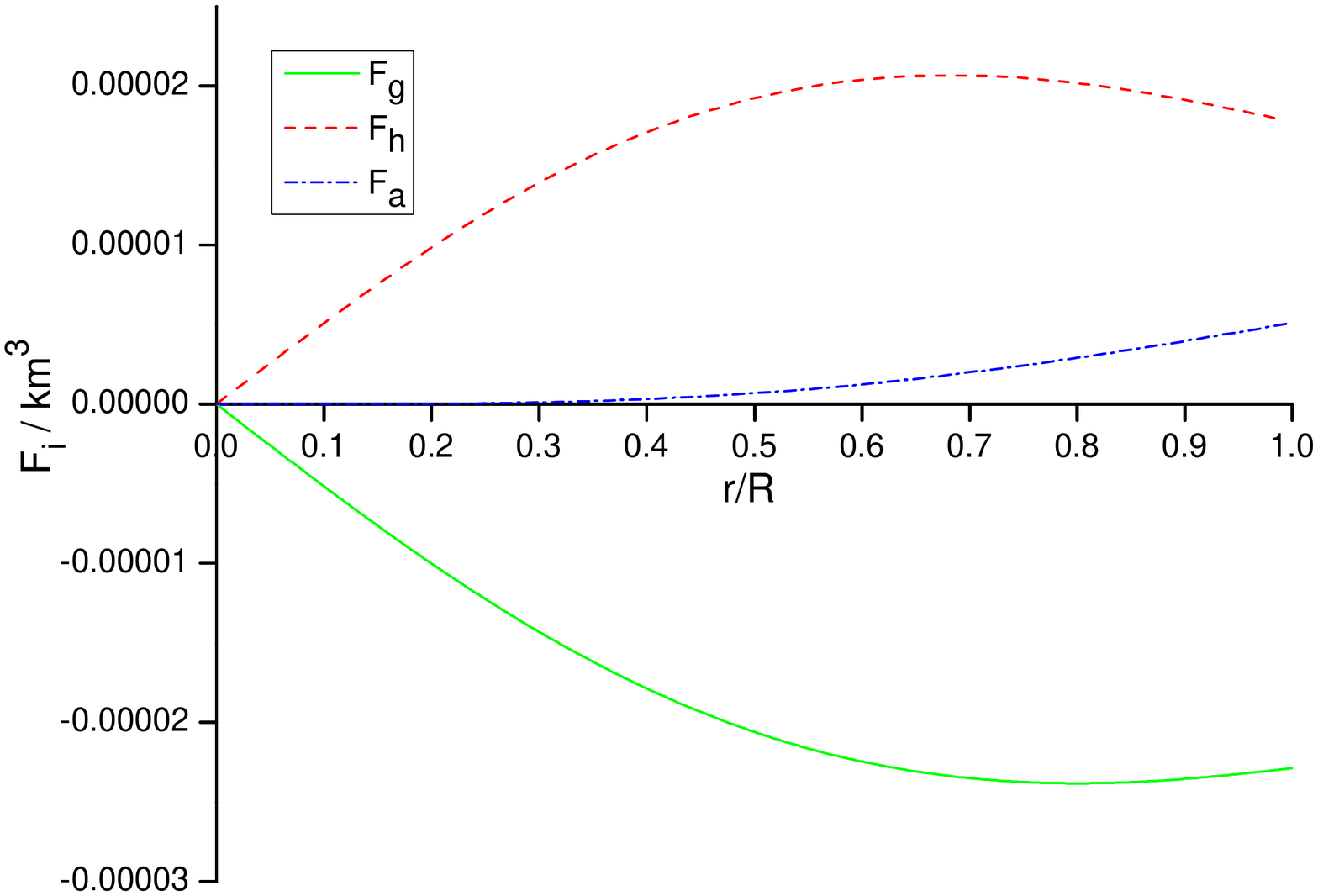}\includegraphics[width=5cm]{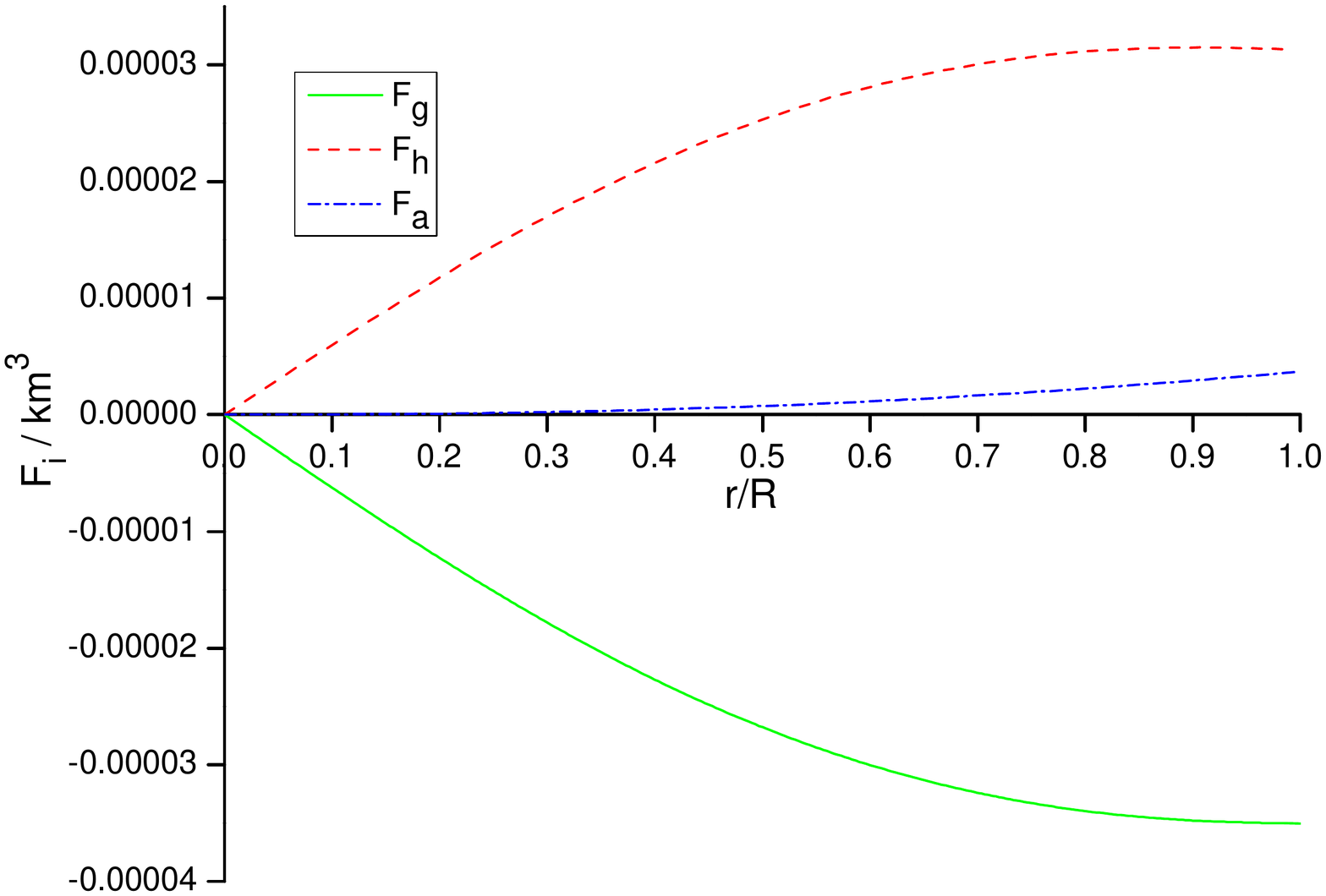}
 \caption{Behaviour of hydrostatic, anisotropic and gravitational forces vs. fractional radius $ r/R $ for case I(first four graph), case II(middle four graph) and case III(last four graph).For plotting this figure we have employed data set values of physical parameters and constants which are the same as used in Fig.\ref{f1}.}\label{f4}
 \end{center}
 \end{figure}

Now we are interested in checking the effect of three different forces, viz. gravitational, hydrostatic and anisotropic force,on our given system. The governing generalized Tolman-Oppenheimer-Volkoff(TOV) equation \cite{35,36} is given by
\begin{eqnarray}
 -\frac{M_G(\rho+p_r)}{r^2}e^{\frac{\lambda-\nu}{2}}-\frac{dp_r}{dr}+\frac{2(p_t-p_r)}{r} =0,\label{34} 
 \end{eqnarray}
where $M_G$ is the effective gravitational mass given by:
\begin{eqnarray}
M_G(r)=\frac{1}{2}r^2 \nu^{\prime}e^{(\nu - \lambda)/2}.\label{35}
\end{eqnarray}
From the equation(\ref{35}) the value of $M_G(r)$ in equation (\ref{34}), we get
\begin{eqnarray}
-\frac{\nu'}{2}(\rho+p_r)-\frac{dp_r}{dr}+\frac{2(p_t-p_r)}{r} =0,  \label{36}
\end{eqnarray}
The equation (\ref{36}) can be expressed into three different elements gravitational $(F_g)$, hydrostatic $(F_h)$ and anisotropic force $(F_a)$ from an equilibrium point of view, which are defined as: 
\begin{eqnarray}
F_g=-\frac{\nu'}{2}(\rho+p),~~~~F_h=-\frac{dp_r}{dr}~~~~\textrm{and}~~~~ F_a =\frac{2(p_t-p_r)}{r}=\frac{2\Delta}{C} \label{37}
\end{eqnarray}
Here, we can derive the hydrostatic force $(F_h)$ from eqns.(\ref{28}),(\ref{30}), (\ref{32}) and anisotropic force $(F_a) = \Delta_i$, where $ i=1, 2, 3 $ for case I, case II and case III. Now the expression of gravitational force $ F_g $ is written in an explicit form:\\
\textbf{Case I}
\begin{eqnarray}
F_g = \frac{Cr}{1-K}\left[ \frac{1-K}{2(1+Cr^2)}-\frac{\sqrt{\frac{K-1}{K+Cr^2}}}{\sqrt{\frac{K+Cr^2}{K-1}}+(B_1/A_1)} \right]\label{38}
\end{eqnarray}
\textbf{Case II}
\begin{eqnarray}
F_g = \frac{Cr}{1-K}\left[ \frac{1-K}{2(1+Cr^2)}-\Phi\sqrt{\frac{K-1}{K+Cr^2}}\left(\frac{e^{\Phi\sqrt{\frac{K+Cr^2}{K-1}}}-(B_2/A_2)\,e^{-\Phi\sqrt{\frac{K+Cr^2}{K-1}}}}{e^{\Phi\sqrt{\frac{K+Cr^2}{K-1}}}+(B_2/A_2)\,e^{-\Phi\sqrt{\frac{K+Cr^2}{K-1}}}}\right) \right]\label{39}
\end{eqnarray}
\textbf{Case III}
\begin{eqnarray}
F_g = \frac{Cr}{1-K}\left[ \frac{1-K}{2(1+Cr^2)}-\Phi\sqrt{\frac{K-1}{K+Cr^2}}F(r) \right]\label{40}
\end{eqnarray}
Figure (\ref{f4}) represents the behavior of the generalized
TOV equations. We observe from these figures that the system
is counterbalanced by the components the gravitational
force $(F_g)$, hydrostatic force$(F_h)$ and electric
force $(F_e)$ and the system attains a static equilibrium.
\section{Energy conditions}
\begin{figure}[]
 \begin{center}
 \includegraphics[width=5cm]{DEN1CH7.pdf}\includegraphics[width=5cm]{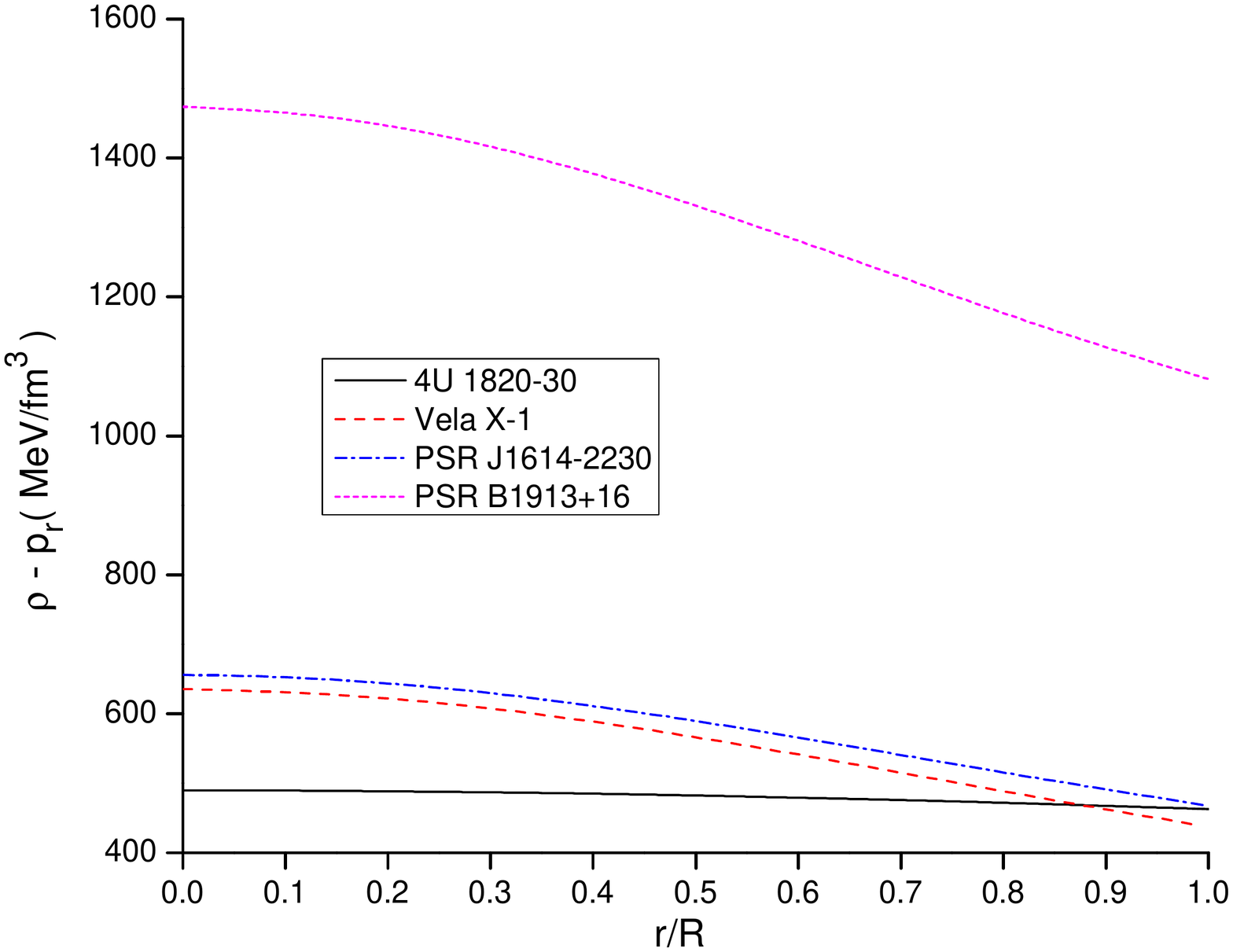}\includegraphics[width=5cm]{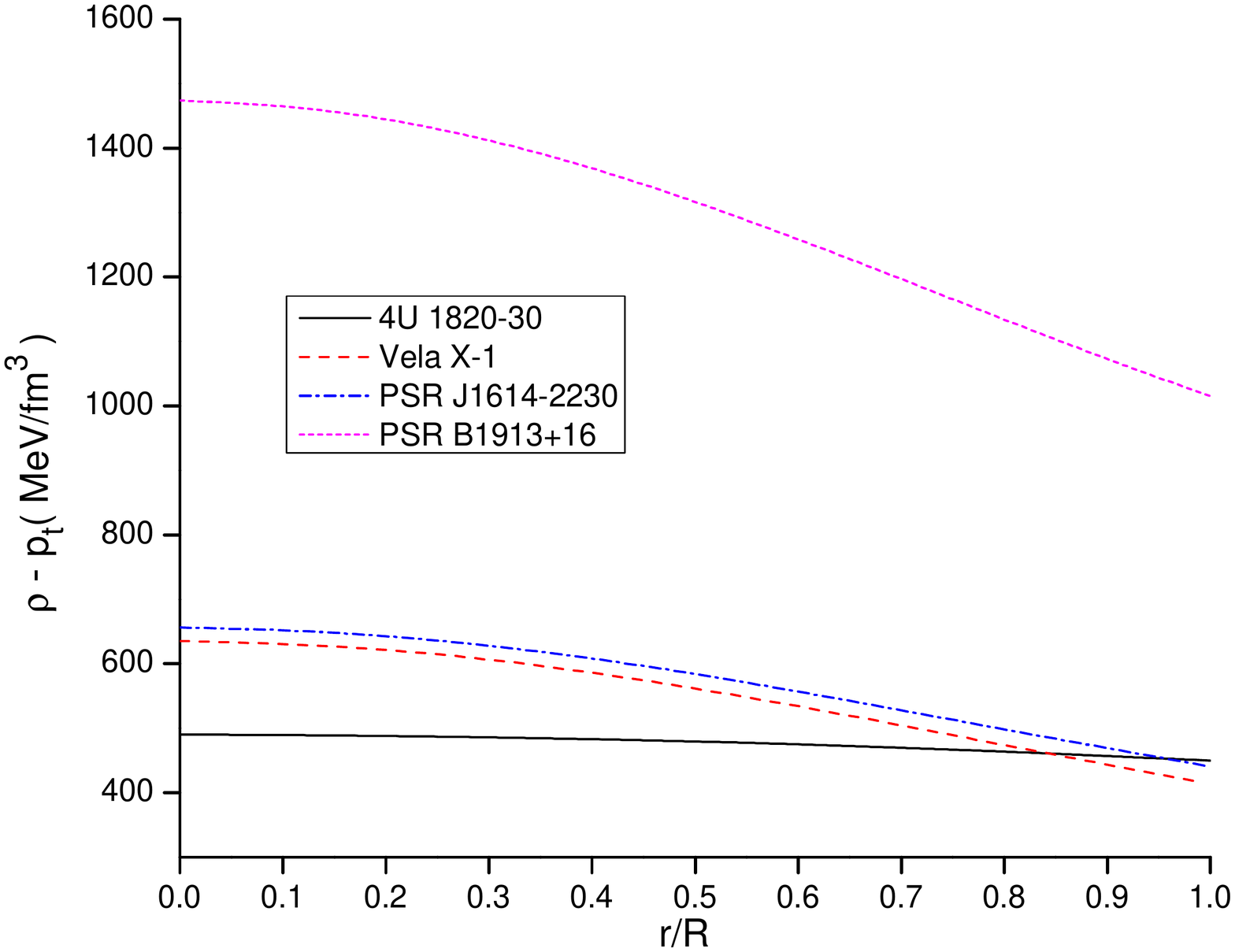}\\\includegraphics[width=5cm]{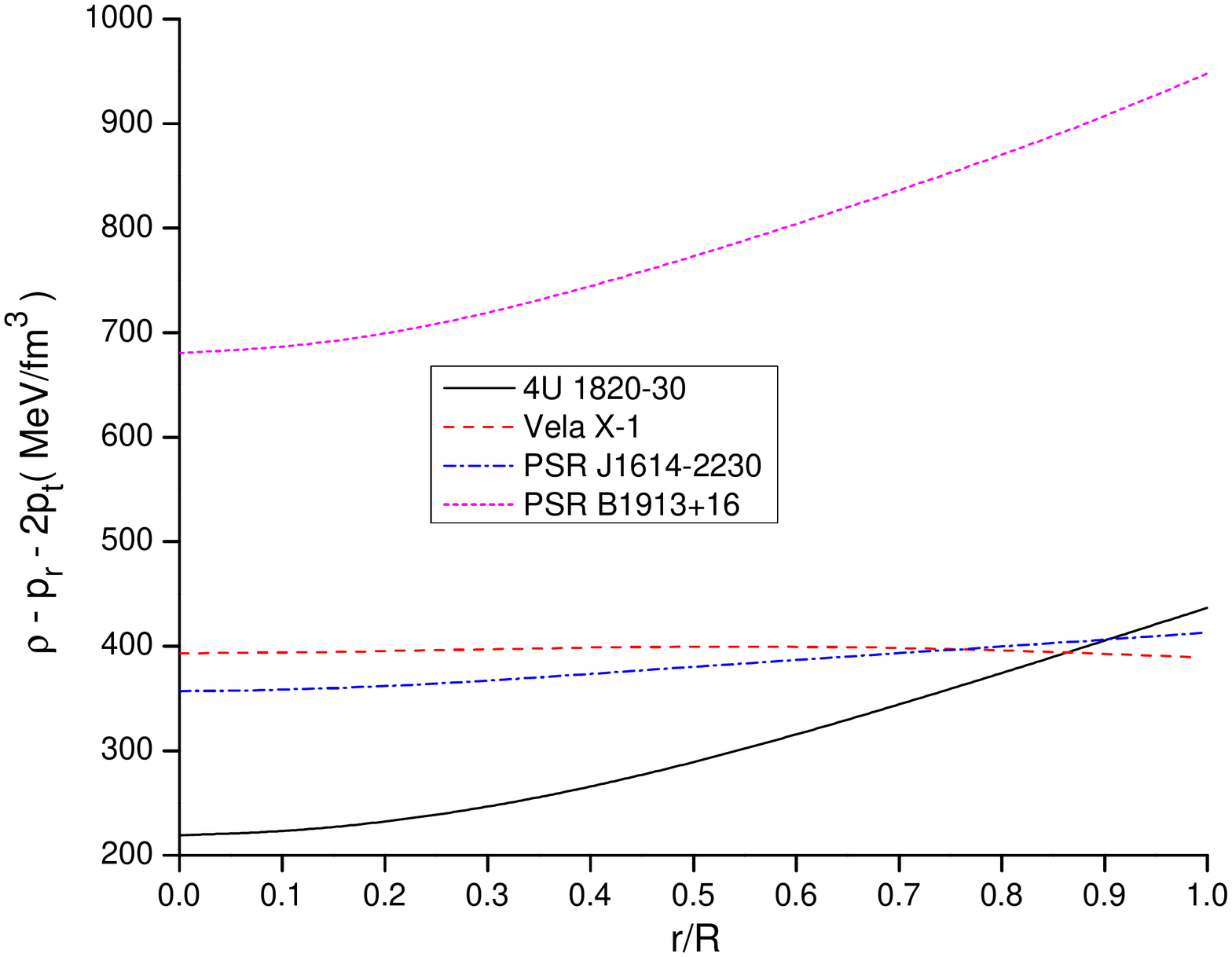}\includegraphics[width=5cm]{DEN2CH7.pdf}\includegraphics[width=5cm]{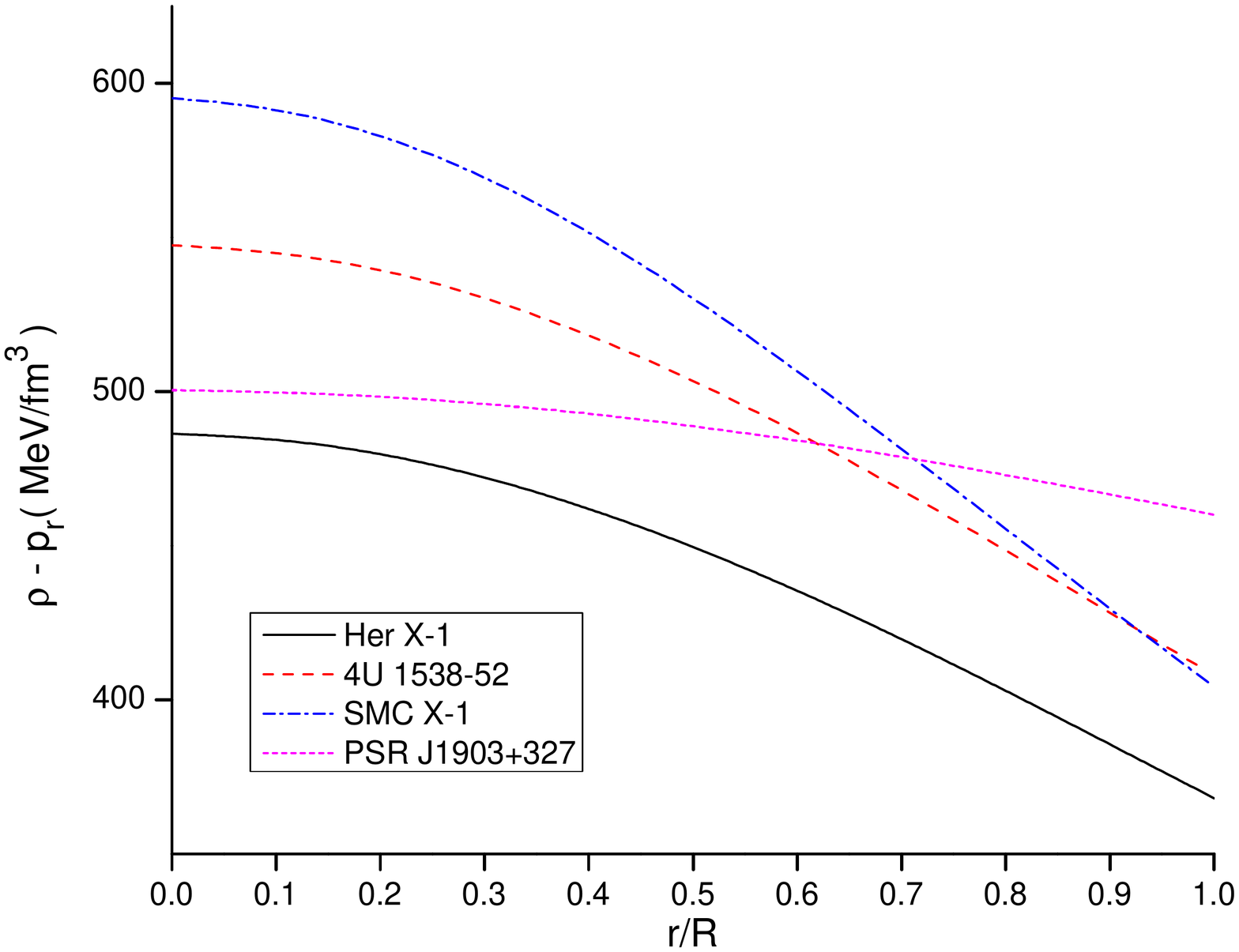}\\\includegraphics[width=5cm]{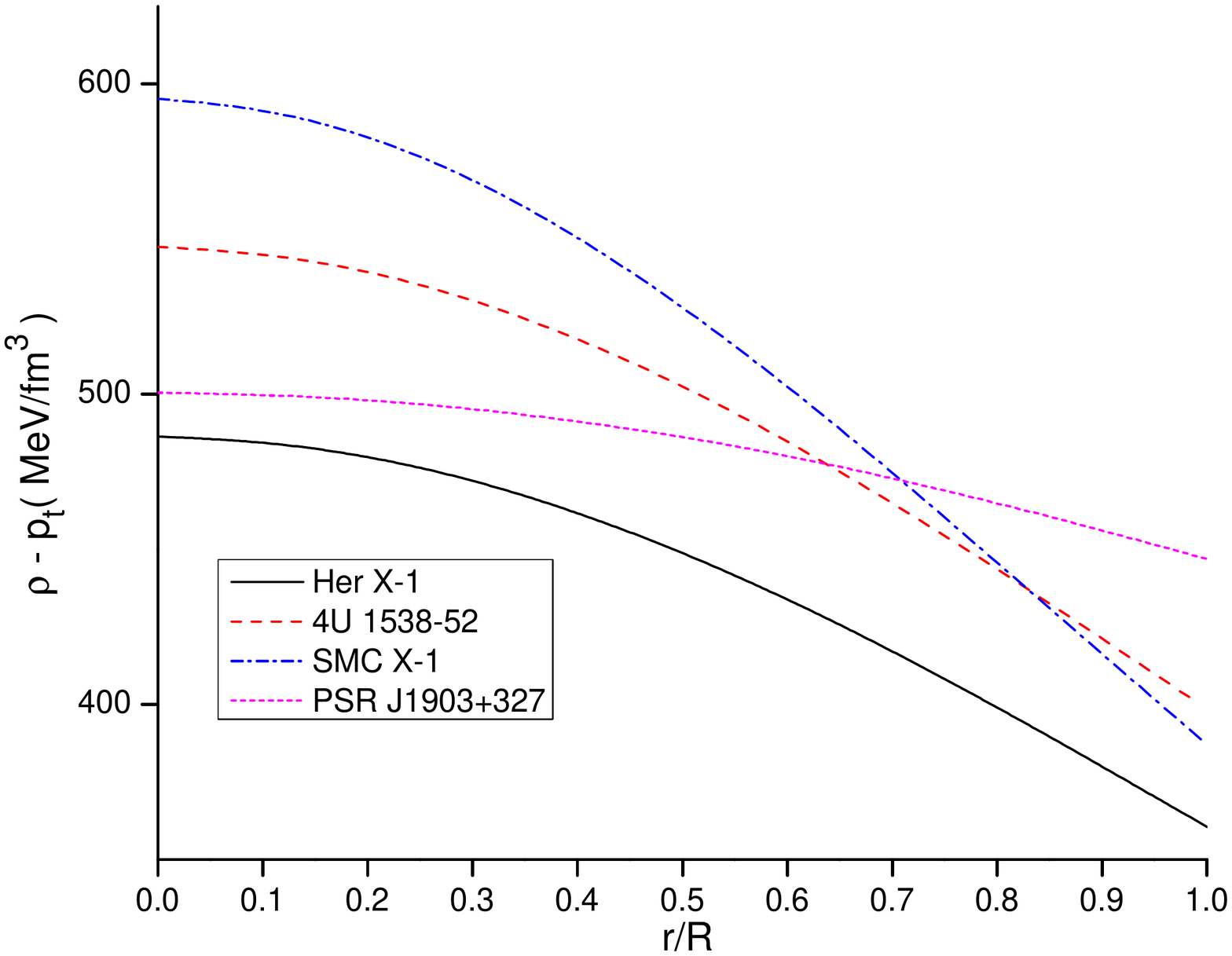}\includegraphics[width=5cm]{SC1CH7.pdf}\includegraphics[width=5cm]{DEN3CH7.pdf}\\\includegraphics[width=5cm]{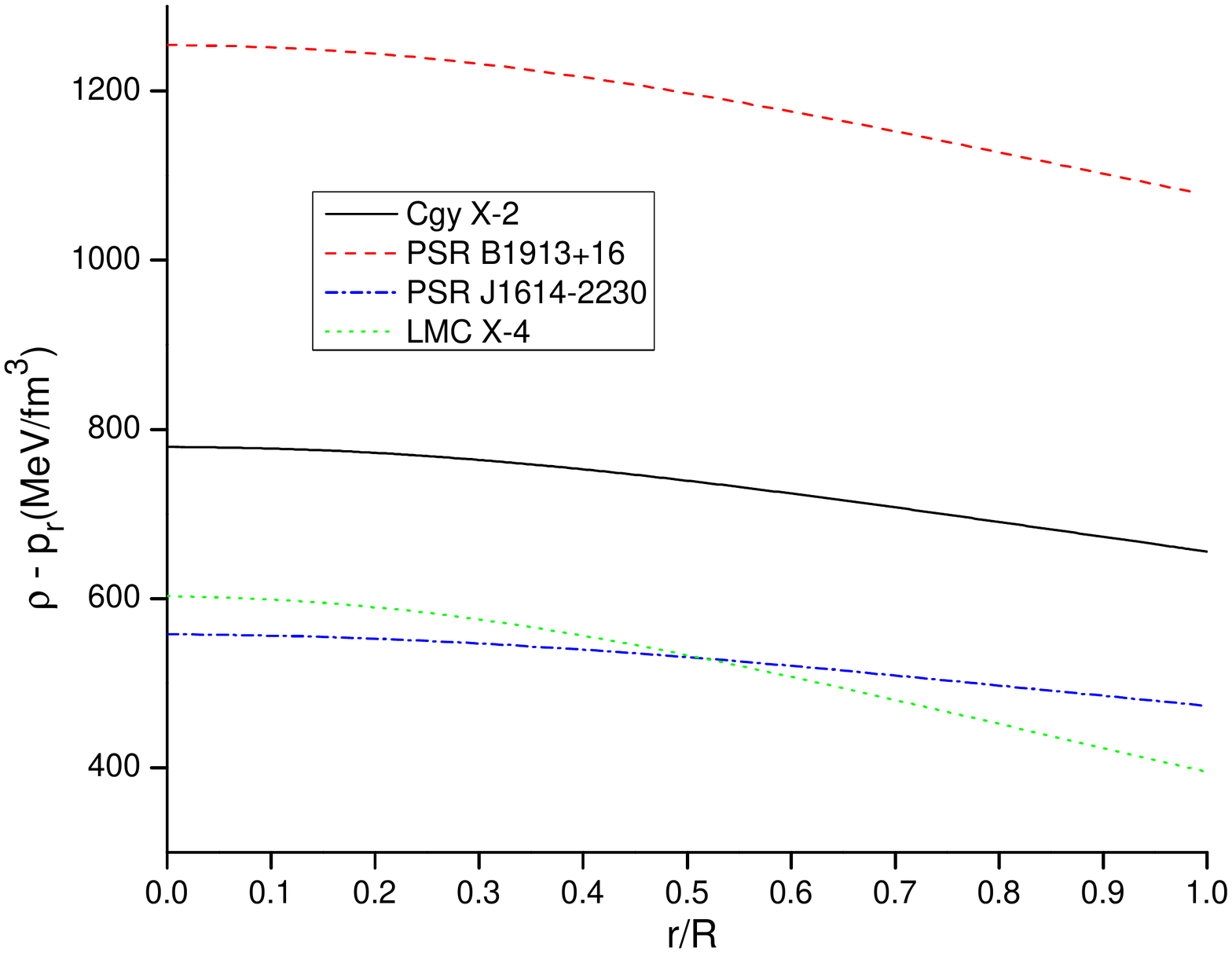}\includegraphics[width=5cm]{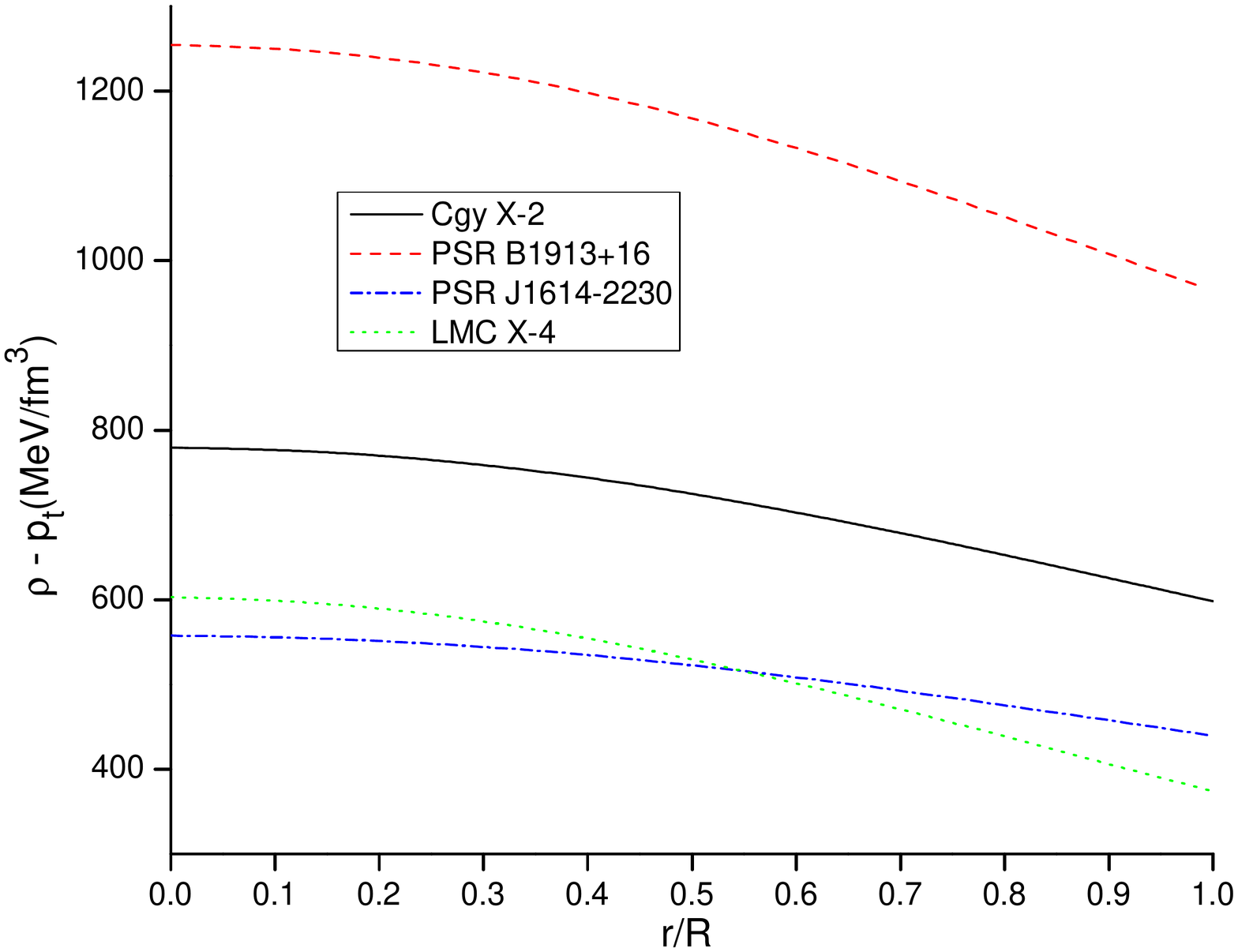}\includegraphics[width=5cm]{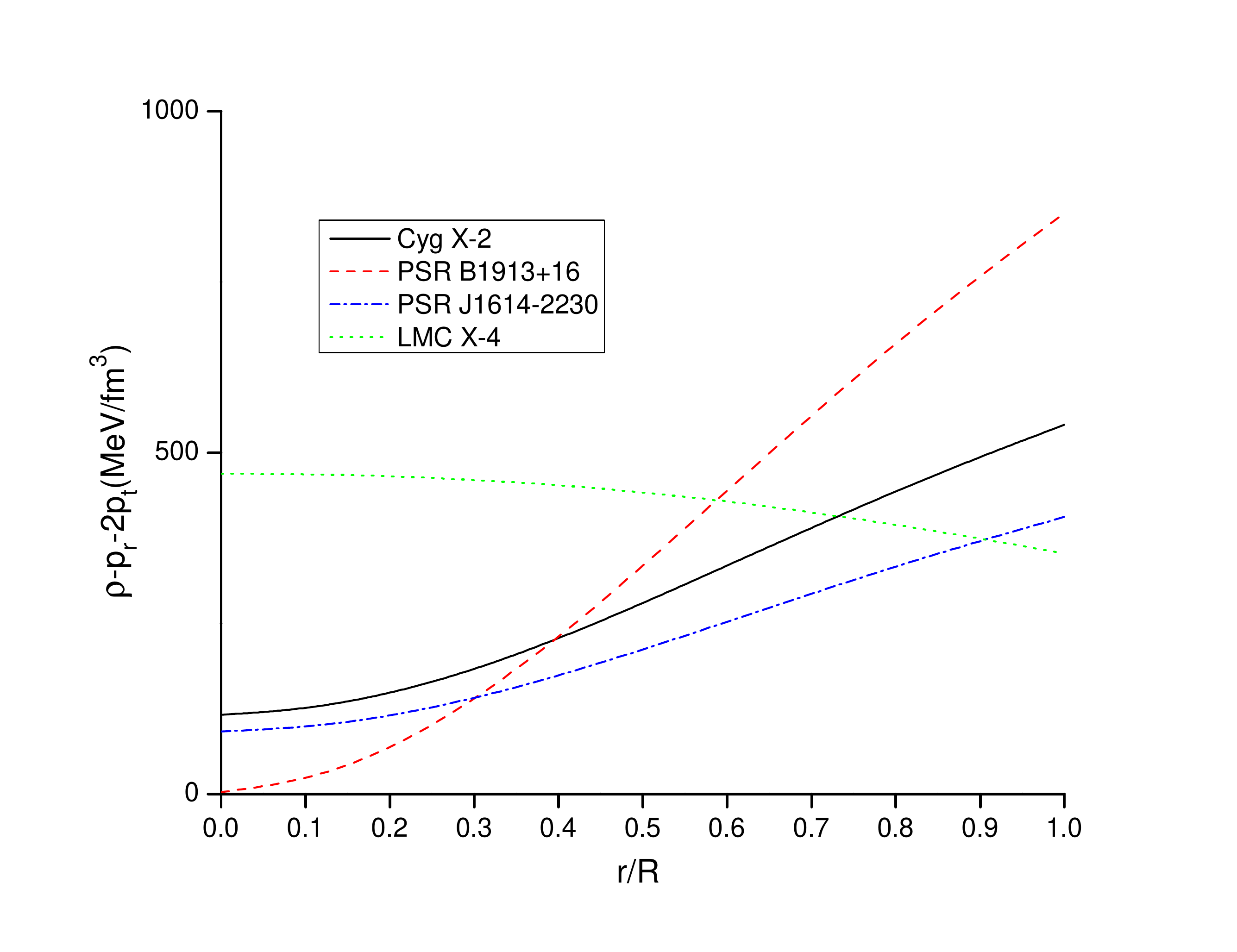}
 \caption{Behaviour of energy conditions vs. fractional radius $ r/R $ for case I(first four graph), case II(middle four graph) and case III(last four graph).For plotting this figure we have employed data set values of physical parameters and constants which are the same as used in Fig.\ref{f1}.}\label{f5}
 \end{center}
 \end{figure}

Inside the anisotropic matter distribution, we analyze the energy conditions according to relativistic classical field theories of gravitation. In the context of GR the energy conditions should be positive. Here we will focus  on (i) the null energy
condition (NEC), (ii) the weak energy condition (WEC),
and (iii) the strong energy condition (SEC)\cite{Ponce,Visser}. In summary:
\begin{enumerate}
\item (NEC) $\rho \geq 0 $
\item (WECr) $\rho - p_r\geq 0 $
\item (NECt) $\rho - p_t\geq 0 $
\item (SEC) $\rho - p_r - 2p_t\geq 0 $
\end{enumerate}
Figure (\ref{f5})demonstrated that all the above inequalities
are fulfilled inside the spherical object. In this way we
have a well-behaved stress-energy tensor.
\begin{table}
\caption{Numerical values of $ R $,  M($ M_{\odot} $), $ C,\, K, $ $ \alpha $ and mass-radius ratio of compact star candidates.}
\label{t1}
\begin{center}
\begin{tabular}{lcccccc}
\hline  Compact star & $ R $($km$) & M($M_{\Theta})$ & $ K $ & $ C $($km ^{-2} $) & $ \alpha $ & M/R \\
\hline 
  4U 1820-30 & 9.316 & 1.58 & -0.4995 & 0.0023 & 0 & 0.25016\\ 
 Vela X-1 & 9.561 & 1.77 & -1.097 & 0.0044 & 0 & 0.27306\\ 
 PSR J1614-2230 & 9.691 & 1.97 & -0.901 & 0.0042 & 0 & 0.29979\\
 PSR B1913+16 & 6.6 & 1.44 & -0.798 & 0.0092 & 0 & 0.3218\\
 Cgy X-2 & 8.31 & 1.73 & -0.8196 & 0.0055 & -0.64 & 0.30718 \\ 
 PSR B1913+16 & 6.6 & 1.44 & -0.82 & 0.0094 & -0.81 & 0.3218\\
 PSR J1614-2230 & 9.691 & 1.97 & -0.8202 & 0.0039 & -0.36 & 0.29979\\
 LMC X-4 & 8.301 & 1.04 & -2.977 & 0.0055 & -2.25 & 0.1847\\
 Her X-1 & 8.1 & 0.85 & -1.9779 & 0.0039 & 0.0081 & 0.15475\\
 4U 1538-52 & 7.866 & 0.87 & -1.9779 & 0.0045 & 0.0064 & 0.16315\\
 SMC X-1 & 8.831 & 1.29 & -1.9776 & 0.0052 & 0.049 & 0.21547\\
 PSR J1903+327 & 9.436 & 1.667 & -0.6 & 0.0027 & 0.16 & 0.26057\\
\hline 
\end{tabular} 
\end{center}
\end{table}

\begin{table}
 \caption{Energy densities and central pressure for different compact star candidates for the above parameter
 values of Table \ref{t1}.}
 \label{t2}
 \begin{center}
 \begin{tabular}{lcccc}
 \hline  Compact star & Central density & Surface density & Central pressure & \\
 & $ (gm/cm^3) $ & $ (gm/cm^3) $ & $ (dyne/cm^2) $& Cases\\
 \hline 
   4U 1820-30 & $ 1.1111 \times 10^{15} $ & $ 8.2301 \times 10^{14} $ & $ 2.1599 \times 10^{35} $& Case I \\ 
  Vela X-1 & $ 1.3433 \times 10^{15} $ & $ 7.7679 \times 10^{14} $ & $ 2.9132 \times 10^{35} $ &  Case I \\ 
  PSR J1614-2230 & $ 1.4324 \times 10^{15} $ & $ 8.3109 \times 10^{14} $ & $ 2.3899 \times 10^{35} $ & Case I \\
  PSR B1913+16 & $ 3.3229 \times 10^{15} $ & $ 1.9214 \times 10^{15} $ & $ 6.3267 \times 10^{35} $ & Case I \\
  Cgy X-2 & $ 1.9754 \times 10^{15} $ & $ 1.1652 \times 10^{15} $ & $ 5.2986 \times 10^{35} $ & Case II  \\ 
  PSR B1913+16 & $ 3.3421 \times 10^{15} $ & $ 1.9142 \times 10^{15} $ & $ 9.9917 \times 10^{35} $ & Case II \\
  PSR J1614-2230 & $ 1.4049 \times 10^{15} $ & $ 8.4066 \times 10^{15} $ & $ 3.7181 \times 10^{35} $ & Case II \\
  LMC X-4 & $ 1.1903 \times 10^{15} $ & $ 7.0236 \times 10^{15} $ & $ 1.057 \times 10^{35} $ & Case II\\
  Her X-1 & $ 0.9538 \times 10^{15} $ & $ 6.5385 \times 10^{14} $ & $ 0.8032 \times 10^{35} $ & Case III \\
  4U 1538-52 & $ 1.0957 \times 10^{15} $ & $ 7.2461 \times 10^{14} $ & $ 0.9751 \times 10^{35} $ & Case III \\
  SMC X-1 & $ 1.2434 \times 10^{15} $ & $ 7.1815 \times 10^{14} $ & $ 1.6685 \times 10^{15} $ & Case III \\
  PSR J1903+327 & $ 1.1683 \times 10^{15} $ & $ 8.1755 \times 10^{14} $ & $ 2.5073 \times 10^{15} $ & Case III \\
 \hline 
 \end{tabular} 
\end{center}
 \end{table}
\section{Adiabatic index}
\begin{figure}[]
 \begin{center}
 \includegraphics[width=5cm]{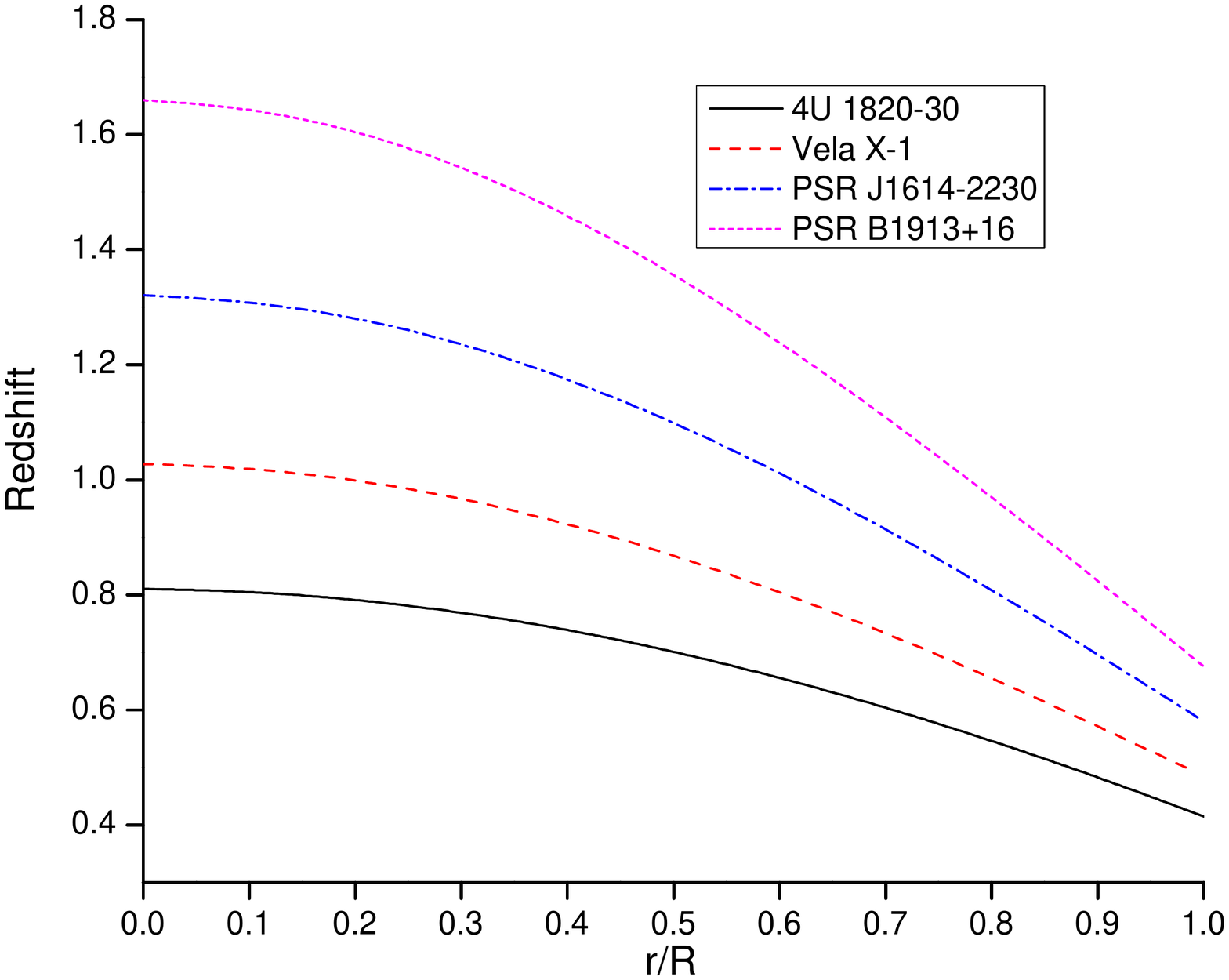}\includegraphics[width=5cm]{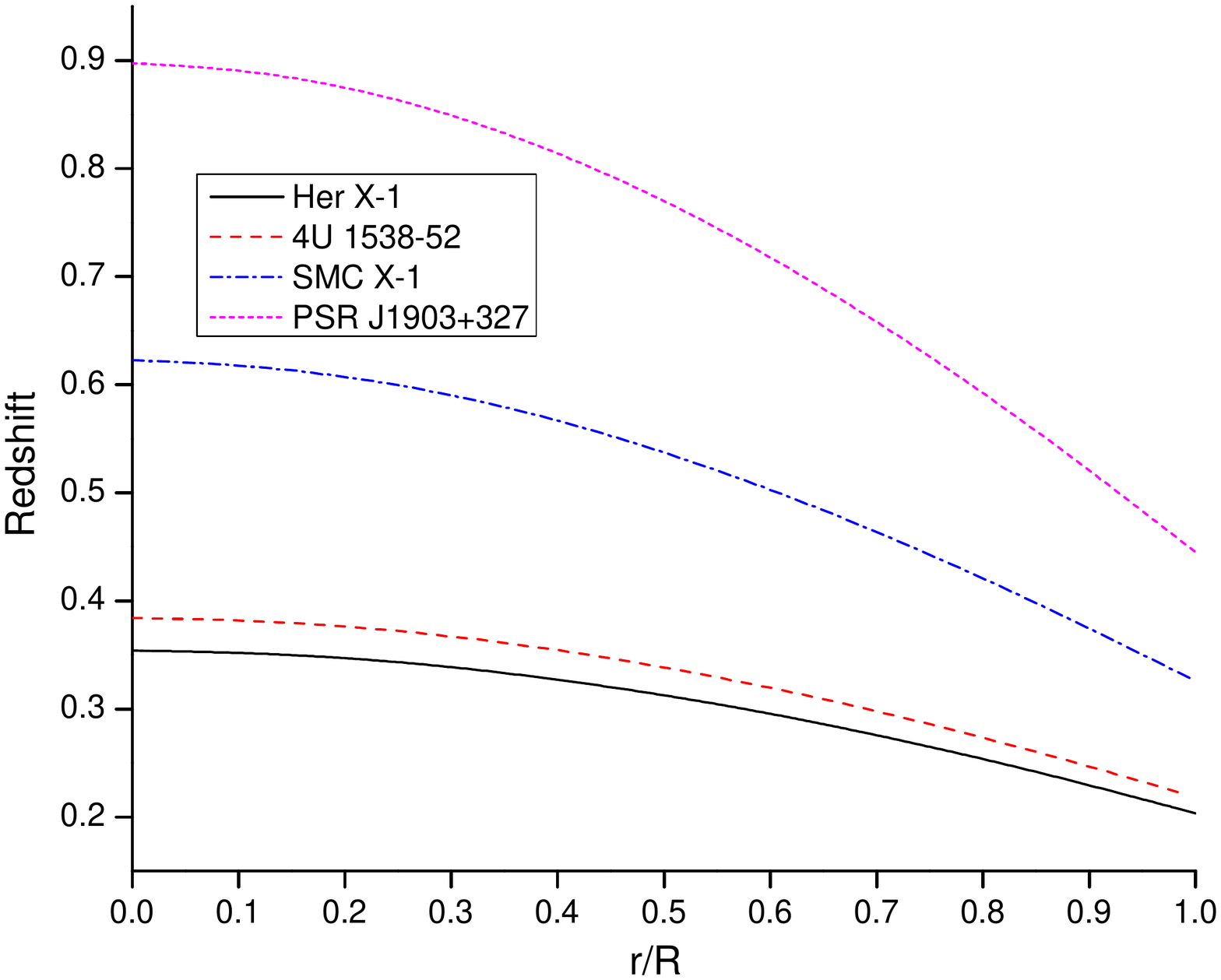}\includegraphics[width=5cm]{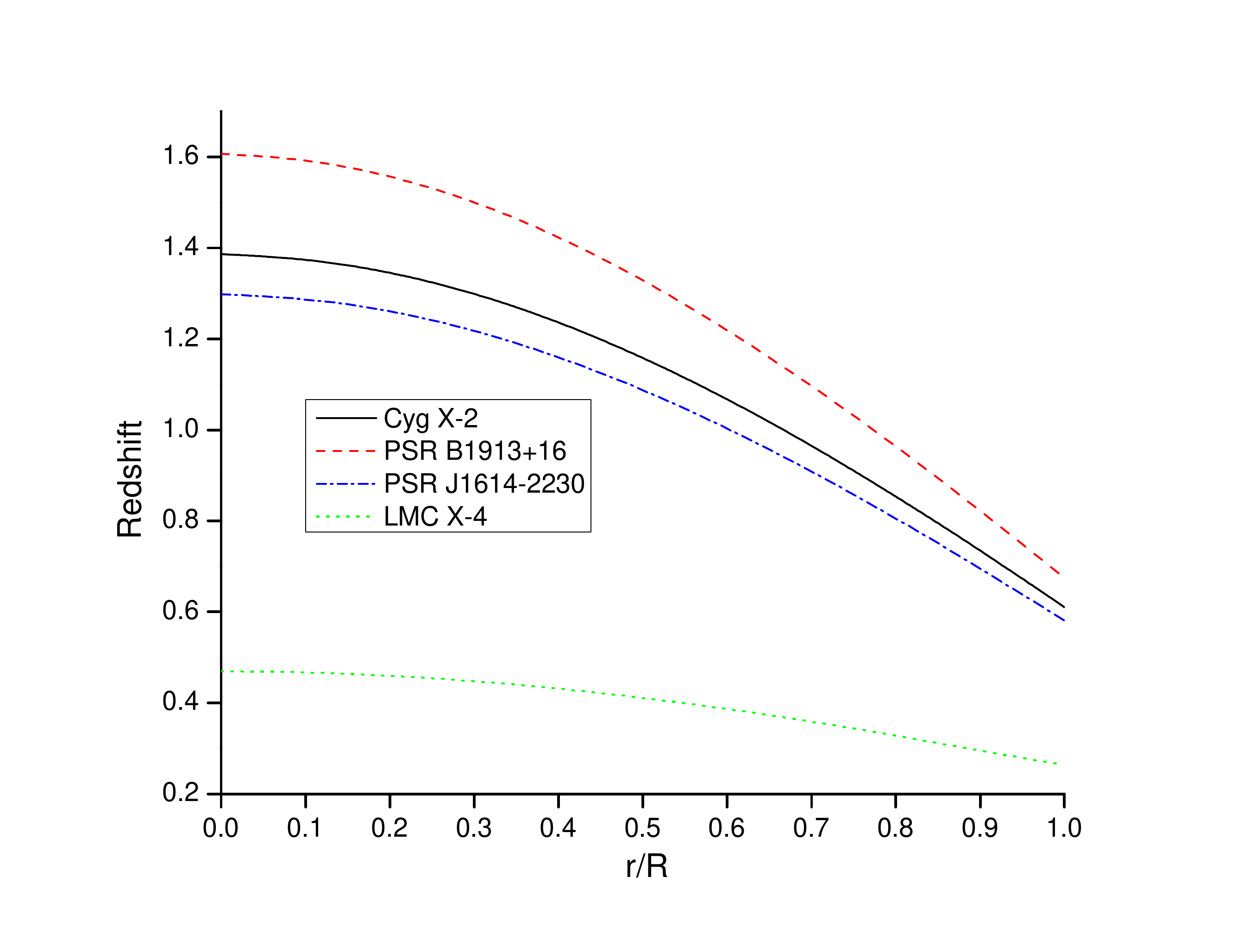}\\\includegraphics[width=5cm]{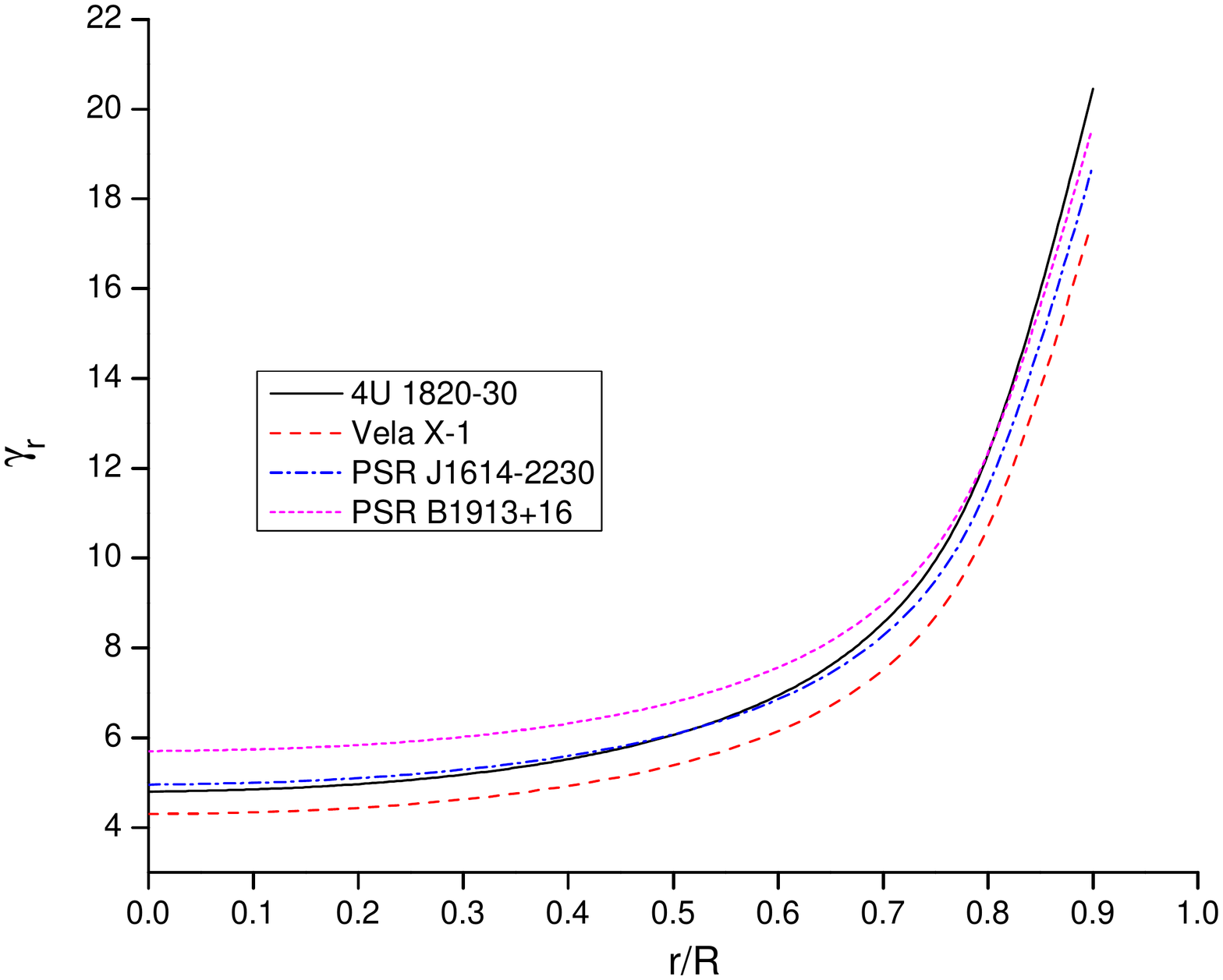}\includegraphics[width=5cm]{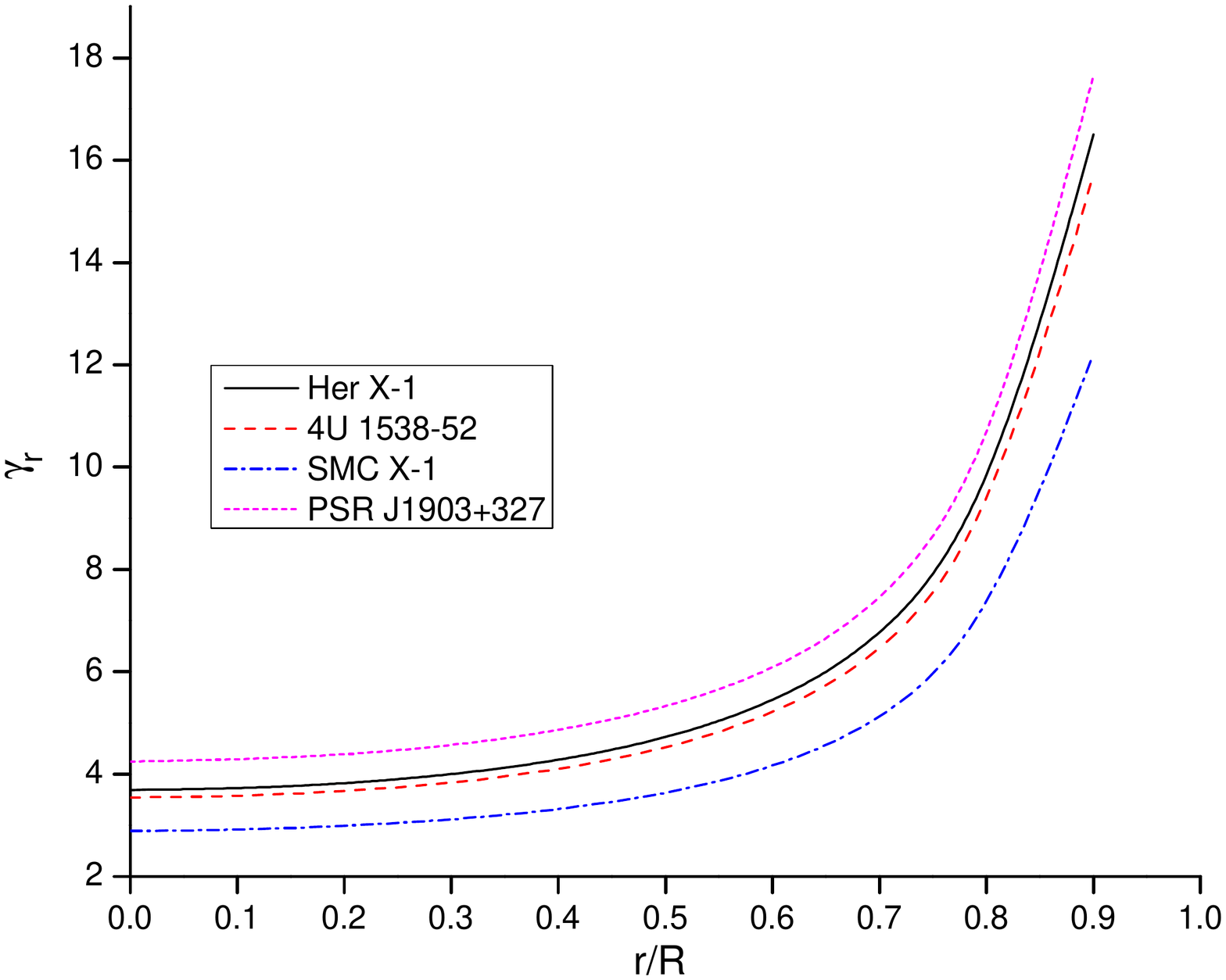}\includegraphics[width=5cm]{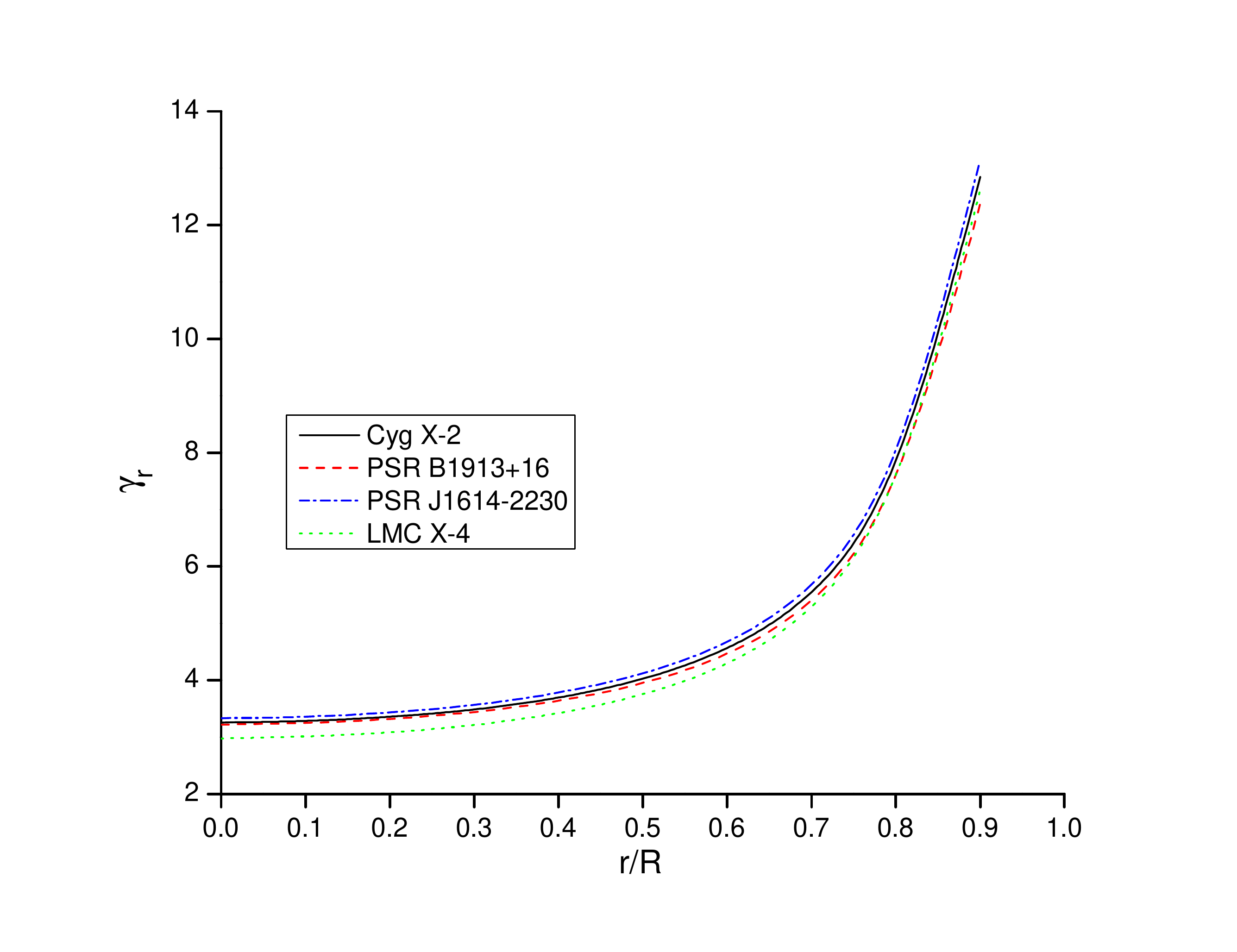}
 \caption{Behaviour of redshift and adiabatic index vs. fractional radius $ r/R $ for case I(first column), case II(second column) and case III(third column).For plotting this figure we have employed data set values of physical parameters and constants which are the same as used in Fig.\ref{f1}. }\label{f6}
 \end{center}
 \end{figure}

The stability of the relativistic as well as non relativistic compact object
likewise relies on the adiabatic index $ \gamma $. Heintzmann and
Hillebrandt \cite{Heintzmann} recommended that the adiabatic index must be more than $ \frac{4}{3} $ at all interior point of relativistic compact object. In other side the non-relativistic and isotropic case (Newtonian fluids), neutron spherical object system has no upper
mass limit for the adiabatic index $ \gamma > \frac{4}{3}$ \cite{Bondi}. So, the relativistic radial adiabatic index is defined by\cite{Chan} 
\begin{equation*}
\gamma_r = \Big(\dfrac{\rho + p_r}{p_r}\Big)\dfrac{dp_r}{d\rho}
\end{equation*}
Chan et al.\cite{Chan} obtained the collapsing condition for the ratio of specific heat for an anisotropic fluid distribution as 
\begin{equation*}
	\gamma \leq \frac{4}{3} \Bigg[1- max\Bigg( \dfrac{p_r-p_t}{r\mid p_r\mid}\Bigg)\Bigg]
\end{equation*}
From above expression, we can say that the unstable profile of $ \gamma $ depend on the anisotropic factor i.e, $ \gamma $ will increase when $ \Delta>0 $ and decrease when $ \Delta<0 $. Also it is possible that sign of $ \Delta $ may be change within the configuration. If it is happen then there is existence of both stable and unstable profile of  the object which could lead to fragmentation of the sphere\cite{Chan}.
We can see from Fig.\ref{f6} that the  value of radial adiabatic index is more than $ \frac{4}{3} $ at all interior points for each different compact
star model.
 \section{ Harrison–Zeldovich–Novikov stability criterion}
 For any solution representing stable compact stars, like PSR J1614-2230, Her X-1, Cgy X-2 etc., have to fulfill the stability criterion\cite{harrison,zeldovich}.  In this criterion, it is postulate that the any stellar configuration has an increasing mass with increasing central density, i.e. $ dM/d\rho_{c} > 0 $ represents stable configuration and vice versa. If the mass remains constant with increasing central density, i.e. $ dM/d\rho_{c} = 0 $ we get the turning point between stable and unstable region. For this model, we obtained $ M(R) $ and $ dM/d\rho_{c} $ as follows-
 \begin{figure}[h!]
 \begin{center}
 \includegraphics[width=5cm]{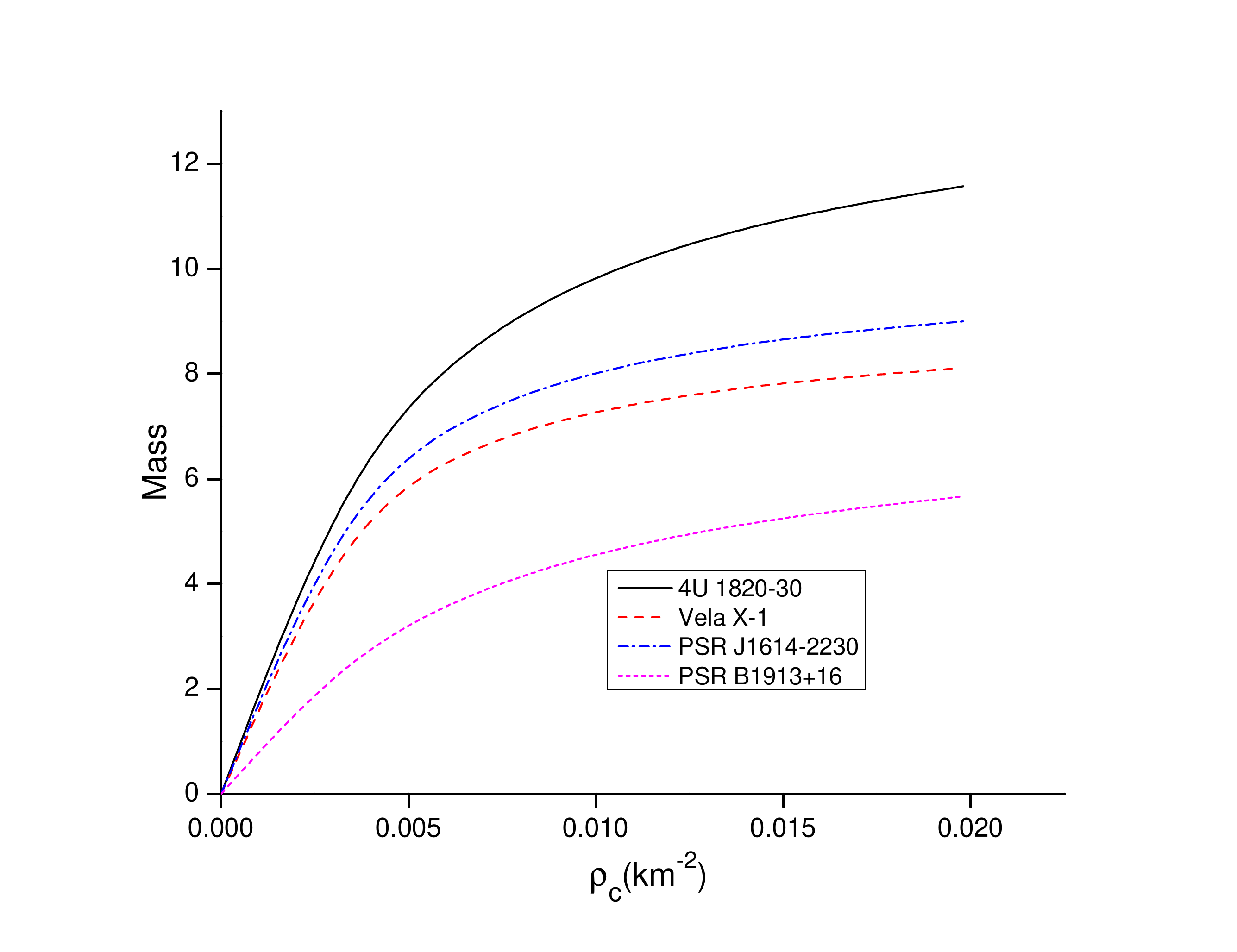}~\includegraphics[width=5cm]{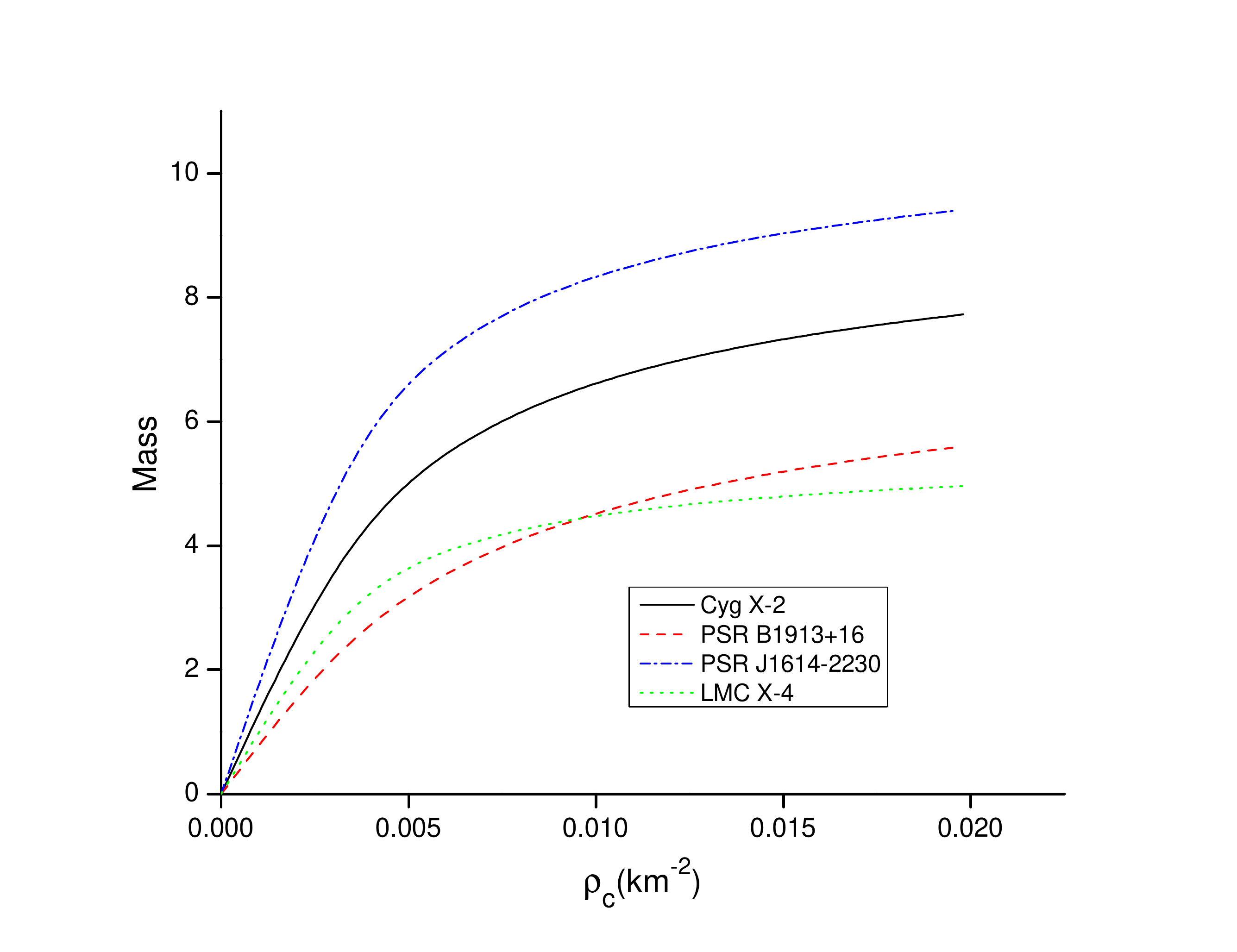}\includegraphics[width=5cm]{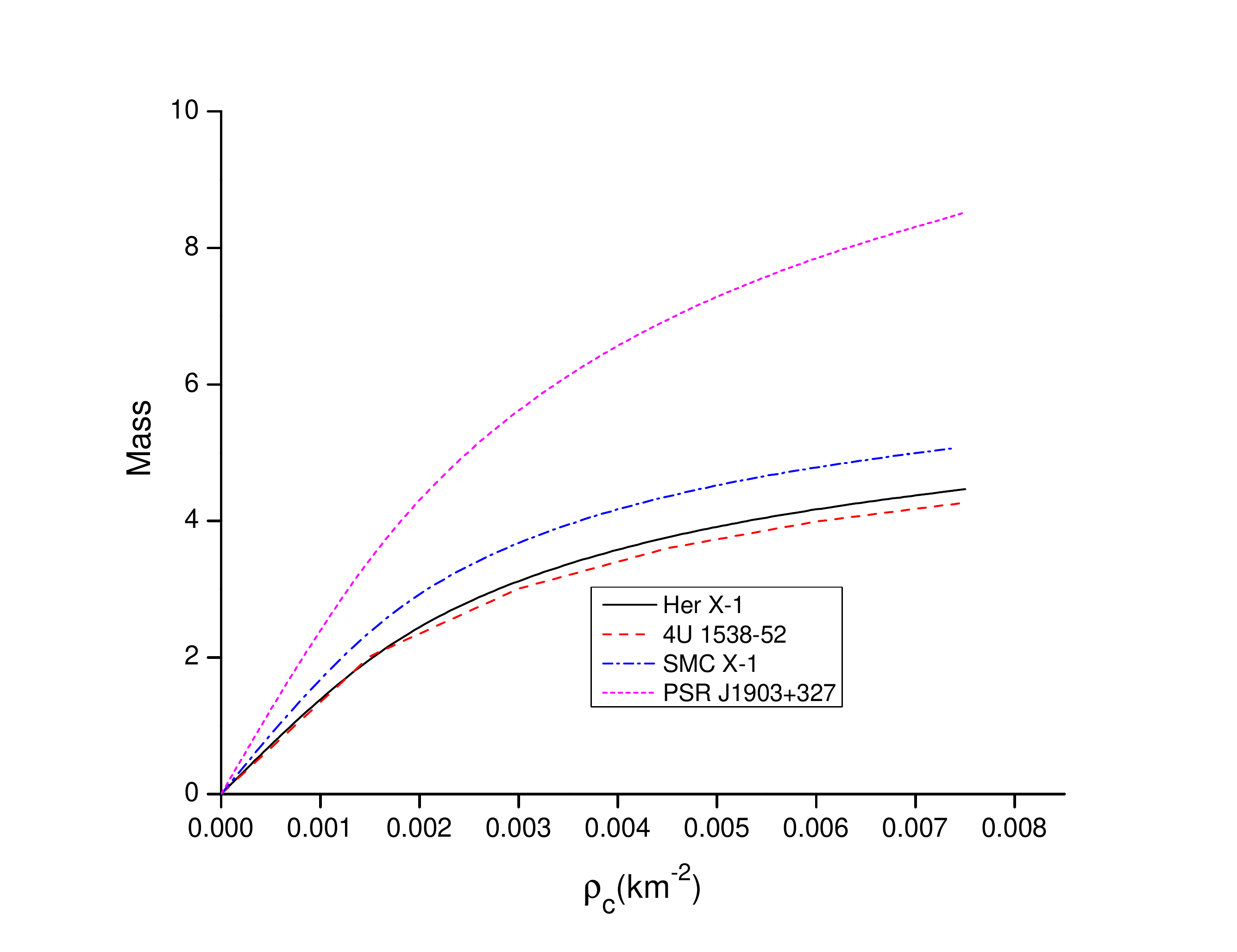}\\\includegraphics[width=5cm]{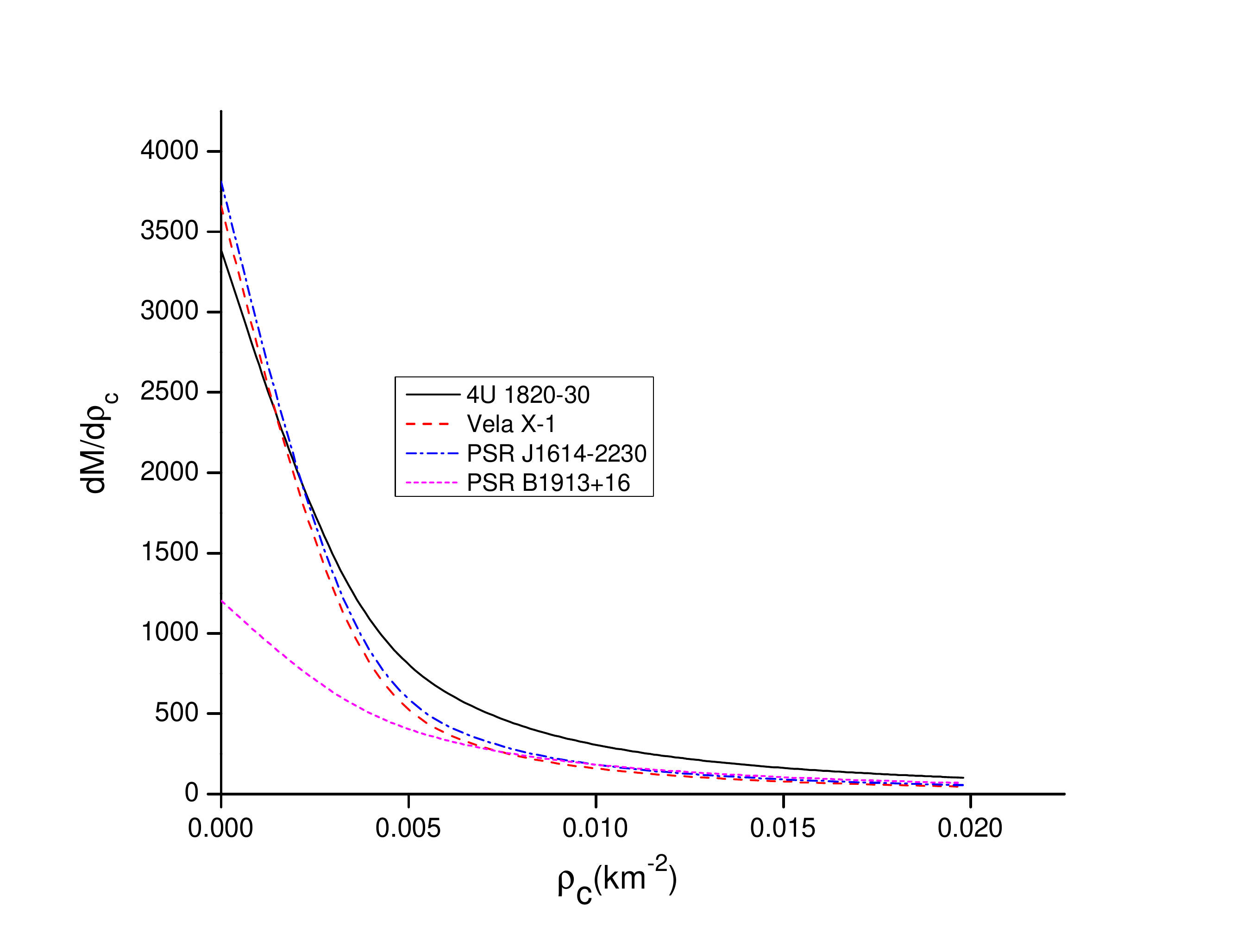}\includegraphics[width=5cm]{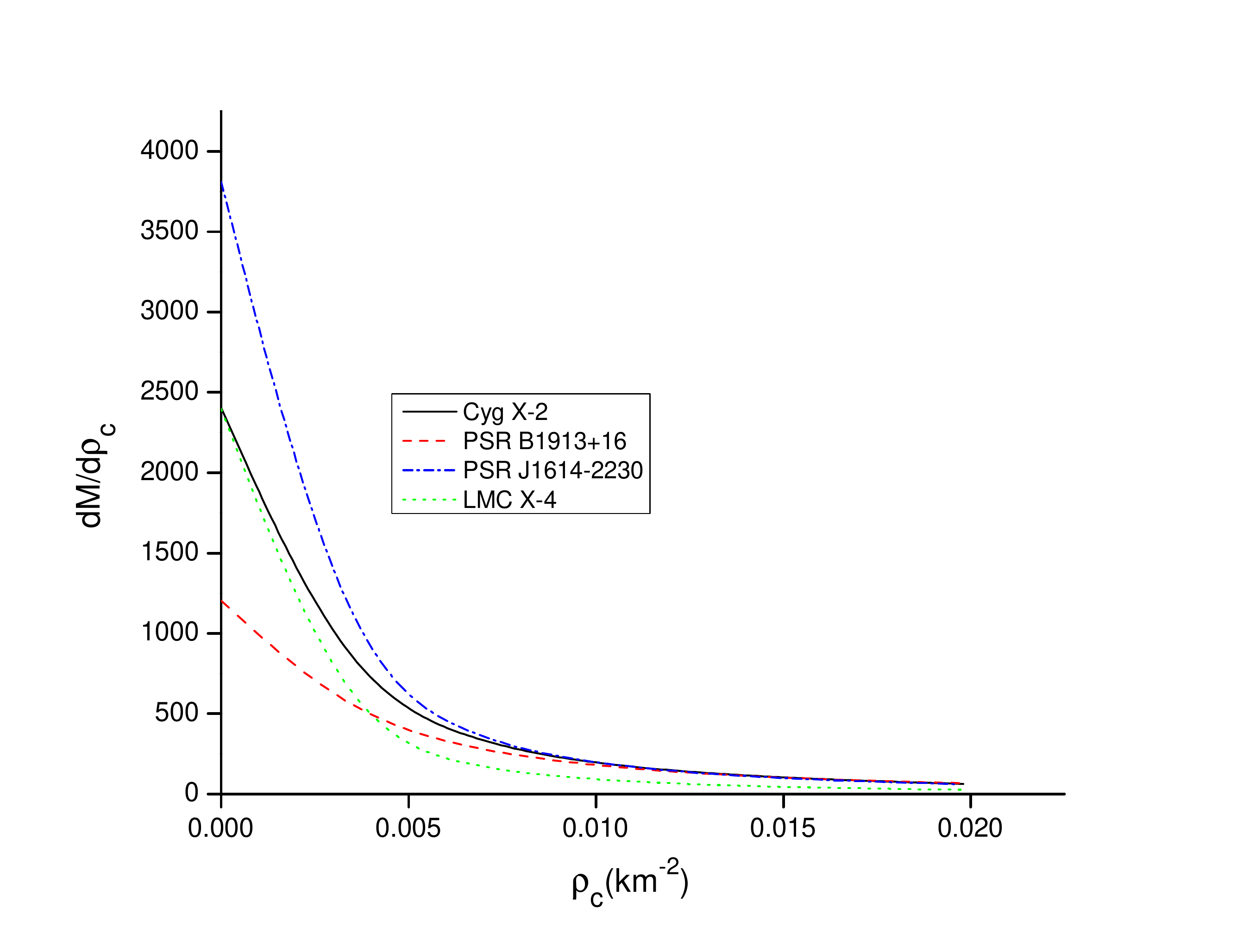}~\includegraphics[width=5cm]{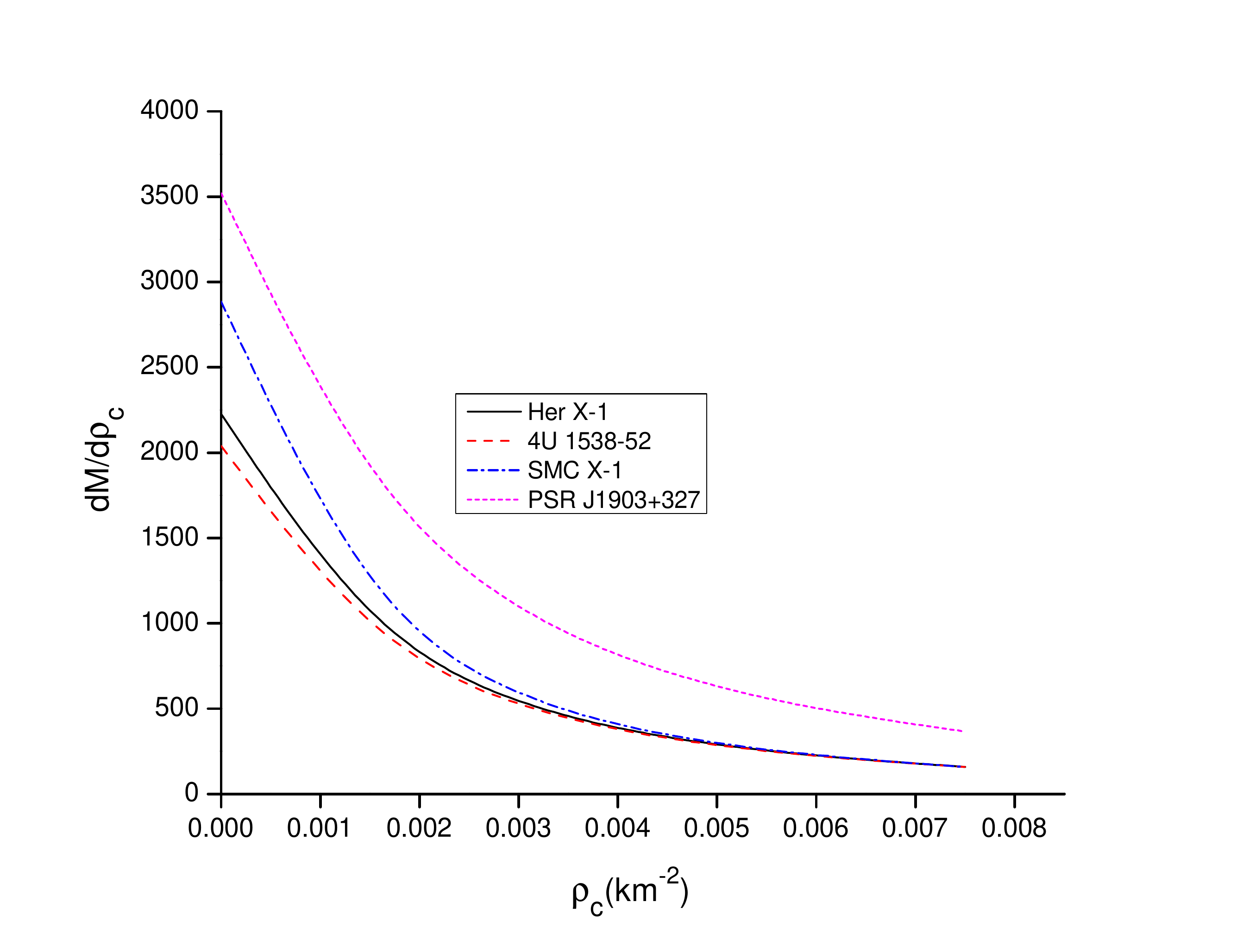}
 \caption{Behaviour of Mass vs. central density and $ dM/d\rho_{c} $ vs. central density for case I(first column), case II(second column) and case III(third column). For plotting this figure we have employed data set values of physical parameters and constants which are the same as used in Fig.\ref{f1}.}\label{f7}
 \end{center}
 \end{figure}
 \begin{equation}
 M(R)=\frac{4\,\pi\,\rho_c\, R^3\,(K-1)}{(3(K-1)+8\,\pi\,K\,\rho_c\,R^2)}~~~~~\textrm{and}~~~~\frac{dM}{d\rho_c}=\frac{12\,\pi\,R^3\,(K-1)^2}{(3(K-1)+8\,\pi\,K\,\rho_c\,R^2)^2}\label{41}
 \end{equation}
 Hence from Fig.\ref{f7}, we can conclude that presenting model represents static stable configuration.

\section{Conclusion}
In this article, we obtained a new class of well behaved anisotropic compact star models after prescribing Buchdahl metric potential and anisotropic factor. In the present stellar model, we demonstrate that the energy density $ \rho $, the radial pressure $ p_r  $ and the tangential pressure $ p_t $  are completely finite and
positive quantities within the stellar configuration, which is shown in Fig.\ref{f1}. So from this figure, we can say that our stellar system is completely free of any physical and geometric singularities. The anisotropy of the stellar model is represented in Fig.\ref{f2}, which exhibit, that
anisotropy increases as the radius increases. For instance, the anisotropy is minimum, i.e., zero at the origin and maximum on the surface of the stellar system. 
 The ratios of pressure and density  $ \dfrac{p_r}{\rho },$   $ \dfrac{p_t}{\rho },$ are monotonically decreasing towards the surface which is outlined in Fig. \ref{f2}. The
model is satisfy the following dominant energy conditions(see Fig.\ref{f5}) (i) strong energy condition(SEC) (ii) weak energy condition(WEC) and (iii) null energy condition(NEC). In current study the redshift is also decreasing from the center to surface (see Fig.\ref{f6}).  From the point of view of causality condition the stellar model is completely stable, because the
square of sound speed is less than 1 everywhere within the
star (Fig.\ref{f3}), besides there is no change in sign $ v^{2}_{r}- v^{2}_{t} $ and stability factor ($ v^{2}_{t}- v^{2}_{r} $) lies between -1 and 0 for stable configuration and 0 to 1 for unstable configuration (Fig.\ref{f3}).The static stable criterion also hold good by
our solutions(see Fig.\ref{f7}) Moreover, the relativistic adiabatic index $ \gamma_r $ is greater than $ \frac{4}{3} $ and monotonically increasing towards the surface. On the other hand, the 
compact stellar model is in equilibrium under three different
forces, namely  the gravitational
force $ F_g $, the hydrostatic force $ F_h $,  and the anisotropic force $ F_a $ (Fig.\ref{f4}). The latest one causes a repulsive force that counteracts the gravitational gradient, this is so because we are in the presence of a positive anisotropy factor $ \Delta_1\,,\Delta_2\,, \Delta_3 $ as can be seen in Fig \ref{f2}.
Based on physical requirements, we match the interior solution to an exterior vacuum Schwarzschild spacetime on the boundary surface at $ r = R, $ and from the comparison of
both side metrics, all constants are determined which are
listed in Table \ref{t1}. By using these constants in our investigation
for several analogue objects with similar mass and radii,
namely, PSR J1614-2230, PSR B1913+16, Cgy X-1, LMC X-4, Vela X-1, 4U 1820-30, 4U 1538-52, SMC X-1, PSR J1903+327 and Her X-1.  Whereas, Table \ref{t2} contains the central pressure, central and surface energy density is within the parameter outlined in Table \ref{t1}.

\end{document}